\documentclass[journal]{IEEEtran}  

\usepackage{graphicx}
\usepackage{amsmath}
\usepackage{mathtools}
\usepackage{amssymb}
\usepackage{amsfonts}
\usepackage{upgreek}
\usepackage{hyperref}
\usepackage{cite}
\usepackage{algorithm}
\usepackage{algorithmic}
\usepackage{mathdots}
\usepackage{bm}
\usepackage{textcomp}
\usepackage{cases}
\usepackage[font=footnotesize]{caption}
\usepackage[font=footnotesize]{subcaption}
\usepackage{color}
\usepackage{soul}
\usepackage[normalem]{ulem}
\usepackage{graphbox}
\usepackage{hyperref}
\usepackage{wasysym}
\usepackage{xcolor}
\usepackage{tikz}
\usepackage{pgfplots}
\pgfplotsset{compat=1.18}
\usetikzlibrary{intersections}
\usetikzlibrary{patterns}
\usepgfplotslibrary{fillbetween}
\usetikzlibrary{arrows.meta}
\usetikzlibrary{pgfplots.colormaps}
\usetikzlibrary{pgfplots.colorbrewer}
\usepackage{comment}
\usepackage{lipsum}
\usepackage{booktabs}
\usepackage{enumitem}

\newtheorem{definition}{Definition}
\newtheorem{lemma}{Lemma}
\newtheorem{remark}{Remark}
\newtheorem{theorem}{Theorem}
\newtheorem{assumption}{Assumption}

\newtheorem{example}{Example}

\newcommand{\R}{\mathbb{R}}
\newcommand{\Rpos}{\mathbb{R}_{\geq 0}}
\newcommand{\Rspos}{\mathbb{R}_{>0}}
\newcommand{\ru}{\overline{\rho}}
\newcommand{\rl}{\underline{\rho}}
\newcommand{\dru}{\dot{\bar{\rho}}}
\newcommand{\drl}{\dot{\underline{\rho}}}
\newcommand{\rank}{\mathrm{rank}}
\newcommand{\alphb}{\bar{\alpha} }

\newcommand{\alphbstr}{\bar{\alpha}^{\ast} }

\newcommand{\alphastr}{\alpha^{\ast} }
\newcommand{\dalphastr}{\dot{\alpha}^{\ast} }
\newcommand{\epialph}{\varepsilon_{\alpha}}
\newcommand{\epialphb}{\bar{\varepsilon}_{\alpha}}
\newcommand{\epialphbo}{\bar{\varepsilon}_{\alpha,1}}
\newcommand{\epialphbt}{\bar{\varepsilon}_{\alpha,2}}
\newcommand{\depialph}{\dot{\varepsilon}_{\alpha} }

\newcommand{\alphah}{\hat{\alpha} }

\newcommand{\ralph}{\rho_{\alpha}}
\newcommand{\dralph}{\dot{\rho}_{\alpha}}

\newcommand{\C}{\mathcal{C}}

\newcommand{\T}{\mathcal{T}}
\newcommand{\I}{\mathcal{I}}

\newcommand{\Hes}{\mathcal{H}}
\newcommand{\xb}{\bar{x}}
\newcommand{\Ob}{\bar{\Omega}}
\newcommand{\Ox}{\Omega_{x_{1}}}
\newcommand{\Oxs}{\Omega_{x_1}^s}
\newcommand{\Oxsp}{\Omega_{x_1}^{s^\prime}}
\newcommand{\Oz}{\Omega_z}
\newcommand{\Og}{\Omega_\nabla}

\newcommand{\xtil}{\tilde{x}}

\newcommand{\bh}{\bar{h}}
\newcommand{\ealph}{e_{\alpha}}
\newcommand{\ealphl}{\underline{e}_{\alpha}}
\newcommand{\ealphb}{\bar{e}_{\alpha}}
\newcommand{\dealph}{\dot{e}_{\alpha}}
\newcommand{\eh}{\hat{e}}
\newcommand{\etil}{\tilde{e}}
\newcommand{\lambdamin}{\lambda_{\textit{min}}}
\newcommand{\mch}{\mu_{\chi}}

\newcommand{\cl}{\mathrm{cl}}
\newcommand{\gradxalph}{\nabla_{x_1} \alpha}
\newcommand{\gradxtilalph}{\nabla_{\tilde{x}_1} \alpha}

\newcommand{\col}{\mathrm{col}}
\newcommand{\diag}{\mathrm{diag}}

\newcommand{\taum}{\tau_{\mathrm{max}}}

\newcommand{\epi}{\varepsilon}
\newcommand{\depi}{\dot{\varepsilon}}

\newcommand{\epibar}{\bar{\varepsilon}}

\title{\LARGE \bf
Low-Complexity Control for a Class of Uncertain MIMO Nonlinear Systems under Generalized Time-Varying Output Constraints 
\\[3pt]
(extended version)
}

\author{Farhad Mehdifar, Lars Lindemann, Charalampos P. Bechlioulis, and Dimos V. Dimarogonas
\thanks{This work is supported by ERC CoG LEAFHOUND, the KAW foundation, and the Swedish Research Council (VR).}
\thanks{F. Mehdifar and D. V. Dimarogonas are with the Division of Decision and Control Systems, KTH Royal Institute of Technology, Stockholm, Sweden.   {\tt\small mehdifar@kth.se; dimos@kth.se}}%
\thanks{Lars Lindemann is with Thomas Lord Department of Computer Science, University of Southern California, Los Angeles, CA, USA.
	{\tt\small llindema@usc.edu}}
\thanks{C. P. Bechlioulis is with the Division of Systems and Control of the Department of Electrical and Computer Engineering at University of Patras, Patra, Greece. {\tt\small chmpechl@upatras.gr}}%
}

\begin{document}

\maketitle
\thispagestyle{empty}
\pagestyle{empty}


\begin{abstract}
	This paper introduces a novel control framework to address the satisfaction of multiple time-varying output constraints in uncertain high-order MIMO nonlinear control systems. Unlike existing methods, which often assume that the constraints are always decoupled and feasible, our approach can handle coupled time-varying constraints even in the presence of potential infeasibilities. First, it is shown that satisfying multiple constraints essentially boils down to ensuring the positivity of a scalar variable, representing the signed distance from the boundary of the time-varying output-constrained set. To achieve this, a single consolidating constraint is designed that, when satisfied, guarantees convergence to and invariance of the time-varying output-constrained set within a user-defined finite time. Next, a novel robust and low-complexity feedback controller is proposed to ensure the satisfaction of the consolidating constraint. Additionally, we provide a mechanism for online modification of the consolidating constraint to find a least violating solution when the constraints become mutually infeasible for some time. Finally, simulation examples of trajectory and region tracking for a mobile robot validate the proposed approach. 
\end{abstract}



\section{Introduction}
\label{sec:intro}

In the last decade, the control of nonlinear systems under constraints has gained significant interest, driven by both practical necessities and challenging theoretical aspects. Constraints play a pervasive role in the design of controllers for practical nonlinear systems, often representing critical performance and safety requirements. Their violation can lead to performance deterioration, system damage, and potential hazards. Over the years, a variety of approaches have emerged to address different types of constraints within control systems, such as model predictive control, reference governors, control barrier functions, funnel control, prescribed performance control, and barrier Lyapunov functions as documented in \cite{mayne2014model,garone2017reference,ames2016control, tee2009barrier, tee2011control, ilchmann2002tracking, ilchmann2007tracking, bechlioulis2008robust, bechlioulis2014low}.

This work focuses on (closed-form) feedback control designs under time-varying output constraints, a crucial area in nonlinear control systems driven by the need to ensure tracking and stabilization performance as well as safety requirements \cite{tee2011control, ilchmann2002tracking, berger2021funnel, bechlioulis2014low, bechlioulis2008robust, theodorakopoulos2015low}. Existing closed-form feedback control approaches for addressing time-varying output constraints fall into three primary categories: Funnel Control (FC) \cite{ilchmann2002tracking,ilchmann2007tracking}, Prescribed Performance Control (PPC) \cite{bechlioulis2008robust,bechlioulis2014low}, and Time-Varying Barrier Lyapunov Function (TVBLF) methods \cite{tee2011control}. Typically, control designs based on FC, PPC, and TVBLF are commonly employed to achieve user-defined transient and steady-state performance for tracking and stabilization errors. These designs restrict the evolution of errors within user-defined time-varying funnels, serving as the sole output constraints. For instance, constraints of the form $-\rho_i(t) < e_i = x_i - x_i^d(t) < \rho_i(t)$ are frequently used for independent tracking errors, where $x_i$ signifies independent state variables, $x_i^d(t)$ denotes desired trajectories, and $\rho_i(t)$ signifies bounded, strictly positive time-varying functions modeling the evolving behavior of these constraints. To ensure the desired transient and steady-state performance of tracking errors, $\rho_i(t)$ is often chosen as a strictly positive exponentially decaying function that approaches a small neighborhood of zero \cite{bechlioulis2008robust,ilchmann2002tracking}.

In recent years, significant advancements have emerged in the utilization of FC, PPC, and TVBLF methods. These developments encompass a broad spectrum of applications, including the control of high-order systems \cite{bechlioulis2014low, theodorakopoulos2015low,  berger2018funnel, berger2021funnel, chowdhury2019funnel}, output feedback \cite{liu2020adaptive, dimanidis2020output}, multi-agent systems \cite{macellari2016multi, bechlioulis2016decentralized, liang2019prescribed , stamouli2020multi, mehdifar2020prescribed, mehdifar20222, lu2022fixed, LEE2022110276, min2023low, lee2023edge}. Additionally, there are works addressing considerations such as unknown control directions \cite{wang2015prescribed, zhang2019lowM,  zhang2021global}, control input constraints \cite{yong2020flexible, ji2021saturation, berger2024input, trakas2023robust, fotiadis2023input, hu2023novel, mishra2023approximation2}, actuator faults \cite{jin2017adaptive, zhao2023unifying}, discontinuous output tracking \cite{fotiadis2020prescribed}, event-triggered control \cite{ji2023event}, asymptotic tracking \cite{lee2019asymptotic, liu2022asymptotic, verginis2023asymptotic, min2022funnel}, and signal temporal logic specifications \cite{lindemann2021funnel, sewlia2022cooperative, liu2022compositional,zhou2022robust}, among others. Furthermore, researchers dedicated efforts to crafting specific designs for time-varying boundary functions of funnel constraints to guarantee finite/fixed time (practical) tracking and stabilization \cite{yin2019appointed, yin2020robust}, considering asymmetric funnel constraints \cite{jin2018adaptive}, introducing monotone tube boundaries to enhance control precision \cite{SHI2024111304}, and addressing compatibility between output and state constraints \cite{cao2023prescribed}. In the same direction, works on hard and soft constraints \cite{mehdifar2022funnel} and reach-avoid specifications \cite{das2024prescribed} have also been conducted. For recent surveys on FC and PPC see \cite{berger2023funnel,bu2022prescribed}.

While FC, PPC, and TVBLF approaches have demonstrated success in various applications and developments, they still face limitations when it comes to handling couplings between multiple time-varying constraints. These methods primarily focus on time-varying funnel constraints applied to independent states or error signals, which inherently remain decoupled from each other. In other words, these methods implicitly assume that the satisfaction of one funnel constraint does not impact the satisfaction of the others. To be more precise, the funnel constraints considered in FC, PPC, and TVBLF methods can be liked to time-varying box constraints in the system's output or error space \cite{berger2021funnel, bechlioulis2014low, jin2018adaptive}. In addition, it is also known that these methods are restricted to systems with the same number of inputs and outputs. However, in various practical applications, such as those involving general safety considerations \cite{glotfelter2017nonsmooth} and general spatiotemporal specifications \cite{lindemann2021funnel}, there is a need to address arbitrary and potentially coupled multiple time-varying output constraints. Consequently, it becomes crucial to develop control methodologies for uncertain nonlinear systems that can handle a more general class of time-varying output constraints. Recently, \cite{mehdifar2022funnel} proposed a low-complexity feedback control law under both hard and soft funnel constraints. Nevertheless, even when there are couplings between these hard and soft funnel constraints, the approach in \cite{mehdifar2022funnel} treats all hard constraints as independent funnels, adhering to the established conventions.  Additionally, \cite{das2024prescribed}, inspired by \cite{mehdifar2022funnel}, introduced funnel-based control for reach-avoid specifications but does not directly address couplings between multiple time-varying constraints.
 
In this paper, we present a novel feedback control law that aims at satisfying potentially coupled, time-varying output constraints for uncertain high-order MIMO nonlinear systems. Drawing inspiration from the approach introduced in \cite{lindemann2021funnel}, our control design revolves around consolidating all time-varying constraints into a carefully crafted single constraint. To ensure the satisfaction of this consolidating constraint, we introduce a new low-complexity robust control strategy inspired by \cite{bechlioulis2014low}. Notably, the approach does not rely on approximations or parameter estimation schemes to handle system uncertainties. Additionally, we demonstrate that by adaptively adjusting the consolidating constraint online, we can achieve a least violating solution for the closed-loop system when the constraints become infeasible during an unknown time interval.

Unlike existing FC, PPC, and TVBLF methods that mainly impose symmetric funnel constraints on system outputs, our approach includes both generic asymmetric funnel constraints and one-sided (time-varying) constraints on system outputs. This allows us to consider a more general range of spatiotemporal specifications. Furthermore, while the aforementioned control methods require all output constraints to be met initially, our method achieves convergence to the time-varying output-constrained set within a user-defined finite time, even if the constraints are not initially satisfied. Specifically, our control method ensures convergence to and invariance of the time-varying output-constrained set within the specified finite time. Overall, our results broaden the scope of feedback control designs for nonlinear systems, accommodating a wider range of time-varying output constraints. Notably, closed-form feedback control designs for reference tracking with prescribed performance and handling time-invariant output constraints in nonlinear systems become special cases of our results.

In connection with our methodology presented in this paper, related works in \cite{glotfelter2017nonsmooth, 10156245, molnar2023composing} share a common approach of constructing a single time-invariant Control Barrier Function (CBF) to satisfy multiple time-invariant constraints. It is worth noting that time-varying CBFs can also be employed for controlling nonlinear systems under time-varying output constraints, as studied in \cite{xu2018constrained, lindemann2018control, Safari2023Arxiv}. However, traditional control synthesis using the CBF concept typically necessitates precise knowledge of the system dynamics and involves solving an online Quadratic Programming problem, which may not be favorable in certain applications. In contrast, our work offers a computationally tractable (optimization-free) and robust (model-free) feedback control law.

The preliminary findings of this study were outlined in \cite{mehdifar2023control}, focusing exclusively on first-order nonlinear MIMO systems with a time-invariant output map. This paper builds upon the foundation laid in \cite{mehdifar2023control}, extending our research to encompass high-order nonlinear MIMO systems with a time-varying output map. This expansion broadens the range of time-varying constraints that our method can effectively handle. Notably, in contrast to the single funnel constraint employed in \cite{mehdifar2023control}, we utilize a one-sided consolidating constraint in this work, which further simplifies the controller design and tuning process. Additionally, this paper addresses the challenge of potential constraint infeasibilities, a consideration not addressed in \cite{mehdifar2023control}.

\textbf{Notations:} $\R^n$ is the real $n$-dimensional space and $\mathbb{N}$ is the set of natural numbers. $\Rpos$ and $\Rspos$ represent non-negative and positive real numbers. A vector $x \in \R^n$ is an $n \times 1$ column vector, and $x^\top$ is its transpose. The Euclidean norm of $x$ is $\|x\|$. The concatenation operator is $\col(x_i) \coloneqq [x_1^\top, \ldots, x_m^\top]^\top \in \R^{mn}$, where $x_i \in \R^n, i = \{1,\ldots,m\}$. The space of real $n \times m$ matrices is $\mathbb{R}^{n \times m}$. For a matrix $A \in \mathbb{R}^{n \times m}$, $A^\top$ is the transpose, $\lambdamin(A)$ is the minimum eigenvalue, and $\|A\|$ is the induced matrix norm. The operator $\diag(\cdot)$ constructs a diagonal matrix from its arguments. The absolute value of a real number is $|\cdot|$. For a set $\Omega$, $\partial \Omega$ is the boundary, and $\cl(\Omega)$ is the closure. $\otimes$ represents the Kronecker product and $\mathbf{0}_n \in \R^n$ and $\mathbf{1}_n \in \R^n$ are the vectors of zeros and ones, respectively. The set of $n$-times continuously differentiable functions is $\C^n$. $\I_i^j = \{i,\ldots,j\}$, is the index set, where $i,j \in \mathbb{N}$ and $i\le j$.

\section{Problem Formulation} 
\label{sec:problemformulation}

Consider a class of general high-order MIMO nonlinear systems described by the following dynamics:
\begin{equation} \label{eq:sys_dynamics_highorder}
	 \!\!\!\!\!\!\!\! \begin{cases}
	\dot{x}_i = f_i(t,\xb_i) + G_i(t,\xb_i) x_{i+1}, \; i \in \I_1^{r-1}, \\
	\dot{x}_r = f_r(t,\xb_r) + G_r(t,\xb_r) u, \\
		y = h(t,x_1),
	\end{cases} \!\!\!\!\!\!\!\!\!\!\!\!\!\!\!\!\!\!\!
\end{equation}
where $x_i \coloneqq [x_{i,1}, x_{i,2}, \ldots, x_{i,n}]^\top \in \R^{n}$, $\xb_i \coloneqq [x_1^\top, \ldots, x_i^\top]^\top \in \R^{ni}, i \in \I_1^r$, $r \in \mathbb{N}$, and $x \coloneqq \xb_r \in \R^{nr}$ is the state vector. Moreover, $u \in \R^n$ and  $y = [y_1, y_2, \ldots, y_m]^\top \in \R^{m}$ denote the control input and output vectors, respectively. In addition, $f_i: \Rpos \times  \R^{ni} \rightarrow \R^{n}, i \in \I_1^r$ denote the vectors of nonlinear functions that are locally Lipschitz in $\xb_i$ and piece-wise continuous in $t$. Moreover, $G_i: \Rpos \times \R^{ni} \rightarrow \R^{n \times n}, i \in \I_1^r$ stand for the control coefficient matrices whose elements are locally Lipschitz in $\xb_i$ and piece-wise continuous in $t$. Finally, $h: \Rpos \times \R^n \rightarrow \R^{m}$ is a $\C^2$ map in $x_1$ and $\C^1$ in $t$. In particular, let $h(t,x_1) = [h_1(t,x_1), h_2(t,x_1), \ldots, h_m(t,x_1)]^{\top}$, so that $y_i =h_i(t,x_1), i \in \I_1^m$. Let $x(t;x(0),u)$ denote the solution of the closed-loop system \eqref{eq:sys_dynamics_highorder} under the control law $u$ and the initial condition $x(0)$. For brevity in the notation, from now on we will use $x(t;x(0))$ instead of $x(t;x(0),u)$. Moreover, consider $x_1(t;x(0))$ as the partial solution of the closed-loop system \eqref{eq:sys_dynamics_highorder} with respect to states $x_1$ under the initial condition $x(0)$ and the control input $u$.

In this paper, we pose the following technical assumptions for \eqref{eq:sys_dynamics_highorder}. Note that these assumptions do not restrict the applicability of our results, as they are relevant to the high-order practical mechanical systems under consideration. 
\begin{assumption} \label{assum:uncertain_f}
	The functions $f_i(t,\xb_i), i \in \I_1^r$, are unknown and there exist continuous functions $\bar{f}_i:\R^{ni} \rightarrow \R^{n}, i \in \I_1^r$, with unknown analytical expressions such that $\|f_i(t,\xb_i)\| \leq \bar{f}_i(\xb_i)$, for all $t \geq 0$, and all $\xb_i \in \R^{ni}$. 
\end{assumption}

\begin{assumption} \label{assum:symm_g}
	The matrices $G_i(t,\xb_i), i \in \I_1^r$ are unknown and (A) there exist continuous functions $\bar{g}_i:\R^{ni} \rightarrow \R, i \in \I_1^r$, with an unknown analytical expression such that $\|G_i(t,\xb_i)\| \leq \bar{g}_i(\xb_i)$, $\forall t \geq 0$; (B) the symmetric components denoted by $G_i^s(t,\xb_i) \coloneqq \frac{1}{2} \left( G_i^\top(t,\xb_i) + G_i(t,\xb_i) \right), i \in \I_1^r$, are uniformly sign-definite with known signs. Without loss of generality, we assume that all $G_i^s(t,\xb_i)$ are uniformly positive definite, implying  the existence of strictly positive constants $\underline{\lambda}_i > 0, i \in \I_1^r$, such that $\lambdamin (G_i^s(t,\xb_i)) \geq \underline{\lambda}_i > 0$, for all $\xb_i \in \R^{ni}$ and all $t \geq 0$.
\end{assumption}

Part (B) of Assumption \ref{assum:symm_g} establishes a global controllability condition for \eqref{eq:sys_dynamics_highorder}. Furthermore, Assumptions \ref{assum:uncertain_f} and \ref{assum:symm_g} suggest that while the entries of $f_i(t,\xb_i)$ and $G_i(t,\xb_i)$, $i \in \I_1^r$, can grow arbitrarily large due to variations in $\xb_i$, they cannot do so as a result of increase in $t$. The following regularity assumptions solely pertain to the output map of \eqref{eq:sys_dynamics_highorder}.
\begin{assumption} \label{assum:output_map_jacob}
	There exist continuous functions $\varkappa_i: \R^n \rightarrow \R, i \in \I_1^3$, such that $\|J(t,x_1)\| \leq \varkappa_1(x_1)$, $\|J_{x_1}(t,x_1)\| \leq \varkappa_2(x_1)$, and $\|J_{t}(t,x_1)\| \leq \varkappa_3(x_1)$, where $J(t,x_1) \coloneqq \frac{\partial h(t,x_1)}{\partial x_1}$ is the Jacobian of the output map, $J_{x_1}(t,x_1) \coloneqq \frac{\partial}{\partial x_1} \left( J(t,x_1) \right)$ and $J_{t}(t,x_1) \coloneqq \frac{\partial}{\partial t} \left( J(t,x_1) \right)$.
\end{assumption}

\begin{assumption} \label{assum:output_map_elements}
	There exist continuous functions $\kappa_i,\bh_i: \R^n \rightarrow \R, i \in \I_1^m$, such that $|h_i(t,x_1)| \leq \bh_i(x_1)$ and $|\frac{\partial h_i(t,x_1)}{\partial t}| \leq \kappa_i(x_1)$, respectively.
\end{assumption}

Assumptions \ref{assum:output_map_jacob} and \ref{assum:output_map_elements} ensure that the entries of the Jacobian matrix $J(t,x_1)$ and their partial derivatives, as well as $h_i(t,x_1)$ and $\frac{\partial h_i(t,x_1)}{\partial t}$ for $i \in \I_1^m$, may grow arbitrarily with $x_1$ but not with $t$. Finally, although the upper-bound functions introduced in Assumptions \ref{assum:output_map_jacob} and \ref{assum:output_map_elements} are known and can be computed, they are used solely for the stability analysis in this paper and do not enter the proposed control design.

\begin{remark}
	Assumptions \ref{assum:output_map_jacob} and \ref{assum:output_map_elements} can be omitted if $h(t,x_1)$ in \eqref{eq:sys_dynamics_highorder} does not explicitly depend on time, i.e., $h(x_1)$, with $h(x_1)$ only requiring to be $\C^2$. Additionally, if $f_i(t,\xb_i)$ and $G_i(t,\xb_i), i \in \I_1^r$ in \eqref{eq:sys_dynamics_highorder}, are replaced by $f_i(\xb_i)$ and $G_i(\xb_i)$, then Assumptions \ref{assum:uncertain_f} and \ref{assum:symm_g} simplify to standard requirements. These requirements then only necessitate $f_i$ and $G_i$ to be locally Lipschitz and $G_i^s(\xb_i) = \frac{1}{2} \left( G_i^\top(\xb_i) + G_i(\xb_i) \right)$ to be positive definite for all $\xb_i \in \R^{ni}$. Moreover, If $r=1$ in \eqref{eq:sys_dynamics_highorder}, it suffices that $h(t,x_1)$ is $\C^1$ in both $x_1$ and $t$, and the requirements on $J_{x_1}$ and $J_t$ in Assumption \ref{assum:output_map_jacob} can be omitted.
\end{remark}

Let the outputs of \eqref{eq:sys_dynamics_highorder} be subject to the following class of time-varying constraints:
\begin{equation}\label{eq:outputconst}
	\!\!\! \rl_i(t) < h_i(t,x_1) < \ru_i(t), \;\; i \in  \I_1^m, \;\; \forall t \geq 0, \!
\end{equation}
where $\rl_i, \ru_i: \Rpos \rightarrow \R \cup \{\pm \infty\}, i \in \I_1^m$. We assume for each $i \in \I_1^m$, that \textit{at least} one of $\ru_i(t)$ and $\rl_i(t)$ is a bounded $\C^1$ function of time with a bounded derivative. In other words, we allow $\rl_i(t) = - \infty$ (resp. $\ru_i(t) = + \infty$) when $\ru_i(t)$ (resp. $\rl_i(t)$) is bounded for all $t \geq 0$. In this respect, \eqref{eq:outputconst} can either represent \textit{Lower Bounded One-sided} (LBO) time-varying constraints in the form of $\rl_i(t) < h_i(t,x_1)$, \textit{Upper Bounded One-sided} (UBO) time-varying constraints in the form of $h_i(t,x_1) < \ru_i(t)$, as well as (time-varying) \textit{funnel constraints} in the form of $\rl_i(t) < h_i(t,x_1) < \ru_i(t)$, for which both $\ru_i(t)$ and $\rl_i(t)$ are bounded. 

Without loss of generality, we assume that the first $p$ constraints in \eqref{eq:outputconst}, i.e., for $i \in \I_1^p, 0 \leq p \leq m$, are funnel constraints, $q$ LBO constraints are indexed by $i \in \I_{p+1}^{p+q}, 0 \leq q \leq m - p$ in \eqref{eq:outputconst}, and the remaining $m-p-q$ constraints represent UBO constraints for which $i \in \I_{p+q+1}^{m}$ in \eqref{eq:outputconst}. We make the assumption that each funnel constraint in \eqref{eq:outputconst} is well-defined, i.e., for every $i \in \I_1^p$, there exists a positive constant $\epsilon_i$ such that $\ru_i(t) - \rl_i(t) \geq \epsilon_i, \forall t\geq0$. This condition guarantees that the $p$ funnel constraints are \textit{independently feasible}.

\begin{remark} \label{rem:applicability}
	Note that the output constraints specified in \eqref{eq:outputconst} depend on $x_1$, which signifies the spatial coordinates (positions) of mechanical systems. Furthermore, as discussed in the introduction, while previous works primarily address $m = n$ decoupled funnel constraints by considering $y = x_1$, we account for $m \geq n$ generalized system outputs denoted as $y = h(t,x_1)$ in \eqref{eq:sys_dynamics_highorder}. This allows for possible couplings between different output constraints presented in \eqref{eq:outputconst}.
\end{remark}

We emphasize that, in this paper, the output map $h(t,x_1)$ is primarily employed to incorporate various types of constraints into the nonlinear dynamics described by \eqref{eq:sys_dynamics_highorder}. Specifically, we utilize $h(t,x_1)$ in conjunction with the time-varying functions $\rl_i(t)$ and $\ru_i(t)$ in \eqref{eq:outputconst} to represent various forms of spatiotemporal constraints for \eqref{eq:sys_dynamics_highorder}. Note that, $h(t,x_1)$ is not necessarily related to the available measurements of the system. As we will further elaborate, we assume that the states of \eqref{eq:sys_dynamics_highorder} are available for the measurement and will be utilized for designing the control input $u(t,x)$.

\begin{definition}
	An output of \eqref{eq:sys_dynamics_highorder}, $y_i = h_i(t, x_1)$, is called \textit{separable} if it can be expressed as the sum of a component solely varying with time and another component dependent solely on $x_1$, i.e., $h_i(t, x_1) = h_i^{x_1}(x_1) + h_i^t(t)$. Otherwise, it is called \textit{inseparable}.
\end{definition}

Note that, if $y_i = h_i(t, x_1)$ is separable, the time-varying term $h_i^t(t)$ can be incorporated into the time-varying bounds $\ru_i(t)$ and $\rl_i(t)$ in \eqref{eq:outputconst}. Thus, one can consider the time-independent output $h_i^{x_1}(x_1)$ instead of $h_i(t, x_1)$ in \eqref{eq:sys_dynamics_highorder}. For example, let \eqref{eq:sys_dynamics_highorder} model a moving vehicle with position $[x_{1,1}, x_{1,2}]^\top$, and the objective is to track a $\C^1$ reference trajectory characterized by $x_1^d(t) = [x_{1,1}^d(t), x_{1,2}^d(t)]^\top$ under the funnel constraints $-\rho_i(t) < h_i(t, x_1) = x_{1,i} - x_{1,i}^d(t) < \rho_i(t), i \in \I_1^2$, where $\rho_i(t)$ are positive functions decaying to a small neighborhood of zero (tracking under a prescribed performance, see \cite{bechlioulis2008robust}). Here $h_i(t, x_1), i\in \I_1^2$ represent the tracking errors and the constraints can be written as $x_{1,i}^d(t) - \rho_i(t) < x_{1,i} < \rho_i(t) + x_{1,i}^d(t), i\in \I_1^2$. On the other hand, if the vehicle is tasked with reaching a moving target defined by its position $x_1^d(t)$, then a single constraint can be imposed: $-\rho(t) < h(t, x_1) = (x_{1,1} - x_{1,1}^d(t))^2 + (x_{1,2} - x_{1,2}^d(t))^2 < \rho(t)$, where $\rho(t)$ is a positive function decaying to a neighborhood of zero and $h(t, x_1)$ represents the squared distance error, which has inseparable time-varying terms. From this observation, a separable output can be regarded as equivalent to a time-independent output map. Thus, without loss of generality, we will use $h_i(t,x_1)$ to denote an inseparable output map in the sequel.

Finally, let us define the output constrained set $\Ob(t)$ based on \eqref{eq:outputconst} as:
\begin{equation} \label{eq:omeg_bar_t}
	\Ob(t) \coloneqq \{ x_1 \in \R^n \mid  \rl_i(t) < h_i(t,x_1) < \ru_i(t), i \in \I_1^m\}.
\end{equation}

\textbf{Objective:} In this paper, our goal is to design a  low-complexity continuous robust feedback control law $u(t,x)$ for \eqref{eq:sys_dynamics_highorder} such that $x_1(t;x(0))$ satisfies the time-varying output constraints \eqref{eq:outputconst} $\forall t > T \geq 0$, where $T$ is a user-defined finite time after which the output constraints are satisfied for all time (i.e., $x_1(t;x(0)) \in \Ob(t), \forall t > T \geq 0$). Note that this problem  reduces to establishing only forward invariance of $\Ob(t)$ for all $t \geq 0$, if $x_1(0) \in \Ob(0)$ ($T = 0$). On the other hand, having $x_1(0) \notin \Ob(0)$ indicates establishing: (i) finite time convergence to $\Ob(t)$ at $t=T$, and (ii) ensuring forward invariance of $\Ob(t)$, for all $t > T$. Furthermore, when $\Ob(t)$ becomes infeasible (empty) for an unknown time interval, we aim to enhance the control scheme such that $u(t,x)$ drives the closed-loop system trajectory towards a \textit{least violating solution} for \eqref{eq:sys_dynamics_highorder} under \eqref{eq:outputconst} (see Section \ref{sec:const_infeasibility} for more details).

\section{Main Results}

In this section, inspired by \cite{lindemann2021funnel}, we first introduce a novel scalar variable, which is the signed distance from the boundary of the time-varying output constrained set \eqref{eq:omeg_bar_t}. This variable serves as a metric of both feasibility and satisfaction of the output constraints. Next, we propose a robust and low-complexity controller design for \eqref{eq:sys_dynamics_highorder}, which ensures the ultimate positivity of the aforementioned variable. This, in turn, leads to the satisfaction of the output constraints.

\subsection{Satisfying Constraints using a Scalar Variable}
\label{subsec:scalar_metric}

Notice that the $m$ output constraints in \eqref{eq:outputconst} can be re-written in the following format:
\begin{subequations}\label{eq:predicate_const_rep}
	\begin{align} 
		&\resizebox{.93\hsize}{!}{$\!\!\!\!\! \begin{cases}
				\psi_{2i-1}(t,x_1) = h_i(t,x_1) - \rl_i(t) > 0, \; \text{(funnel constraints)} \\
				\psi_{2i}(t,x_1) = \ru_i(t) - h_i(t,x_1) > 0,  \;  i \in \I_1^p
			\end{cases}$} \!\!\!\!\!\!\!\!\!\!\!\! \label{eq:predicate_funnel_rep} \\
		&\resizebox{.93\hsize}{!}{$\!\!\!\!\! \begin{cases}
				\psi_{i}(t,x_1) = h_j(t,x_1) - \rl_j(t) > 0, \quad \text{(LBO constraints)} \\
				\; i \in \I_{2p+1}^{2p+q}, \; j \in \I_{p+1}^{p+q}, \\
				\psi_{i}(t,x_1) = \ru_j(t) - h_j(t,x_1) > 0, \quad \text{(UBO constraints)} \\ 
				\; i \in \I_{2p+q+1}^{m+p}, \;  j  \in \I_{p+q+1}^{m}.
			\end{cases}$} \!\!\!\!\!\!\!\!\!\!\!\! \label{eq:predicate_onesided_rep}   
	\end{align}
\end{subequations}
Now, without loss of generality, consider all these $m+p$ constraints in \eqref{eq:predicate_const_rep} as:
\begin{equation} \label{eq:psi_constr}
	\psi_i(t,x_1) > 0, \quad i \in \I_{1}^{m+p},
\end{equation}
where $\psi_i: \Rpos \times \R^{n} \rightarrow \R$ are $\C^2$ in $x_1$ and $\C^1$ in $t$. As a result, one can re-write \eqref{eq:omeg_bar_t} as: 
\begin{equation} \label{eq:omega_y}
	\Ob(t) = \{ x_1 \in \R^n \mid \psi_i(t,x_1) > 0, \forall i \in \I_1^{m+p} \}.
\end{equation}

\begin{figure}[!tbp]
	\centering
	\begin{subfigure}[t]{0.45\linewidth}
		\centering
		\includegraphics[width=\linewidth]{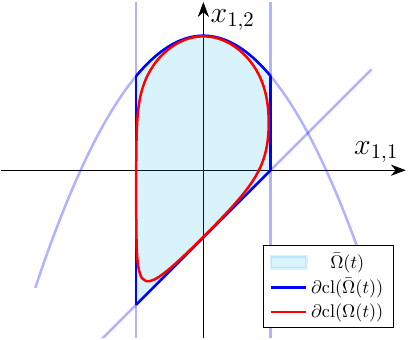}
		\caption{}
		\label{fig:omega_y_bar_and_Omega_y}
	\end{subfigure}%
	~
	\begin{subfigure}[t]{0.45\linewidth}
		\centering
		\includegraphics[width=\linewidth]{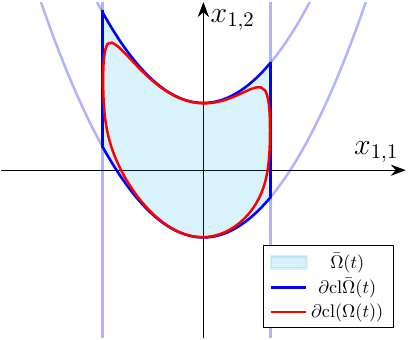}
		\caption{}
		\label{fig:condition II}
	\end{subfigure}
	\begin{subfigure}[t]{0.90\linewidth}
		\centering
		\includegraphics[width=\linewidth]{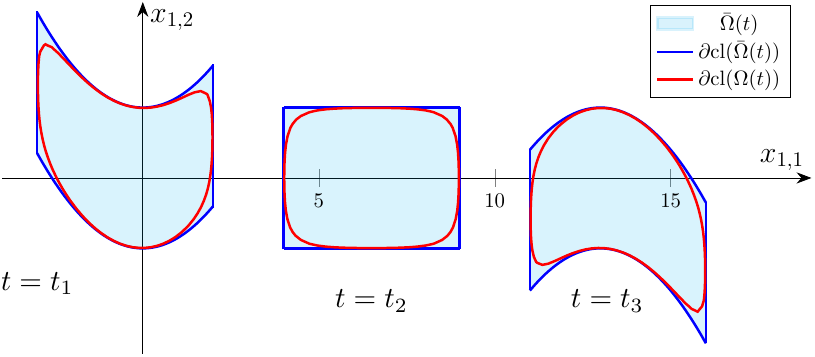}
		\caption{\vspace{-0.1cm}}
		\label{fig:time-varying_output}
	\end{subfigure}%
	\caption{Snapshots of $\Ob(t)$ and its corresponding inner-approximation under \eqref{smooth_alph} for three different examples. \vspace{-0.4cm}}
	\label{fig:examples}
\end{figure}

Now, define the scalar function $\alphb: \Rpos \times \R^n \rightarrow \R$, as:
\begin{equation} \label{eq:metric}
	\alphb(t,x_1) \coloneqq \min \{ \psi_1(t,x_1), \ldots , \psi_{m+p}(t,x_1) \},
\end{equation}
where $\alphb(t,x_1)$ represents the signed (minimum) distance from $\partial \cl(\Ob(t))$. In this respect, one can re-write \eqref{eq:omega_y} as the zero super level set of $\alphb(t,x_1)$:
\begin{equation} \label{eq:omega_alpha_bar}
	\Ob(t) = \{x_1 \in \R^n \mid \alphb(t,x_1) > 0 \}.
\end{equation}
Note that if $\alphb(t^{\prime},x_1)<0$, then \textit{at least} one constraint is not satisfied at $t = t^{\prime}$, while $\alphb(t,x_1)>0, \forall t \geq 0$ means that all constraints are satisfied for all time. Owing to the $\min$ operator in \eqref{eq:metric}, in general, $\alphb(t,x_1)$ is a continuous but nonsmooth function; therefore, to facilitate the controller design and stability analysis, we will consider the smooth under-approximation of $\alphb(t,x_1)$ using the log-sum-exp function \cite{gilpin2020smooth}:
\begin{subequations}\label{smooth_alph}
\begin{align}
	\alpha(t,x_1) &\coloneqq -\frac{1}{\nu} \ln \Big( \sum_{i=1}^{m+p} e^{- \nu  \, \psi_i(t,x_1)} \Big) \label{smooth_alph_def} \\
	&\leq \alphb(t,x_1) \leq \alpha(t,x_1) + \frac{1}{\nu} \ln(m+p), \label{smooth_alph_ineq}
\end{align}
\end{subequations}
where $\nu>0$ is a tuning coefficient whose larger values give a closer (under) approximation (i.e., $\alpha(t,x_1) \rightarrow \alphb(t,x_1)$ as $\nu \rightarrow \infty$). Note that, $\alpha(t,x_1)$ provides the signed distance from the boundary of a \textit{smooth inner-approximation} of $\cl(\Ob(t))$. Therefore, ensuring $\alpha(t,x_1) > 0, \forall t \geq 0$ guarantees $\alphb(t,x_1) > 0, \forall t \geq 0$ and thus the satisfaction of \eqref{eq:psi_constr} (equivalently \eqref{eq:outputconst}). Define $\Omega(t) \subset \Ob(t)$ as the smooth inner-approximation of the set $\Ob(t)$, given by:
\begin{equation} \label{eq:omega_alpha}
	\Omega(t) \coloneqq \{x_1 \in \R^n \mid \alpha(t,x_1) > 0 \}.
\end{equation} 
Note that we have $x_1 \in \Omega(t) \Rightarrow x_1 \in \Ob(t)$, and when $\Ob(t)$ is bounded, then $\Omega(t)$ is also bounded. Fig.~\ref{fig:examples} depicts snapshots of $\Ob(t)$ and $\Omega(t)$ with $\nu = 2$ in \eqref{smooth_alph} for the following examples:

\begin{example} \label{ex:concave_example}
	Consider $h(x_1) = [h_1(x_1), h_2(x_1), h_3(x_1)]^{\top}$, where $h_1(x_1) = x_{1,1}$, $h_2(x_1) = - x_{1,1} + x_{1,2} $, and $h_3(x_1) = 0.3 x_{1,1}^2 + x_{1,2}$ and let the output constraints be $\rl_1(t) < h_1(x_1) < \ru_1(t)$ (funnel constraint), $\rl_2(t) < h_2(x_1)$ (LBO constraint), and $h_3(x_1) < \ru_3(t)$ (UBO constraint), respectively. Fig.~\ref{fig:omega_y_bar_and_Omega_y} depicts a snapshot of the time-varying output constrained set and its smooth inner approximation, for which $-\rl_1(t)=\ru_1(t) =2$, $\rl_2(t) = -2$, and $ \ru_3(t)= 4$.  
\end{example}

\begin{example}\label{ex:independent_funnels_example}
	Consider $h(x_1) = [h_1(x_1), h_2(x_1)]^{\top}$, with $h_1(x_1) = x_{1,1}$ and $h_2(x_1) = 0.3 x_{1,1}^2 - x_{1,2}$ and let the output constraints be $\rl_1(t) < h_1(x_1) < \ru_1(t)$ and $\rl_2(t) < h_2(x_1) < \ru_2(t)$ (two funnel constraints), respectively. Fig.~\ref{fig:condition II} depicts a snapshot of the time-varying output-constrained set and its smooth inner-approximation, for which $\rl_1(t)=-3, \ru_1(t) =2$, and $-\rl_2(t) = \ru_2(t) = 2$.
\end{example} 

\begin{example} \label{ex:independent_funnels_time-varying_example}
	Consider the constraints of Example 2, however, this time we modify the second output such that $h_2(t,x_1) = c_1(t) (x_{1,1} - o_1(t))^2 - x_{1,2}$, where $c_1(t)$ and $o_1(t)$ are bounded continuously differentiable time-varying functions. Fig.~\ref{fig:time-varying_output} depicts three snapshots of the time-varying output-constrained set and its smooth inner-approximations, for which $\rl_1(t_1)=-3, \ru_1(t_1) =2, \rl_1(t_2)= 4, \ru_1(t_2) = 9, \rl_1(t_3)=11, \ru_1(t_3) =16$, and $-\rl_2(t) = \ru_2(t) = 2, \forall t \in \{t_1, t_2, t_3\}$, where $t_1<t_2<t_3$. Moreover, $c_1(t_1) = 0.3, c_1(t_2) = 0, c_1(t_3) = -0.3$ and $o_1(t_1) = 0, o_1(t_2) = 6, o_1(t_3) = 13$. Note that different from Example \ref{ex:independent_funnels_example}, $o_1(t)$ and $c_1(t)$ in $h_2(t,x_1)$ can contribute in shifting and changing the boundaries of the time-varying constrained set simultaneously at different time instances.
\end{example}
\begin{assumption} \label{assum:coercive_alphabar}
The function $-\alphb(t,x_1)$ is coercive (radially unbounded) in $x_1$ and uniformly in $t$, i.e., $-\alphb(t,x_1) \rightarrow +\infty$ as $\|x_1\| \rightarrow +\infty, \forall t \geq 0$. 
\end{assumption}

Note that, the focus of this work is the satisfaction of \eqref{eq:outputconst}. On the other hand, it is also essential to design $u(t,x)$ such that $\|x_i(t)\|, i =\I_1^r$ remain bounded $\forall t\geq 0$. In this respect, if the output-constrained set $\Ob(t)$ is well-posed (i.e., it is bounded), the satisfaction of the constraints inherently leads to the boundedness of $\|x_1(t)\|$. Assumption \ref{assum:coercive_alphabar} serves as a necessary and sufficient condition for ensuring the boundedness of $\Ob(t)$ (and $\Omega(t)$) for all $t \geq 0$. The following lemma establishes this. 

\begin{lemma} \label{lem:Omegab_bounded}
	Under Assumption \ref{assum:coercive_alphabar}, $\Ob(t)$ (resp. $\Omega(t)$) is a bounded set for all $t \geq 0$. 
\end{lemma}
\begin{IEEEproof}
	See Appendix \ref{appen:proof_lemma_Omegab_bounded}.
\end{IEEEproof}

Note that, Assumption \ref{assum:coercive_alphabar} implies that for any time instant $-\alphb(t,x_1)$ should approach $+\infty$ along \textit{any path} within $\R^n$ on which $\|x_1\|$ tends to infinity. Define $h_f(t,x_1) \coloneqq \col(h_i(t,x_1)) \in \R^p, i \in \I_1^p$, $h_{\mathrm{L}}(t,x_1) \coloneqq \col(h_i(t,x_1)) \in \R^q, i \in \I_{p+1}^{p+q}$, and $h_{\mathrm{U}}(t,x_1) \coloneqq \col(h_i(t,x_1)) \in \R^{m-p-q}, i \in \I_{p+q+1}^{m}$, as the stacked vectors of system outputs associated with funnel, LBO, and UBO constraints in \eqref{eq:outputconst}, respectively. The following lemma provides explicit conditions on $h_i(t,x_1), i \in \I_1^m$, to ensure that $-\alphb(t,x_1)$ (resp. $-\alpha(t,x_1)$) is coercive. 

\begin{lemma}\label{lem:alphb_coercive_h}
	The function $-\alphb(t,x_1)$ (resp. $-\alpha(t,x_1)$) is coercive in $x_1$ for all $t\geq 0$ if and only if, for each time instant $t$, at least one of the following conditions holds: 
	\begin{itemize}
		\item[(I)] $\|h_f(t,x_1)\| \rightarrow +\infty$;
		\item[(II)] one or more elements of $h_{\mathrm{L}}(t,x_1)$ approaches $-\infty$;
		\item[(III)] one or more elements of $h_{\mathrm{U}}(t,x_1)$ approaches $+\infty$;
	\end{itemize} 
	along any path in $\R^n$ as $\|x_1\| \rightarrow +\infty$.
\end{lemma}
\begin{IEEEproof}
	See Appendix \ref{appen:proof_lemma_alphb_coercive_h}.
\end{IEEEproof}

In Example \ref{ex:concave_example}, we observe that $h_f(t,x_1) = h_1(x_1)$, $h_L(t,x_1) = h_2(x_1)$, $h_U(t,x_1) = h_3(x_1)$. It can be verified that the condition in Lemma \ref{lem:alphb_coercive_h} is satisfied along any path in $\R^2$ as $\|x_1\| \rightarrow +\infty$. Hence, $-\alphb(t,x_1)$ (resp. $-\alpha(t,x_1)$) is coercive, implying that $\Ob(t)$ (resp. $\Omega(t)$) is bounded based on Lemma \ref{lem:Omegab_bounded} (See Fig.~\ref{fig:omega_y_bar_and_Omega_y}). However, if we remove the LBO constraint $\rl_2(t) < h_2(x_1) = x_{1,2} - x_{1,1}$, $\Ob(t)$ (and $\Omega(t)$) will not be bounded since along the path where $x_{1,1} = 0$ and $x_{1,2}\rightarrow -\infty$ we get $h_f(x_1) = 0$, $h_U(x_1) \rightarrow -\infty$, which do not satisfy any of the conditions in Lemma \ref{lem:alphb_coercive_h}. In Example \ref{ex:independent_funnels_example}, we have $h_f(t,x_1) = [h_1(x_1), h_2(x_1)]^{\top}$. One can verify that $\|h_f(t,x_1)\| \rightarrow +\infty$ along any path in $\R^2$ as $\|x_1\| \rightarrow +\infty$. In Example \ref{ex:independent_funnels_time-varying_example}, we observe that $h_f(t,x_1) = [h_1(x_1), h_2(t,x_1)]^{\top}$. It is not difficult to see that Condition I in Lemma \ref{lem:alphb_coercive_h} holds for all time instances and as depicted in Fig.~\ref{fig:time-varying_output}, $\Ob(t)$ (resp. $\Omega(t)$) remains bounded $\forall t \geq 0$.

Verifying the boundedness of $\Ob(t)$ as per Lemma \ref{lem:Omegab_bounded} can be challenging, especially when dealing with time-dependent outputs (i.e., $h_i(t,x_1)$ instead of $h_i(x_1)$). In such cases, one must check the condition in Lemma \ref{lem:alphb_coercive_h} for all time instances. However, note that ensuring the boundedness of $\Ob(t)$ is merely a technical requirement in this paper. To meet this requirement, one approach is to introduce an auxiliary output, denoted as $h_{\mathrm{aux}}(x_1)$, under the UBO constraint: $h_{\mathrm{aux}}(x_1) \coloneqq \|x_1\|^2 < c_{\mathrm{aux}}$, where $c_{\mathrm{aux}}>0$ is a sufficiently large constant. This constraint represents a large ball around the origin in the $x_1$ space, encompassing all other time-varying constraints in \eqref{eq:outputconst}. Notably, this constraint guarantees the satisfaction of Lemma \ref{lem:alphb_coercive_h}'s condition at all times, regardless of the choice of other system outputs, i.e., $h_i(t,x_1), i \in \I_1^m$.

Note that, Assumption \ref{assum:coercive_alphabar} also guarantees the existence of at least one global maximizer for $\alphb(t,x_1)$ (resp. $\alpha(t,x_1)$) $\forall t \geq 0$, refer to \cite[Theorem 1.4.4, p. 27]{peressini1988mathematics} or \cite[Proposition 2.9]{Grippo2023intro}. Thus, for each instant $t$, we can define:
\begin{equation} \label{eq:alphab_opt}
	\alphbstr(t) \coloneqq \max_{x_1 \in \R^n} \alphb(t,x_1),
\end{equation} 
where $\alphbstr(t)$ is bounded and denotes the maximum value of $\alphb(t,x_1)$ at time $t$. It is clear that if $\alphbstr(t^\prime) \geq 0$ then the time-varying output constraints are feasible at time $t = t^\prime$, whereas $\alphbstr(t^\prime) < 0$ indicates that the constraints are infeasible at time $t = t^\prime$, thus impossible to be satisfied. Similarly, for a given $\nu$ in \eqref{smooth_alph} we can define:
\begin{equation}\label{eq:alpha_opt}
	\alphastr(t) \coloneqq \max_{x_1 \in \R^n} \alpha(t,x_1) \leq \alphbstr(t).
\end{equation}
From \eqref{eq:alpha_opt} and \eqref{smooth_alph}, one can conclude that having $\alphastr(t^\prime) > 0$ is \textit{sufficient} for the feasibility of the time-varying output constraints \eqref{eq:outputconst} at time $t = t^\prime$. In addition, notice that $\alphastr(t^\prime) < 0$, does not necessarily imply that the actual output constrained set $\Ob(t^\prime)$ in \eqref{eq:omega_alpha_bar} is empty, i.e., $\alphbstr(t^\prime) < 0$ in \eqref{eq:alphab_opt}. In fact, from \eqref{smooth_alph_ineq} and the fact that $\alpha(t,x_1) \leq \alphastr(t)$ for all $t \geq 0$ and all $x_1 \in \mathbb{R}^n$, we can deduce that $\alphastr(t^\prime) < - \frac{1}{\nu} \ln(m+p)$ provides a sufficient condition for the infeasibility of $\Ob(t^\prime)$.

\subsection{Consolidating Multiple Constraints into One Constraint}
\label{subsec:single_const}

As discussed in Subsection \ref{subsec:scalar_metric}, satisfying \eqref{eq:outputconst} can be achieved by maintaining the positivity of  $\alpha(t,x_1(t;x(0)))$. Therefore, the main challenge in designing the control law outlined in Section \ref{sec:problemformulation} is to determine $u(t,x)$ for \eqref{eq:sys_dynamics_highorder} such that if $\alpha(0,x_1(0)) > 0$, then $\alpha(t,x_1(t;x(0))) > 0$ for all $t\geq 0$, and if $\alpha(0,x_1(0)) \leq 0$, then $\alpha(t,x_1(t;x(0))) > 0$ for all $t\geq T$. To achieve this objective, we propose ensuring the following single \textit{consolidating constraint} for \eqref{eq:sys_dynamics_highorder}:
\begin{equation} \label{eq:consuli_const}
	\ralph(t) < \alpha(t,x_1(t;x(0))), \quad \forall t\geq 0,
\end{equation}
where $\ralph: \Rpos \rightarrow \R$ is a properly designed bounded and continuously differentiable function of time with a bounded derivative. Before proposing a design for $\ralph(t)$ we emphasize that, in general, any appropriate $\ralph(t)$ in \eqref{eq:consuli_const} has to satisfy:
\begin{itemize}[left = 55pt]
	\item[{\bf Property (i):}]  $\alphastr(t) - \ralph(t) \geq \varsigma > 0, \forall t \geq 0$, where $\varsigma$ is a positive constant that may be unknown;
\end{itemize}
\begin{itemize}[left = 57.5pt]
	\item[{\bf Property (ii):}]  $\ralph(0) < \alpha(0,x_1(0))$.
\end{itemize}
Due to \eqref{eq:alpha_opt} and Assumption \ref{assum:coercive_alphabar}, we have $\alpha(t,x_1) \leq \alphastr(t)$ for all $t \geq 0$ and $x_1 \in \R^n$. Thus, $\alpha(t,x_1(t;x(0)))$ in \eqref{eq:consuli_const} is implicitly upper bounded by $\alphastr(t)$ for all time. Hence, \eqref{eq:consuli_const} is a valid constraint when Property (i) holds. Additionally, for controller design, detailed in Section \ref{subsec:control_design}, it is essential to design $\ralph(t)$ to ensure Property (ii) holds, meaning \eqref{eq:consuli_const} must be satisfied at $t=0$. This requirement does not impose significant restrictions since we assume that the initial condition of system \eqref{eq:sys_dynamics_highorder} is available for the controller design.

\subsection{Design of $\ralph(t)$ under Feasibility of the Constraints}
\label{subsec:ralph_design_simpl}

In order to guarantee the fulfillment of the time-varying output constraints specified in \eqref{eq:outputconst} through enforcing \eqref{eq:consuli_const}, one needs to properly design $\ralph(t)$ to enforce the positivity of $\alpha(t,x_1(t;x(0)))$ while respecting   Properties (i) and (ii) mentioned in Subsection \ref{subsec:single_const}. It turns out that it is straightforward to design $\ralph(t)$ in accordance with the following assumption:

\begin{assumption}\label{assu:feasible_output_constr}
	There exists $\epsilon_{f}>0$ such that $\alphastr(t) \geq \epsilon_{f}  > 0, \forall t\geq 0$, i.e., $\Omega(t)$ is non-empty (feasible) for all time.
\end{assumption}

This assumption implies that all output constraints in \eqref{eq:outputconst} are mutually satisfiable for all time. Under Assumption \ref{assu:feasible_output_constr} one can design $\ralph(t)$ through the following strategy:
\begin{itemize}
	\item[(a)] If $\alpha(0,x_1(0)) >  0$ (i.e., the constraints are initially satisfied), set $\ralph(t) = 0, \forall t \geq 0$;
	\item[(b)] If $\alpha(0,x_1(0)) \leq 0$, design $\ralph(t)$ such that $\ralph(0) < \alpha(0,x_1(0)) \leq 0$ and $\ralph(t \geq T) = 0$.
\end{itemize} 
Note that in the second case, the lower bound in equation \eqref{eq:consuli_const} needs to be increased over time to ensure that $\alpha(t, x(t;x_1(0)))$ becomes and remains positive for all $t \geq T > 0$. To achieve this, inspired by \cite[Remark 4]{yin2020robust}, we can design:
\begin{equation}\label{eq:alpha_lower_bound}
	\ralph(t) = \begin{cases}
		\left( \frac{T-t}{T} \right)^{\frac{1}{1-\beta}} (\rho_0 - \rho_{\infty}) + \rho_{\infty}, & 	0 \leq t < T,\\
		\rho_{\infty}, & t \geq T,
	\end{cases}
\end{equation}
where $\beta \in (0,1)$ is a constant, $\rho_0, \rho_{\infty},$ are constants such that $\rho_0 \leq \rho_{\infty}$, and $T > 0$ is the user-defined appointed finite time for constraints satisfaction. Note that \eqref{eq:alpha_lower_bound} is an increasing function and we have $\ralph(0) = \rho_0$ and $\ralph(t \geq T) = \rho_{\infty}$. Moreover, for case (a) above, we set $\rho_0 = \rho_{\infty} = 0$, while for case (b), we set $\rho_0$ such that $\ralph(0) = \rho_0 < \alpha(0,x_1(0)) < 0$ and $\rho_{\infty} = 0$. Finally, note that, the proposed design of $\ralph(t)$ ensures feasibility of \eqref{eq:consuli_const} since owing to $\rho_{\infty} = 0$ and Assumption \ref{assu:feasible_output_constr}, we get $\alphastr(t) - \ralph(t) \geq \varsigma = \epsilon_f > 0, \forall t \geq 0$. 

\begin{remark} \label{rem:compute_alpha_opt}
	We highlight that, when Assumption \ref{assu:feasible_output_constr} holds and $\rho_{\infty}=0$ in \eqref{eq:alpha_lower_bound}, no information about the solution of the time-varying optimization problem \eqref{eq:alphab_opt} is required for designing $\ralph(t)$ in \eqref{eq:consuli_const}. However, taking $\rho_{\infty} > 0$ requires $\rho_{\infty} < \inf_{\forall t\geq 0}(\alphastr(t))$ to hold for ensuring the feasibility of \eqref{eq:consuli_const}.   
\end{remark}

\begin{remark} \label{rem:robustness_dgree}
	Under Assumption \ref{assu:feasible_output_constr}, the choice of a larger $\rho_{\infty}$, with the condition $0 < \rho_{\infty} < \inf_{\forall t\geq 0}(\alphastr(t))$, dictates the extent to which the time-varying output constraints are satisfied. Specifically, when $\rho_{\infty} > 0$ is increased, it leads to a more robust enforcement of $\alpha(t,x_1(t;x(0)))$ for all $t \geq T$. Consequently, the satisfaction of \eqref{eq:consuli_const} leads to $x_1(t;x(0))$ being further confined away from the boundary of $\cl(\Omega(t))$ for all $t \geq T$, effectively pushing it deeper inside $\Omega(t)$.
\end{remark}

\subsection{Low-Complexity Controller Design and Stability Analysis}
\label{subsec:control_design}

Now similarly to the PPC method in \cite{bechlioulis2014low}, we design a model-free low-complexity robust state feedback controller for \eqref{eq:sys_dynamics_highorder} to ensure the satisfaction of the consolidating constraint \eqref{eq:consuli_const}. Due to the lower triangular structure of \eqref{eq:sys_dynamics_highorder}, we employ a backstepping-like design scheme. The process begins by creating an intermediate (virtual) control input $s_1(t,x_1)$ for the dynamics of $x_1$ in \eqref{eq:sys_dynamics_highorder}, ensuring the fulfillment of \eqref{eq:consuli_const}. Subsequently, we design a second intermediate control $s_2(t,\bar{x}_2)$ for the dynamics of $x_2$, making certain that $x_2$ follows the trajectory set by $s_1(t,x_1)$. This iterative top-down approach to design intermediate control laws $s_i(t,\bar{x}_i), i\in \I_1^r$, continues until we obtain the actual control input of the system, $u(t,x)$. The controller design is summarized in the following steps:

\textbf{Step 1-a.} Given \(x_1(0)\), compute \(\alpha(0,x_1(0))\) and select \(\rho_0\) in \eqref{eq:alpha_lower_bound} such that $\rho_0 = \ralph(0) < \alpha(0,x_1(0))$ is satisfied (ensuring \textit{Property (ii)}), as detailed in Subsection \ref{subsec:ralph_design_simpl}.

\textbf{Step 1-b.} Define:
\begin{equation} \label{eq:e_alph}
	\ealph(t,x_1) \coloneqq  \alpha(t,x_1) - \ralph(t),
\end{equation}
and consider the following nonlinear transformation:
\begin{equation} \label{eq:mapped_alpha}
	\epialph(t,x_1) = \T_{\alpha}(\ealph) \coloneqq  \ln \left( \frac{\ealph}{\upsilon} \right),
\end{equation}
where $\upsilon > 0$ is a constant and $\T_{\alpha}: (0, +\infty) \rightarrow (-\infty, +\infty)$ is a smooth, strictly increasing bijective mapping satisfying $\T_{\alpha}(\upsilon) = 0$. Note that maintaining the boundedness of $\epialph$ enforces $\ealph \in (0, \infty)$, thus satisfying \eqref{eq:consuli_const}. We call $\epialph \in (-\infty, +\infty)$ the \textit{unconstrained transformed signal} of $\ealph$.

\textbf{Step 1-c.} To design the first intermediate (virtual) control law we proceed as follows: first, define $V_1(\epialph) \coloneqq \frac{1}{2} \epialph^2$, which is a positive definite and radially unbounded (implicitly time-varying) \textit{barrier function} associated with the consolidating constraint in \eqref{eq:consuli_const}. Note that $V_1(0) = 0$ and as $\alpha(t,x_1)$ approaches $\ralph(t)$ (i.e., as $\ealph$ approaches zero) we get $V_1(\epialph) \rightarrow +\infty$. Next, from \eqref{eq:mapped_alpha}, with a slight abuse of notation one may consider ${V}_1(t, x_1)$, and design the first intermediate (gradient-based) control law as:
\begin{equation} \label{eq:1st_intermed_ctrl}
	s_1(t,x_1) \coloneqq - k_1 \, \nabla_{x_1} V_1(t,x_1),
\end{equation}
where $k_1 > 0$ is a control gain and $\nabla_{x_1}$ denotes the gradient with respect to $x_1$. Applying the chain rule in \eqref{eq:1st_intermed_ctrl} gives:
\begin{align} \label{eq:1st_intermed_ctrl_explicit}
	s_1(t,x_1) &= - k_1 \, \left( \dfrac{\partial V_1(\epialph)}{\partial \epialph} \, 
	\dfrac{\partial \epialph(\ealph)}{\partial \ealph} \,
	 \dfrac{\partial \ealph(t,\alpha)}{\partial \alpha} \,  \dfrac{\partial \alpha(t,x_1)}{\partial x_1} 
	 \right)^{\top}& \nonumber \\
	&= -k_1 \, \nabla_{x_1} \alpha(t,x_1) \, \frac{\epialph}{\ealph}.
\end{align} 

\textbf{Step $\mathbf{i}$-a ($\mathbf{2 \leq i \leq r}$).} Define the $i$-th intermediate error vector as: 
\begin{equation}\label{eq:intermediate_err}
	e_i = \col(e_{i,j}) \coloneqq x_i - s_{i-1}(t,\bar{x}_{i-1}),
\end{equation}
where $e_i  \in \R^n$. Now the objective is to design the $i$-th intermediate (virtual) control law $s_{i}(t,e_i)$ for \eqref{eq:sys_dynamics_highorder} to compensate $e_{i,j}(t,\bar{x}_i), j \in \I_1^n$, by enforcing the following narrowing \textit{intermediate funnel constraints}:
\begin{equation} \label{eq:i-th_inter_funnels}
	-\vartheta_{i,j}(t)  < e_{i,j}(t,\bar{x}_i) <  \vartheta_{i,j}(t), \quad j \in \I_1^n, 
\end{equation}
for all $t\geq 0$, where $\vartheta_{i,j}: \Rpos \rightarrow \Rspos$, are continuously differentiable \textit{strictly positive performance functions} that are decaying to a neighborhood of zero. One choice for $\vartheta_{i,j}(t)$ is: 
\begin{equation}\label{exponential_performance_fun}
	\vartheta_{i,j}(t) \coloneqq (\vartheta_{i,j}^0 - \vartheta_{i,j}^{\infty}) \exp(-l_{i,j} t) +  \vartheta_{i,j}^{\infty},
\end{equation} 
where $l_{i,j}, \vartheta_{i,j}^{\infty}, \vartheta_{i,j}^0$ are user-defined positive constants. Moreover, one should choose $\vartheta_{i,j}^0 > |e_{i,j}(0,\bar{x}_{i}(0))|$ to ensure $e_{i,j}(0,\bar{x}_{i}(0)) \in (-\vartheta_{i,j}(0), \vartheta_{i,j}(0)), j \in \I_1^n$.

\textbf{Step $\mathbf{i}$-b ($\mathbf{2 \leq i \leq r}$).} Now define the diagonal matrix $\varTheta_i(t) \coloneqq \diag(\vartheta_{i,j}(t)) \in \R^{n \times n}$, and consider
\begin{equation}\label{eq:normalized_err_vec}
	\eh_i(t,e_i)= \col(\eh_{i,j})  \coloneqq \varTheta_i^{-1}(t) \, e_i,
\end{equation}
as the vector of normalized errors, whose elements are:
\begin{equation} \label{eq:normal_e_i_j}
	\eh_{i,j}(t,e_{i,j}) = \frac{e_{i,j}}{\vartheta_{i,j}(t)}, \quad j \in \I_1^n.
\end{equation}
Note that, $\eh_{i,j} \in (-1,1)$ if and only if $e_{i,j} \in (-\vartheta_{i,j}(t), \vartheta_{i,j}(t))$. Next we introduce the following nonlinear transformations:
\begin{equation} \label{eq:mapping_fun}
	\epi_{i,j}(t,e_{i,j}) = \T(\eh_{i,j}) \coloneqq \ln \left( \frac{1+\eh_{i,j}}{1 - \eh_{i,j}} \right), \quad j \in \I_1^n, 
\end{equation}
where $\epi_{i,j}$ represents the unconstrained transformed signal of $e_{i,j}(t,\bar{x}_i)$ and $\T: (-1 , 1) \rightarrow (-\infty, +\infty)$ is a smooth strictly increasing bijective mapping, which satisfies $\T(0) = 0$. Note that enforcing the boundedness of $\varepsilon_{i,j}$ ensures that $\eh_{i,j}$ remains within the range of $(-1, 1)$, leading to the satisfaction of \eqref{eq:i-th_inter_funnels}.

\textbf{Step $\mathbf{i}$-c ($\mathbf{2 \leq i \leq r}$).} Finally, similarly to \textit{Step 1-c} we can design $s_{i}(t,e_i)$. In particular, define $\epi_i \coloneqq \col(\epi_{i,j}) \in \R^n$ and let $V_i(\epi_i) \coloneqq \frac{1}{2} \,\epi_i^\top \epi_i$, which is a positive definite and radially unbounded (implicitly time-varying) \textit{composite barrier function} associated with the intermediate funnel constraints in \eqref{eq:i-th_inter_funnels}. Note that $V_i(\mathbf{0}_n) = 0$ and for all $j \in \I_1^n$ if any $e_{i,j}(t,\bar{x}_i)$ approaches $\pm \vartheta_{i,j}(t)$  (i.e., as $\eh_{i,j}$ approaches $\pm 1$) we get $V_i(\epi_i) \rightarrow +\infty$. From \eqref{eq:mapping_fun}, with a slight abuse of notation, one can consider $V_i(t,e_i)$, and design the $i$-th intermediate control as:
\begin{equation} \label{eq:i-th_intermed_ctrl}
	s_i(t,e_i) \coloneqq - k_i \, \nabla_{e_i} V_i(t,e_i),
\end{equation}
where $k_i > 0$ is a control gain and $\nabla_{e_i}$ denotes the gradient with respect to $e_i$. Consequently, one can obtain $s_i(t,e_i)$ more explicitly by applying the chain rule:
\begin{align} \label{eq:i_th_intermed_ctrl_explicit}
	s_i(t,e_i) &= - k_i \, \left( \dfrac{\partial V_i(\epi_i)}{\partial \epi_i} \, 
	\dfrac{\partial \epi_i(\eh_i)}{\partial \eh_i} \,
	\dfrac{\partial \eh_i(t,e_i)}{\partial e_i}  \right)^{\top}& \nonumber \\
	&= -k_i \, \Xi_i \, \epi_i,
\end{align}
where $\Xi_i \coloneqq \diag(\xi_{i,j}) \coloneqq \frac{\partial \epi_i(\eh_i)}{\partial \eh_i}  \frac{\partial \eh_i(t,e_i)}{\partial e_i} \in \R^{n \times n}$ is a diagonal matrix whose diagonal entries are:
\begin{equation} \label{eq:xi_i,j}
	\xi_{i,j}(t,e_{i,j}) \coloneqq \frac{2}{\vartheta_{i,j}(t) \, (1-\eh_{i,j}^{\, 2})}, \quad j \in \I_1^n.
\end{equation}
Notice that $s_i(t,e_i)$ can be considered as a function of $t$ and $\bar{x}_i$, as $e_i$ itself depends on $\bar{x}_i$ (see \eqref{eq:intermediate_err}) so with a slight abuse of notation one can write $s_i(t,\xb_i)$.

\textbf{Step $\mathbf{r+1}$.} Finally we design the control input $u(t,x)$ as:
\begin{equation} \label{eq:control_law}
	u(t,x) \coloneqq s_r(t,x).
\end{equation}

\begin{remark}
	The proposed control method, similarly to backstepping, aims to make $x_2$ closely track $s_1(t,x_1)$ in the dynamics \eqref{eq:sys_dynamics_highorder}, where $s_1(t,x_1)$ is designed to satisfy \eqref{eq:consuli_const}. We design the second intermediate control $s_2(t,e_1)$ to ensure that all components of the error $e_2 = x_2 - s_1(t,x_1)$, denoted as $e_{2,j}, j \in \I_1^n$, become sufficiently small through the satisfaction of \eqref{eq:i-th_inter_funnels}. This iterative design process continues until we obtain $u(t,x)$ for \eqref{eq:sys_dynamics_highorder}. Importantly, unlike the classical backstepping method, we do not use derivatives of $e_i, i \in \I_2^r$, or any filtering scheme in the design of intermediate control laws $s_i(t,e_i), i \in \I_2^r$ \cite{bechlioulis2014low}. Furthermore, we do not rely on prior knowledge of the system's nonlinearities or any upper/lower bounds on uncertainties in the design of \eqref{eq:control_law}.
\end{remark}

\begin{remark}
	It is important to note that the satisfaction of the proposed consolidating constraint \eqref{eq:consuli_const}, as well as \eqref{eq:i-th_inter_funnels} for the intermediate error signals $e_{i,j}, i \in \I_2^r, j \in \I_1^n$, are ensured merely by keeping $\epialph$ and $\epi_i$ bounded, respectively. This is achieved by applying the designed control input \eqref{eq:control_law} in \eqref{eq:sys_dynamics_highorder}. This key observation will be leveraged in the stability analysis of the closed-loop system.
\end{remark}

Notice that $\nabla_{x_1}\alpha(t,x_1)$ in \eqref{eq:1st_intermed_ctrl_explicit} determines the direction of the first intermediate control law $s_1(t,x_1)$ used to enforce the time-varying constraints. Recall that $\alpha(t,x_1)$ in \eqref{smooth_alph_def} is constructed directly from the constraints in \eqref{eq:predicate_const_rep}. As a result, $s_1(t,x_1)=\mathbf{0}_n$ may occur at undesired (time-varying) critical points of $\alpha(t,x_1)$ that lie outside the (inner-approximated) constraint set $\Omega(t)$, thereby potentially preventing $s_1(t,x_1)$ from ensuring $\alpha(t,x_1)>0$. In contrast, if $\nabla_{x_1}\alpha(t,x_1)=\mathbf{0}_n$ occurs only at points where the constraints are already satisfied, i.e., where $\alpha(t,x_1)>0$ (equivalently, inside $\Omega(t)$), then the proposed control scheme can be guaranteed to render $\Omega(t)$ forward invariant. Therefore, it is essential to preclude the closed-loop trajectories from reaching undesired critical points of $\alpha(t,x_1)$ outside $\Omega(t)$, such as saddle points and local minima. To this end, we introduce the following technical assumption as a sufficient condition to rule out the aforementioned scenario: 

\begin{assumption}\label{assu:alpha_globalmax}
	For all $t \geq 0$ the function $-\alpha(t,x_1)$ is invex, i.e., every critical point of $\alpha(t,x_1)$ is a (time-varying) global maximizer (see \cite[Theorem 2.2]{mishra2008invexity}). 
\end{assumption}

The following lemma gives some \textit{sufficient} conditions for ensuring Assumption \ref{assu:alpha_globalmax}.
\begin{lemma} \label{lem:global_max_suffi}
	The function $-\alpha(t,x_1)$ is invex $\forall t \geq 0$, if at each time instant $t$ one of the following conditions holds:
	\begin{itemize}
		\item [(I)] $\psi_i(t,x_1), \forall i \in \I_1^{m+p}$ in \eqref{eq:psi_constr} are concave in $x_1$.
		\item [(II)] Having only $n$ funnel constraints (i.e., $n = m = p$ in \eqref{eq:outputconst}) such that: (i) the output map $y = h(t,x_1)$ in \eqref{eq:sys_dynamics_highorder} is norm-coercive (i.e., $\|h(t,x_1)\| \rightarrow +\infty \quad \text{as} \quad \|x_1\| \rightarrow +\infty$), and (ii) the Jacobian matrix $J(t,x_1) \coloneqq \frac{\partial h(t,x_1)}{\partial x_1} \in \R^{n \times n}$ is full rank for all $x_1 \in \R^{n}$.
	\end{itemize}
\end{lemma}
\begin{IEEEproof}
	See Appendix \ref{appen:proof_lemma_globalmax}.
\end{IEEEproof}

The concavity of $\psi_i(t, x_1), i=\I_1^{m+p}$ at time $t$ in Lemma \ref{lem:global_max_suffi} can be understood by examining \eqref{eq:predicate_const_rep} in terms of system outputs $h_i(t,x_1), i \in \I_1^{m}$. Specifically, for funnel constraints, the functions $h_i(t,x_1),  i \in \I_1^{p}$, should be an affine function of $x_1$ at time $t$, as $\psi_{2i}(t,x_1)$ and $\psi_{2i-1}(t,x_1)$ are concave only when $h_i(t,x_1)$ and $-h_i(t,x_1)$ are concave, see \eqref{eq:predicate_funnel_rep}. On the other hand, for LBO constraints, $h_i(t,x_1), i \in \I_{p+1}^{p+q}$, should be concave, and for UBO constraints, $h_i(t,x_1),  i \in \I_{p+q+1}^m$, should be convex at time $t$, see \eqref{eq:predicate_onesided_rep}.

It is straightforward to see that Example \ref{ex:concave_example} (illustrated in Fig.~\ref{fig:omega_y_bar_and_Omega_y}) satisfies Condition I of Lemma \ref{lem:global_max_suffi} at all times. Additionally, Example \ref{ex:independent_funnels_example} (shown in Fig.~\ref{fig:condition II}) meets Condition II of Lemma \ref{lem:global_max_suffi} at all times. Specifically, in Example \ref{ex:independent_funnels_example}, we have $n=m=p=2$ and the Jacobian matrix of $h(x_1)$, denoted as $J(x_1) = \left[\begin{smallmatrix} 1 & 0 \\ 0.6x_{1,1} & -1 \end{smallmatrix} \right]$, has full rank for all $x_1 \in \R^2$. Additionally, $h(x_1) = h_f(x_1)$ is norm-coercive. Note that Example \ref{ex:independent_funnels_example} fails to satisfy Condition I of Lemma \ref{lem:global_max_suffi} because $h_2(x_1)$ is not an affine function. Likewise, we can easily verify that Example \ref{ex:independent_funnels_time-varying_example} also meets Condition II of Lemma \ref{lem:global_max_suffi} at all times. It is worth emphasizing that Condition II in Lemma \ref{lem:global_max_suffi} accurately captures the notion of independence between $n$ funnel constraints in $\R^n$. This means that the satisfaction of individual feasible funnel constraints does not interfere with each other, i.e., the funnel constraints are decoupled.

\begin{remark}
	If $\Ob(t)$ is the interior of a \textit{time-varying bounded convex polytope} in $\R^n$, then $\alpha(t,x_1)$ satisfies Assumptions \ref{assum:coercive_alphabar} and \ref{assu:alpha_globalmax}. The former holds as a consequence of the polytope's boundedness assumption and the latter is true because in a convex polytope, all $h_i(t,x_1)$ are affine in $x_1$ for all time (which satisfies Condition I of Lemma \ref{lem:global_max_suffi} for all $t \geq 0$).
\end{remark}

\begin{remark}
	The invexity of $-\alpha(t,x_1)$ is ensured even if conditions I and II of Lemma \ref{lem:global_max_suffi} interchange at different time instances. Unlike Examples 1-3, this case allows for a modified version of Example \ref{ex:concave_example} with $h(t,x_1) = [h_1(x_1), h_2(t,x_1), h_3(t,x_1)]^{\top}$. Here, $h_1(x_1) = x_{1,1}$, $h_2(t,x_1) = c_1(t) x_{1,1}^2 + c_2(t) x_{1,2} + c_3(t) x_{1,1}$, $h_3(x_1) = 0.3 x_{1,1}^2 + c_4(t) x_{1,2}$, and $c_i(t)$, $i \in \I_1^4$, are bounded continuous time-varying functions. Initially, at $t = t_1$, with $c_1(t_1) = 0$, $c_2(t_1) = 1$, $c_3(t_1) = -1$, and $c_4(t_1) = 1$, the constrained set mirrors Fig.~\ref{fig:omega_y_bar_and_Omega_y}, satisfying Condition I in Lemma \ref{lem:global_max_suffi}. Then, as $c_i(t)$, $i \in \I_1^4$, continuously vary over time, at $t = t_2$, where $c_1(t_2) = 0.3$, $c_2(t_2) = -1$, $c_3(t_2) = 0$, and $c_4(t_2) = -1$, we observe $h_2(t_2,x_1) = h_3(t_2,x_1)$. The LBO and UBO constraints for $h_2(t_2,x_1)$ and $h_3(t_2,x_1)$ combine into a single funnel constraint, resulting in a constrained set resembling the one in Example \ref{ex:independent_funnels_example}, depicted in Fig.~\ref{fig:condition II}. Hence, if, for $t \in (t_1,t_2)$, the functions $c_i(t)$, $i \in \I_1^4$, vary in such a way that any condition in Lemma \ref{lem:global_max_suffi} holds (requiring the constrained set in Fig.~\ref{fig:omega_y_bar_and_Omega_y} to transform into a box and then into the one in Fig.~\ref{fig:condition II}), the invexity of $-\alpha(t,x_1)$ is guaranteed for all $t \in [t_1,t_2]$.
\end{remark}

\begin{remark} \label{rem:boundedness_invexity}
	Notice that satisfying Condition I of Lemma \ref{lem:global_max_suffi} alone is not enough to ensure the boundedness of $\Omega(t)$. To guarantee that $\Omega(t)$ is bounded, $h_i(t,x_1), i \in \I_1^m$, functions used in $\psi_{i}(t,x_1)$, $i \in \I_1^{m+p}$ should also meet the condition of Lemma \ref{lem:alphb_coercive_h} (see Lemma \ref{lem:alphb_coercive_h}'s proof). However, for Condition II of Lemma \ref{lem:global_max_suffi}, it is worth noting that since $h(t,x_1)$ is norm-coercive and only funnel-type constraints are considered (i.e., $h(t,x_1) \equiv h_f(t,x_1)$), one can verify that Condition I in Lemma \ref{lem:alphb_coercive_h} is already satisfied. This, in turn, ensures the boundedness of $\Omega(t)$.
\end{remark}

The following theorem summarizes our main result:
\begin{theorem} \label{th:main}
	Consider the MIMO nonlinear system \eqref{eq:sys_dynamics_highorder} subject to time-varying output constraints \eqref{eq:outputconst}. Let the design of $\ralph(t)$ satisfy Properties (i) and (ii) in Subsection \ref{subsec:single_const} and $\dralph(t)$ be bounded. Additionally, select constants $\vartheta_{i,j}^0, i \in \I_2^r, j \in \I_1^n$ in \eqref{exponential_performance_fun} such that $\vartheta_{i,j}^0 > |e_{i,j}(0,\bar{x}_{i}(0))|$ (as explained in \textit{Step i-a} in Subsection \ref{subsec:control_design}). Under Assumptions 1-5 and 7, the feedback control law \eqref{eq:control_law} ensures the satisfaction of the consolidating constraint \eqref{eq:consuli_const}, as well as the boundedness of all closed-loop signals for all time.
\end{theorem}
\begin{IEEEproof}
	See Appendix \ref{appen:proof_theorem}. 
\end{IEEEproof}

\begin{remark} \label{rem:feasibility_assumption_theorem}
		Note that the conclusions of Theorem \ref{th:main} do not directly depend on Assumption \ref{assu:feasible_output_constr}. In particular, Theorem \ref{th:main} remains valid provided that $\ralph(t)$ satisfies \textit{Properties (i)} and \textit{(ii)} stated in Subsection \ref{subsec:single_const}, and that both $\ralph(t)$ and its derivative $\dralph(t)$ are bounded. As discussed in Subsection \ref{subsec:ralph_design_simpl}, Assumption \ref{assu:feasible_output_constr} is mainly invoked to facilitate the design of $\ralph(t)$ in \eqref{eq:alpha_lower_bound} so as to guarantee \textit{Property (i)}. In Section \ref{sec:const_infeasibility}, we propose an adaptive design of $\ralph(t)$ that does not rely on Assumption \ref{assu:feasible_output_constr}.
\end{remark}

\begin{remark}
	Control law \eqref{eq:control_law} ensures that \eqref{eq:consuli_const} is met for all time, but the parameter $\upsilon > 0$ in \eqref{eq:mapped_alpha} significantly shapes $\alpha(t,x_1(t;x(0)))$ concerning $\ralph(t)$ in \eqref{eq:consuli_const}. Specifically, with a very small $\nu$, even a slight increase in $\ealph > 0$ strongly influences $\epialph$ growth. Consequently, the intermediate control law $s_1(t,x_1)$ \eqref{eq:1st_intermed_ctrl_explicit} restricts $\ealph$ growth, keeping $\alpha(t,x_1(t;x(0)))$ closer to $\ralph(t)$. Conversely, a larger $\nu$ relaxes this restriction, allowing $\alpha(t,x_1(t;x(0)))$ more freedom to approach $\alphastr(t)$.
\end{remark}

\begin{remark}
	The tunings of $\ralph(0) = \rho_0 < \alpha(0,x_1(0)) < 0$ in \eqref{eq:alpha_lower_bound} and $\vartheta_{i,j}^0 > |e_{i,j}(0,\bar{x}_{i}(0))|$ in \eqref{exponential_performance_fun} necessitate knowledge of the initial condition $x(0)$. Consequently, the stability results presented in Theorem  \ref{th:main} are semi-global. However, it is possible to eliminate this requirement by incorporating shifting functions in the controller design process, as proposed in \cite{song2018tracking}.
\end{remark}

\textbf{More on Assumption  \ref{assu:alpha_globalmax}:} Assumption \ref{assu:alpha_globalmax} is crucial for Theorem \ref{th:main} and ensures the effectiveness of the proposed control law \eqref{eq:control_law}. However, it places certain limitations on the class of time-varying constraints suitable for \eqref{eq:sys_dynamics_highorder}. 

When the constraint set $\Omega(t)$ is feasible, Assumption \ref{assu:alpha_globalmax} provides a sufficient (though conservative) condition for avoiding $\nabla_{x_1}\alpha(t,x_1)=\mathbf{0}_n$ outside of $\Omega(t)$. Alternatively, one may adopt a less conservative (though less structured) assumption that: \textit{Every non-positive real number is a regular value of $\alpha(t,x_1)$ for all time, i.e., $\nabla_{x_1}\alpha(t,x_1)\neq\mathbf{0}_n$ for all $\alpha(t,x_1) \leq 0$ and all $t \geq 0$}. 

\begin{figure}[tbp]
	\centering
	\begin{subfigure}[t]{0.44\linewidth}
		\centering
		\includegraphics[width=\linewidth]{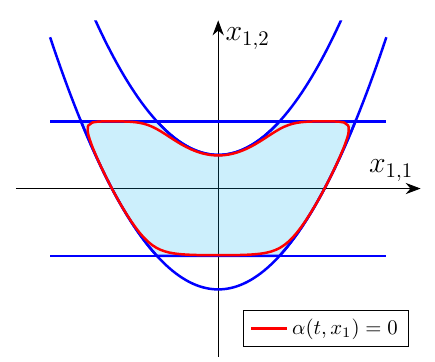}
		\caption{}
		\label{fig:regular}
	\end{subfigure}%
	~
	\begin{subfigure}[t]{0.5\linewidth}
		\centering
		\includegraphics[width=\linewidth]{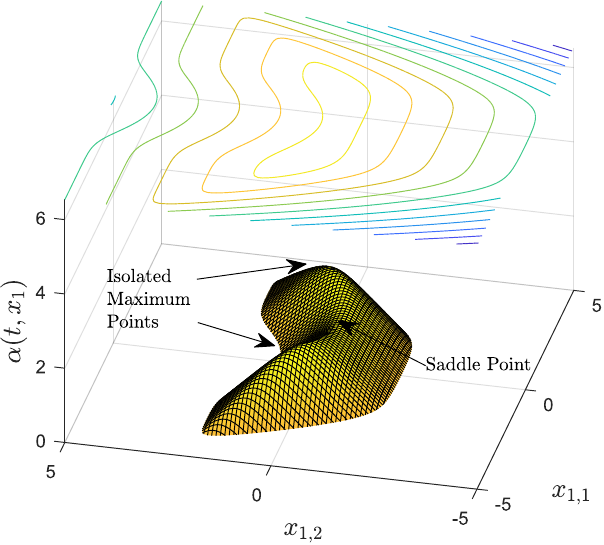}
		\caption{}
		\label{fig:regular_surface}
	\end{subfigure}
	\caption{Illustrative example: regularity of $\alpha(t,x_1)$ on $\Rpos$. (a) Time-varying constraint set. (b) Surface of $\alpha(t,x_1)$.}
	\vspace{-0.5cm}
	\label{fig:regularity_of_alpha_fun}
\end{figure}

To illustrate the above condition, consider the following example. Let $x_1 = [x_{1,1},x_{1,2}]^\top \in \mathbb{R}^2$, and suppose that the constraints are described by the map $h(x_1) = [h_1(x_1),h_2(x_1)]^\top$, where $h_1(x_1)=x_{1,2}$ and $h_2(x_1)=-0.3x_{1,1}^2 + x_{1,2}$. The corresponding constraints are $\rl_1(t) < h_1(x_1) < \ru_1(t)$ and $\rl_2(t) < h_2(x_1) < \ru_2(t)$. Fig.~\ref{fig:regular} shows a snapshot of the time-varying constraint set together with the boundary of the zero superlevel set of the associated function $\alpha(t,x_1)$ (shown in red), which provides a smooth inner approximation of the original constraint set. The surface plot of $\alpha(t,x_1)$ in Fig.~\ref{fig:regular_surface} indicates that $-\alpha(t,x_1)$ is not invex, since $\alpha(t,x_1)$ admits multiple critical points (two isolated maxima and one saddle point). Importantly, all of these critical points lie inside $\Omega(t)$, where $\alpha(t,x_1)>0$. In particular, for this example, it can be readily verified that if $\Omega(t)$ remains non-empty and connected for all time, then every non-positive real number is a regular value of $\alpha(t,x_1)$. Consequently, by the discussion preceding Assumption \ref{assu:alpha_globalmax}, the proposed control law \eqref{eq:control_law} still guarantees satisfaction of the time-varying constraints, since it remains nonzero everywhere outside $\Omega(t)$.

\begin{figure}[!tbp]
	\centering
	\begin{subfigure}[t]{0.44\linewidth}
		\centering
		\includegraphics[width=\linewidth]{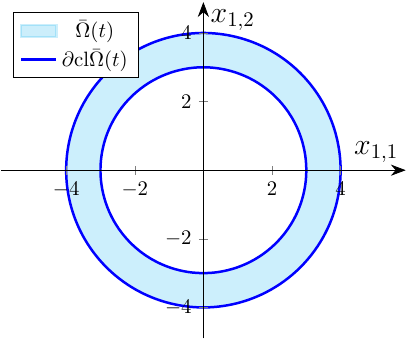}
		\caption{}
		\label{fig:1}
	\end{subfigure}%
	~
	\begin{subfigure}[t]{0.52\linewidth}
		\centering
		\includegraphics[width=\linewidth]{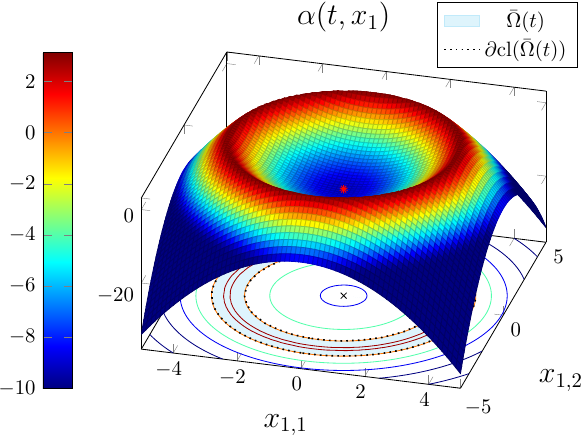}
		\caption{}
		\label{fig:2}
	\end{subfigure}
	\caption{(a) Snapshot of the output constraint $\rl(t) < h(x_1) = x_{1,1}^2 + x_{1,2}^2 < \ru(t)$ for which its corresponding $\alpha(t,x_1)$ does not satisfy Assumption \ref{assu:alpha_globalmax} nor is regular on $\Rpos$ due to the existence of a local minimum at $x_1 = [0, 0]^\top$ at which $\alpha(t,x_1)<0$. (b) surface of $\alpha(t,x_1)$. \vspace{-0.55cm} }
	\label{fig:examp}
\end{figure}

It is worth noting that there are scenarios where \eqref{eq:control_law} can still work effectively without satisfying Assumption \ref{assu:alpha_globalmax} nor the aforementioned condition on the regularity of $\alpha(t,x_1)$. For instance, for simplicity assume the system dynamics is an ideal two-dimensional double integrator system, i.e., $\dot{x}_1 = x_2$ and $\dot{x}_2 = u$ and let $y = h(x_1) = x_{1,1}^2 + x_{1,2}^2$ be the sole output of this system. Implicitly, $h(x_1) \geq 0$ holds, and if we choose $h(x_1) < \ru(t)$ as the output constraint, it is straightforward to verify that this choice meets Condition I of Lemma \ref{lem:global_max_suffi}, implying that Assumption \ref{assu:alpha_globalmax} is satisfied (i.e., $-\alpha(t,x_1)$ is invex). However, if we set $\rl(t) < h(x_1) < \ru(t)$ as the output constraint, where $0 < \rl(t) < \ru(t)$, this violates Assumption \ref{assu:alpha_globalmax}. Fig.~\ref{fig:1} displays a snapshot of the output-constrained set at time $t$ with $\rl(t)= 9$ and $\ru(t) =16$, and Fig.~\ref{fig:2} shows its corresponding $\alpha(t,x_1)$ surface. In Fig.~\ref{fig:2}, $x_{1} = [0,0]^\top$ (denoted by $\times$) represents the local minimum of $\alpha(t,x_1)$, where $s_1(t,\mathbf{0}_2) = \mathbf{0}_2$. It is important to note that, since the control direction of the first intermediate control law, $s_1(t,x_1)$, aligns with the positive gradient of $\alpha(t,x_1)$, any point in the form of $x = [0, 0, \star, \star]^\top$ is repelling for the closed-loop dynamical system in this example. Therefore, the proposed control law \eqref{eq:control_law} can still be effective in satisfying $\rl(t) < h(x_1) < \ru(t)$, with the singular case occurring when the initial condition of the system, i.e., $x(0) = [x_1(0)^\top, x_2(0)^\top]^\top$, is $x(0) = \mathbf{0}_4$. However, this type of singularity are of measure zero, and even the slightest influence of external disturbances or time-varying internal dynamics in the closed-loop system \eqref{eq:sys_dynamics_highorder} can prevent its occurrence.

\section{Dealing with Potential Infeasibilities}
\label{sec:const_infeasibility}

In this section, we propose an adaptive design for $\ralph(t)$ in \eqref{eq:consuli_const} to address the potential infeasibility of the inner-approximated output constrained set $\Omega(t)$ within an unknown time interval $I$, which is captured by having $\alphastr(t) < 0$ in \eqref{eq:alpha_opt} for all $t \in I$. Our objective is to address conflicts that may arise from the couplings between multiple time-varying output constraints, leading to a possible violation of Assumption \ref{assu:feasible_output_constr}, which  renders the proposed design of $\ralph(t)$ in Subsection \ref{subsec:ralph_design_simpl} inapplicable. To resolve this issue, first, we introduce the concept of \textit{Least Violating Solution} (LVS) for \eqref{eq:sys_dynamics_highorder}.

Recall that according to \eqref{eq:alpha_opt}, if $\alphastr(t) < 0$ holds for an unknown time interval $I$, then the inner-approximated output constrained set $\Omega(t)$ in \eqref{eq:omega_alpha} is empty (infeasible) for all $t \in I$.
\begin{definition} \label{def:least_vio_sol}
	 When $\alphastr(t) < 0$, $x(t;x(0))$ is a \textit{least violating solution for \eqref{eq:sys_dynamics_highorder} with a given gap of} $\mu^\ast>0$ if:
	\begin{equation}\label{eq:least_vio_def}
		\alphastr(t) - \mu^\ast < \alpha(t,x_1(t;x(0))), \quad \forall t \in I.
	\end{equation}
\end{definition}

In other words, whenever $\alphastr(t) < 0$, maintaining $\alpha(t, x_1(t; x(0)))$ in a sufficiently small neighborhood below $\alphastr(t)$ establishes an LVS for \eqref{eq:sys_dynamics_highorder} under the constraints in \eqref{eq:outputconst}.
\subsection{Estimating $\alphastr(t)$ via Online Continuous-Time Optimization}
\label{subsec: estimator}

Upon examining \eqref{eq:consuli_const} and \eqref{eq:least_vio_def}, it becomes evident that having knowledge of $\alphastr(t)$ is crucial for effective design of $\ralph(t)$, ensuring the attainment of a least violating solution when $\alphastr(t) < 0$. However, direct access to $\alphastr(t)$ can be limiting in various applications. To overcome this limitation, we introduce $\alphah(t)$ as an online estimate of $\alphastr(t)$ and propose an online continuous-time optimization scheme to estimate $\alphastr(t)$. Recall that $\alphastr(t)$ in \eqref{eq:alpha_opt} does not depend on the dynamical system \eqref{eq:sys_dynamics_highorder} but the behavior and properties of the output constraints in \eqref{eq:outputconst}. To prevent any ambiguity in the notations, henceforth, we distinguish between the state vector $x_1$ of the dynamical system \eqref{eq:sys_dynamics_highorder} and the optimization variable $x_1$ in \eqref{eq:alpha_opt}. Thus, we denote the optimization variable in \eqref{eq:alpha_opt} as $\xtil_1 \in \R^n$, yielding:
\begin{equation}\label{eq:alpha_opt_TVO}
	\alphastr(t) \coloneqq \max_{\xtil_1 \in \R^n} \alpha(t,\xtil_1).
\end{equation}
To obtain $\alphah(t)$, we propose the following first-order continuous-time optimization scheme (some function arguments are dropped for compactness in the notation):
\begin{subnumcases}{\label{eq:first_order_gradaccent}} 
	\dot{\xtil}_1 = k_{\alpha} \, \gradxtilalph - \frac{\gradxtilalph}{\|\gradxtilalph\|^2 + \epsilon_g \, \chi(\|\gradxtilalph\|)} \frac{\partial \alpha}{\partial t} & \label{eq:first_order_gradaccent_dyn}\\
	\alphah(t) = \alpha(t,\xtil_1) & \label{eq:first_order_gradaccent_out}
\end{subnumcases}
where $k_{\alpha}>0$, and $\epsilon_g > 0$ is a sufficiently small constant. Moreover, $\chi: \R \rightarrow [0,1]$ is a $\C^1$ switch function defined as:
\begin{align}
	\chi(z) =
	\begin{cases}
		1 & z < 0  \\
		\dfrac{2}{\mch^3} z^3 - \dfrac{3}{\mch^2} z^2 + 1 & 0 \leq z \leq \mch \\
		0 & z > \mch
	\end{cases}, \label{smooth_switch_fun_estimator}
\end{align}
in which $\mch > 0$ is a sufficiently small tuning parameter. Note that $\xtil_1(0)$ can be chosen arbitrarily in \eqref{eq:first_order_gradaccent_dyn}, and $\alphah(t)$ in \eqref{eq:first_order_gradaccent} represents the continuous-time evaluation of the time-varying cost function $\alpha(t,\xtil_1)$ at each instant $t$ under the updating rule in \eqref{eq:first_order_gradaccent_dyn}. Additionally, it is straightforward to obtain $\gradxtilalph(t,\xtil_1)$ (see \eqref{eq:grad_alph}) and from \eqref{smooth_alph_def} one can obtain:
\begin{flalign} \label{eq:partial_alpha_partial_time}
	&\frac{\partial \alpha}{\partial t} = \frac{1}{\sum_{i=1}^{m+p} e^{-\nu \, \psi_i}} \sum_{i=1}^{m+p} \frac{\partial \psi_i}{\partial t} e^{-\nu \, \psi_i} = \frac{\partial \psi}{\partial t} ^\top \!\! \varpi \, e^{\nu \alpha(t,\xtil_1)},\!\!\!\!\!\!\!\!& 
\end{flalign}
where $\varpi \coloneqq [e^{-\nu \, \psi_1}, \ldots, e^{-\nu \, \psi_{m+p}}]^\top$ and $\psi \coloneqq [\psi_1,\ldots, \psi_{m+p}]^\top$. Recall that, since $\alphastr(t)$ denotes the maximum value of $\alpha(t,\xtil_1)$ for any $\xtil_1$ at each time instant, it is evident that $\alpha(t,\xtil_1) \leq \alphastr(t)$ holds for all $t\geq 0$. Consequently, in \eqref{eq:first_order_gradaccent}, $\alphah(t)$ can only approach $\alphastr(t)$ from below, i.e., $\alphah(t) \leq \alphastr(t)$ for all $t \geq 0$.

Recall that, according to Assumption \ref{assu:alpha_globalmax}, every critical point of \(\alpha(t,\tilde{x}_1)\) is a global maximizer. If \(\alpha^*(t)\) varies slowly over time, it is anticipated that following the gradient of \(\alpha(t,\tilde{x}_1)\) with respect to \(\tilde{x}_1\) can effectively approximate \(\alpha^*(t)\) \cite{simonetto2020time}. While this approach may not guarantee precise convergence to \(\alpha^*(t)\), it is well-suited for our needs in this paper. Specifically, the first term in \eqref{eq:first_order_gradaccent_dyn} represents the standard gradient ascent for updating \(\tilde{x}_1\), while the second term is introduced to counteract the variation of \(\alpha(t,\tilde{x}_1)\) with respect to time at \(\tilde{x}_1\) when $\|\gradxtilalph(t,\xtil_1)\| \geq \mch$. Notably, one can expect that increasing \(k_{\alpha}\) and decreasing \(\epsilon_g\) in \eqref{eq:first_order_gradaccent}, as well as choosing a sufficiently small $\mch$ in \eqref{smooth_switch_fun_estimator}, can significantly improve the estimation of \(\alpha^*(t)\), especially when \(\alpha^*(t)\) has rapid variations over time.

\begin{remark}
	In the realm of continuous-time optimization for time-varying cost functions, second-order gradient flows under a prediction-correction scheme has been proposed for achieving asymptotic convergence to the optimal point \cite{simonetto2020time}. However, this method relies on the Hessian inverse of the time-varying cost function, necessitating the cost function to be strongly concave (or convex). It is important to note that in our work $\alpha(t,\xtil_1)$ does not always satisfy this condition. Since this second-order approach is akin to continuous-time variant of Newton's method, using it in our context does not guarantee convergence to the global optimum of $\alpha(t,\xtil_1)$ and, in the worst-case scenario, could result in divergence. Consequently, we have chosen to employ a first-order optimization scheme \eqref{eq:first_order_gradaccent}, which offers practical convergence to the time-varying optimum of $\alpha(t,\xtil_1)$, provided an appropriate choice of $k_{\alpha}$.
\end{remark}

In Subsection \ref{subsec:ralph_design_simpl}, we proposed a method to design \(\ralph(t)\) effectively, for fulfillment of \eqref{eq:outputconst}, which relied on Assumption \ref{assu:feasible_output_constr}. In the next subsection, we present an adaptive design for \(\ralph(t)\) that does not require this assumption. Instead, we will use available information on \(\alpha^*(t)\) via the estimation scheme \eqref{eq:first_order_gradaccent} to handle potentially infeasible time-varying output constraints. Our goal is to design \(\ralph(t)\) to ensure an LVS whenever \(\alpha^*(t) < 0\), while still preserving Properties (i) and (ii) from Subsection \ref{subsec:single_const}.

\subsection{Design of $\ralph(t)$ for Potentially Infeasible Constraints}
\label{subsec: modif_rhoalph}

Let us first introduce $\varrho(t)$ as a \textit{nominal lower bound} for $\alpha(t,x_1(t;x(0)))$, which determines the nominal behavior of the lower bound in \eqref{eq:consuli_const}. Specifically, $\varrho(t)$ is designed to ensure the satisfaction of output constraints by enforcing $\alpha(t,x_1(t;x(0)))$ to become and remain positive within a user-defined finite time $T$. In this regard, similar to the design of $\ralph(t)$ in Subsection \ref{subsec:ralph_design_simpl}, we can design $\varrho(t)$ as:
\begin{equation}\label{eq:alpha_lower_bound_nomi}
	\varrho(t) \coloneqq \begin{cases}
		\left( \frac{T-t}{T} \right)^{\frac{1}{1-\beta}} (\varrho_0 - \varrho_{\infty}) + \varrho_{\infty}, & 	0 \leq t < T,\\
		\varrho_{\infty}, & t \geq T,
	\end{cases}
\end{equation}
where $\beta \in (0,1)$, $\varrho(0) = \varrho_0 < \alpha(0,x_1(0))$, and $\varrho_{\infty} \geq 0$ is a user-defined arbitrary  non-negative constant. Recall that a larger $\varrho_{\infty}$ enforces how well the output constraints should be satisfied (in the nominal case) after finite time $t = T$ (see Remark \ref{rem:robustness_dgree}). In this respect, we refer to $\varrho_{\infty}$ as the \textit{nominal constraint satisfaction margin}. We now propose an alternative design for $\ralph(t)$ as follows:
\begin{equation} \label{eq:rho_lower_optim_schme}
	\ralph(t) = \iota(t) \varrho(t) + (1 -\iota(t)) ( \alphah(t) - \mu),
\end{equation}
where $\mu>0$ is a user-defined small positive constant, and $\iota: \Rpos \rightarrow [0,1]$ is a $\mathcal{C}^1$ switch function given by:
\begin{align}
	\iota(t) =
	\begin{cases}
		1 & \varphi(t) > \mu  \\
		-\dfrac{2}{\mu^3} \varphi^3(t) + \dfrac{3}{\mu^2} \varphi^2(t) & 0 \leq \varphi(t)\leq \mu \\
		0 & \varphi(t) < 0
	\end{cases}, \label{smooth_switch_fun}
\end{align}
in which $\varphi(t) \coloneqq \alphah(t) - \varrho(t)$. It is important to note that the third-order polynomial in \eqref{smooth_switch_fun} is deliberately designed to ensure that $\iota(t)$ varies smoothly between 1 and 0.
\begin{figure}[!tbp]
	\centering
	\includegraphics[width=0.8\linewidth]{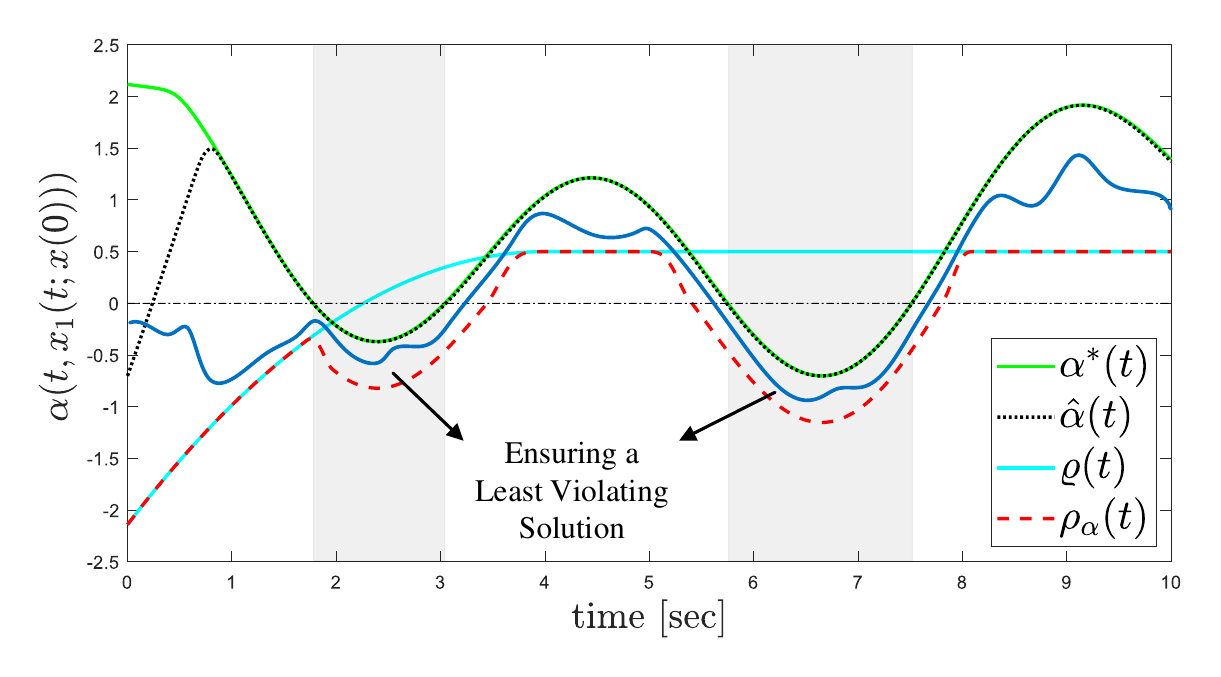}
	\caption{The evolution of $\alpha(t,x_1(t;x(0)))$ under the consolidating constraint \eqref{eq:consuli_const}, where $\ralph(t)$ is determined by \eqref{eq:rho_lower_optim_schme}. The adaptation of $\ralph(t)$ (dashed line) based on the evolution of $\alphah(t)$ in \eqref{eq:first_order_gradaccent} (dotted line) allows for deviations of $\ralph(t)$ from the nominal lower bound function $\varrho(t)$ in \eqref{eq:alpha_lower_bound_nomi}. Consequently, satisfaction of \eqref{eq:consuli_const} during the time intervals when $\alphastr(t) < 0$ (shaded intervals) results in a least violating solution. In this illustrative example,  roughly after one second $\alphah(t)$ provides a reliable estimate of $\alphastr(t)$. \vspace{-0.4cm}}
	\label{fig:illust}
\end{figure}

The logic behind the design in \eqref{eq:alpha_lower_bound_nomi}, \eqref{eq:rho_lower_optim_schme} and \eqref{smooth_switch_fun} is summarized as follows: first, $\varrho(t)$ in \eqref{eq:alpha_lower_bound_nomi} is designed as the nominal lower bound on $\alpha(t,x_1(t;x(0)))$ to address the user's desired specifications regarding the satisfaction of the time-varying output constraints while ignoring whether these constraints are feasible or not for all time. Next, $\ralph(t)$ in \eqref{eq:rho_lower_optim_schme} is designed as a convex combination of two terms such that when $\alphah(t) - \varrho(t) > \mu$, we obtain the nominal lower bound behavior $\ralph(t) = \varrho(t)$. Otherwise, when $\alphah(t) - \varrho(t) < 0$, we get $\ralph(t) = \alphah(t) - \mu$. The transition between these two modes is achieved through the smooth switch \eqref{smooth_switch_fun}. Note that since \eqref{eq:rho_lower_optim_schme} is a convex combination, $\ralph(t)$ always takes a value between $\varrho(t)$ and $\alphah(t) - \mu$ during the transition phase where $0 \leq \varphi(t) \leq \mu$. In particular, by employing \eqref{eq:rho_lower_optim_schme}, we allow the lower bound $\ralph(t)$ in \eqref{eq:consuli_const} to deviate from its nominal behavior $\varrho(t)$ in order to achieve a minimum user-defined gap of $\mu$ with respect to $\alphah(t)$. Fig.~\ref{fig:illust} illustrates the behavior of $\ralph(t)$ in \eqref{eq:rho_lower_optim_schme}.

\begin{lemma} \label{lem:least_violating}
	Let $\etil \coloneqq \alphastr(t) - \alphah(t)$. If $\alphastr(t)< 0 $ and $\varrho(t) \geq 0$ for all $t \in I$, where $I$ is some unknown time interval, then the satisfaction of \eqref{eq:consuli_const} under $\ralph(t)$ given by \eqref{eq:rho_lower_optim_schme} guarantees a least violating solution with the gap of $\mu^\ast = \etil + \mu$.
\end{lemma} 
\begin{IEEEproof}
	First, note that since $\alphah(t) = \alpha(t,\xtil_1) \leq \alphastr(t), \forall t \geq 0$, we always have $\etil \geq 0$. Given the conditions in the lemma it is easy to verify that $\varphi(t) = \alphah(t) - \varrho(t) < 0$ for all $t \in I$. Hence, from \eqref{eq:rho_lower_optim_schme} and \eqref{smooth_switch_fun} we get $\ralph(t) = \alphah(t) - \mu$. Consequently, the satisfaction of \eqref{eq:consuli_const} leads to $\alphah(t) - \mu < \alpha(t,x_1(t;x(0))), \forall t \in I$, which is equivalent to $\alphastr(t) - \mu^\ast < \alpha(t,x_1(t;x(0))), \forall t \in I$, with $\mu^\ast = \etil + \mu$.
\end{IEEEproof}

Lemma \ref{lem:least_violating} clarifies the impact of \(\tilde{e}\) and the tunable constant \(\mu > 0\) in \eqref{eq:rho_lower_optim_schme} on the gap of the obtained LVS when using \eqref{eq:rho_lower_optim_schme} in \eqref{eq:consuli_const}. Specifically, the more accurately \(\hat{\alpha}(t)\) estimates \(\alpha^*(t)\), the smaller the gap \(\mu^\ast\) becomes.

Consider the case where $\alphastr(t) > 0, \forall t \in I$, indicating that the time-varying constrained set $\Omega(t)$ is feasible for all $t \in I$, and further assume that $\alphastr(t) - \varrho(t) > \mu, \forall t \in I$. In this scenario, when $\alphah(t)$ poorly estimates $\alphastr(t)$, a situation may arise where $\varphi = \alphah(t) - \varrho(t) \leq \mu$, or particularly, $\varphi < 0$ in \eqref{smooth_switch_fun}. Consequently, $\ralph(t)$ in \eqref{eq:rho_lower_optim_schme} might not effectively follow the intended behavior designed by $\varrho(t)$ for all $t \in I$. This discrepancy can lead to a certain degree of conservativeness in fulfilling the output constraints. To clarify, even when the constraints are feasible at all time, enforcing \eqref{eq:consuli_const} under \eqref{eq:rho_lower_optim_schme} might not guarantee constraint satisfaction if $\alphah(t)$ has a very poor performance in estimating $\alphastr(t)$. Therefore, it is crucial to properly tune the parameters $k_{\alpha}$, $\epsilon_g$, and $\mch$ in \eqref{eq:first_order_gradaccent} and \eqref{smooth_switch_fun_estimator}, respectively, to enhance the performance of \eqref{eq:first_order_gradaccent} especially when $\alphastr(t)$ does not vary slowly enough with time. 

\begin{figure}[!tbp]
	\centering
	\includegraphics[width=0.93\linewidth]{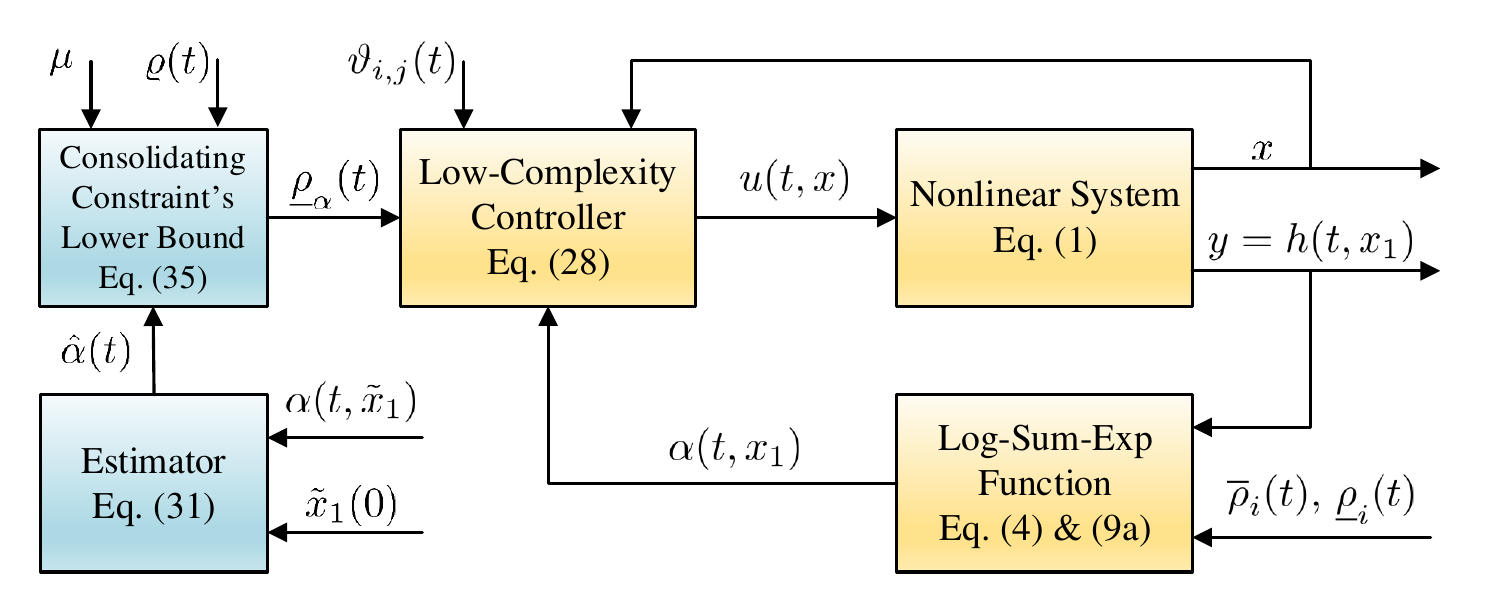}
	\caption{Cascaded control architecture under the estimation scheme \eqref{eq:first_order_gradaccent} and online computation of $\ralph(t)$ in \eqref{eq:rho_lower_optim_schme}. \vspace{-0.3cm}}
	\label{fig:cascade}
\end{figure}

Notice that, for a given $\xtil_1(0)$ the dynamical system \eqref{eq:first_order_gradaccent} runs in parallel with the closed-loop system dynamics \eqref{eq:sys_dynamics_highorder}. It generates $\alphah(t)$ at each time instant $t$, which is then used in the (online) computation of $\ralph(t)$ in \eqref{eq:rho_lower_optim_schme}. Recall that $\ralph(t)$ is utilized in the control law $u(t,x)$, specifically in the first intermediate control \eqref{eq:1st_intermed_ctrl_explicit}. Therefore, estimator dynamics \eqref{eq:first_order_gradaccent} is connected to the closed-loop system in a cascaded form (see Fig.~\ref{fig:cascade}), and thus is independent of \eqref{eq:sys_dynamics_highorder}.

Before concluding this section we show that the results stated in Theorem \ref{th:main} still hold under $\ralph(t)$ given in \eqref{eq:rho_lower_optim_schme}. In this regard, to ensure the boundedness of $\alphah(t)$ in \eqref{eq:first_order_gradaccent}, we require the following technical assumption:
\begin{assumption} \label{assu:estimation_technical}
	Given a sufficiently small $\mch>0$ in \eqref{smooth_switch_fun_estimator} the set $\Og \coloneqq \{\xtil_1 \in \R^n \mid \|\gradxtilalph(t,\xtil_1)\| \leq \mch \} \subset \R^n$ is compact for all $t \geq 0$.
\end{assumption}

Recall that $\alpha(t,\xtil_1)$ is smooth, has compact level curves, and attains only global maxima at which $\|\gradxtilalph(t,\xtil_1)\| = 0$ holds for all time (Assumption \ref{assu:alpha_globalmax}). In this respect  there always exists a sufficiently small $\mch>0$ validating the above assumption. Therefore, Assumption \ref{assu:estimation_technical} is not restrictive in practice.

\begin{theorem} \label{th:refined_rho}
	Consider the estimation scheme \eqref{eq:first_order_gradaccent} with an arbitrary initialization $\xtil_1(0)$ and let $\ralph(t)$ be given by \eqref{eq:rho_lower_optim_schme}. Moreover, suppose that $\varrho_0$ in \eqref{eq:alpha_lower_bound_nomi} is selected such that $\varrho_0 < \alpha(0,x_1(0))$. Under Assumptions \ref{assum:output_map_jacob}, \ref{assum:output_map_elements}, \ref{assum:coercive_alphabar}, \ref{assu:alpha_globalmax}, and \ref{assu:estimation_technical} $\ralph(t)$ attains Properties (i) and (ii) mentioned in Subsection \ref{subsec:single_const}. Moreover, $\dralph(t)$ is bounded. Therefore, under the requirements stated in Theorem \ref{th:main}, the control law \eqref{eq:control_law} ensures the satisfaction of  $\ralph(t) < \alpha(t,x_1(t;x(0)))$ and the boundedness of all closed-loop signals for all time.
\end{theorem}
\begin{IEEEproof}
	See Appendix \ref{appen:proof_th_leastviolating}.
\end{IEEEproof} 

\begin{remark} \label{rem:estimation_err_ultimate_bound}
	Notice that the boundedness of $\alphah(t)$ is established in the proof of Theorem \ref{th:refined_rho} using Assumption \ref{assu:estimation_technical}. Additionally, $\alphastr(t)$ is known to be bounded by construction due to the compact level curves of $\alpha(t,\xtil_1)$. Thus, the boundedness of the estimation error $\etil = \alphastr(t) - \alphah(t) \geq 0$ for all time is straightforward without further analysis. Moreover, by examining the dynamics of $\etil$, one can verify that a larger $k_{\alpha}>0$ and smaller $\epsilon_g > 0$ and $\mch > 0$ in \eqref{eq:first_order_gradaccent_dyn} and \eqref{smooth_switch_fun_estimator} yield a smaller ultimate bound for $\etil$. However, deriving an explicit relation between these parameters and the ultimate bound of the estimation error may not be possible, as it requires knowing the upper bound of $|\dalphastr(t)|$,  which exists but is typically unknown.
\end{remark}

\section{Simulation Results}
\label{sec:simu_results}

In this section, we present two simulation examples to validate the proposed control approach. The first example demonstrates our method's effectiveness in addressing problems that are already solvable using existing approaches, such as the PPC method, where the considered time-varying output constraints are decoupled. Subsequently, we offer an example involving coupled time-varying constraints, which cannot be accommodated by previous approaches.

\begin{figure}[!tbp]
	\centering
	\includegraphics[align=c,width=0.45\linewidth]{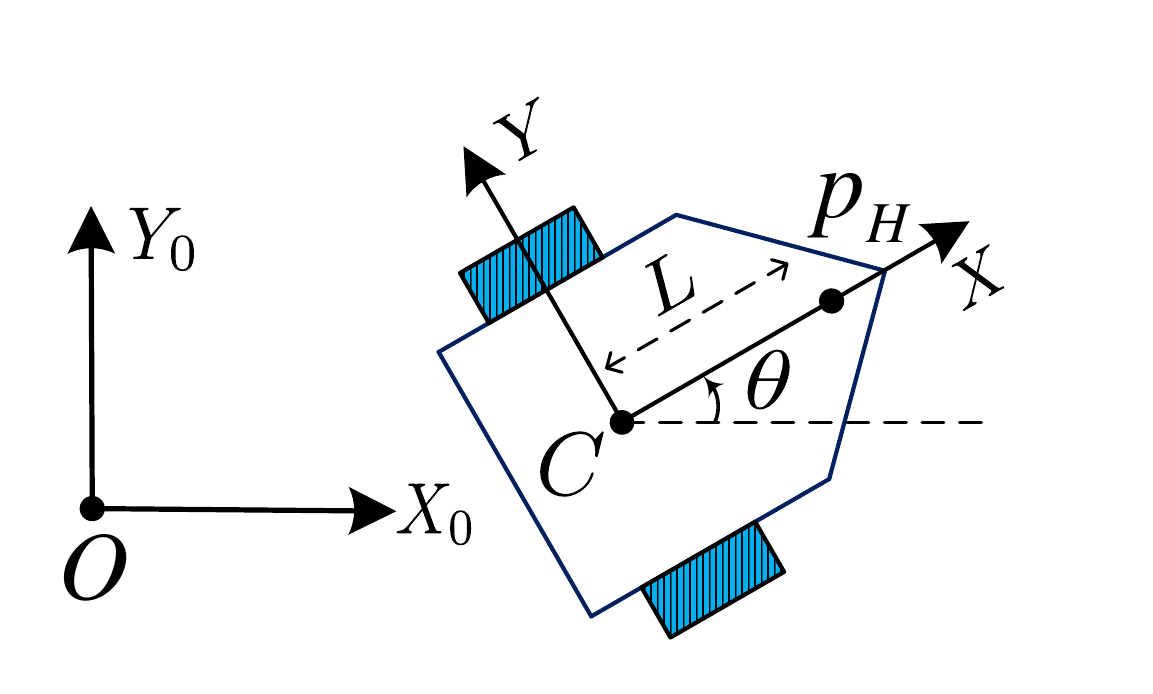}
	\caption{Mobile robot. \vspace{-0.4cm}}
	\label{fig:mobile_robot}
\end{figure}

In the upcoming simulation examples, we will consider a mobile robot operating in a 2-D plane (refer to Fig.~\ref{fig:mobile_robot}) with kinematics and dynamics expressed by the following equations:
\begin{equation} \label{eq:robot_kin_dyn}
	\begin{cases}
		\dot{p}_c = S(\theta)\zeta \\ 
		\bar{M} \dot{\zeta} + \bar{D} \zeta = \bar{u} + \bar{d}(t) 
	\end{cases}\!\!\!\!\!\!, ~
	\begin{aligned}
		S(\theta)\! =\!\! \left[ \!\!\!
		\begin{array}{ccc}
			\cos \theta &\!\! \sin \theta &\!\! 0 \\
			0 &\!\! 0 &\!\! 1
		\end{array}\!\!\!\right]^{\!\!\top}\!\!\!.\!\!\!\!\!\!\!
	\end{aligned}
\end{equation}
Here, $p_c = [x_c, y_c, \theta]^{\top}$ represents the position and orientation of the body frame ${C}$ relative to the reference frame ${O}$. The vector $\zeta = [v_T, \dot{\theta}]^{\top}$ includes the translational speed $v_T$ along the direction of $\theta$ and the angular speed $\dot{\theta}$ about the vertical axis passing through $C$. The matrices involved are defined as follows: $\bar{M} = \diag(m_R,I_R)$, where $m_R$ and $I_R$ represent the mass and moment of inertia of the robot about the vertical axis, respectively. The input $\bar{u}$ denotes the force/torque-level control inputs, $\bar{D} = \diag(\bar{D}_1, \bar{D}_2)$ is a constant damping matrix, and $\bar{d}(t)$ is the vector of bounded external disturbances.

To address the under-actuated nature of \eqref{eq:robot_kin_dyn} and avoid nonholonomic constraints, we transform it with respect to the hand position $p_H \coloneqq [x_c,y_c]^{\top} + L [\cos \theta, \sin \theta]^{\top}$ (as shown in Fig.~\ref{fig:mobile_robot}). This transformation leads to an equivalent Euler-Lagrangian dynamics in state-space form:
\begin{equation}
	\begin{aligned}\label{eq:robot_transformed_dyn}
		\!\! \dot{x}_1 &= x_2, \\ 
		\!\! \dot{x}_2 &= M(x_1)^{-1} \big( - C(x_1, x_2) x_2 - D(x_1) x_2 + u + d(t) \big). \!\!\!
	\end{aligned} 
\end{equation}
Here, $x_1$ corresponds to the hand position of the mobile robot ($x_1 = p_H$), and $x_2$ represents its velocity. The matrices $M(x_1)$, $C(x_1, x_2)$, and $D(x_1)$ are locally Lipschitz continuous functions of their arguments. The relationships between the parameters in \eqref{eq:robot_transformed_dyn} and those in \eqref{eq:robot_kin_dyn} are given by: $M = \Upsilon^\top \bar{M} \Upsilon$, $C = \Upsilon^\top \bar{M} \dot{\Upsilon}$, $D = \Upsilon^\top \bar{D} \Upsilon$, $d(t) = \Upsilon^\top \bar{d}(t)$, and $u = \Upsilon^\top \bar{u}$, where $\Upsilon = \left[\begin{smallmatrix} \cos \theta & \sin \theta \\ -(\sin \theta) / L & (\cos \theta) / L \end{smallmatrix} \right]$ \cite{cai2014adaptive}. It is worth noting that \eqref{eq:robot_transformed_dyn} can be viewed as a specific form of \eqref{eq:sys_dynamics_highorder} with $n = 2$ and $r = 2$ and it is not difficult to verify that Assumptions \ref{assum:uncertain_f}, \ref{assum:symm_g} hold for \eqref{eq:robot_transformed_dyn}. In the simulations we set $m_R = 3.6$, $I_R = 0.0405$, $\bar{D}_1 = 0.3$, $\bar{D}_2 = 0.04$, $L = 0.2$, and $\bar{d}(t) = [0.75\sin(3t + \tfrac{\pi}{3}) + 1.5\cos(t + \tfrac{3\pi}{7}), -2.4 \exp(\cos(t+\tfrac{\pi}{3}) + 1) \sin(t)]^\top$. 

\subsection{Decoupled Time-Varying Constraints}
In our first simulation example, we will focus on trajectory tracking for the mobile robot described by \eqref{eq:robot_transformed_dyn}. The desired trajectory is $x_1^d(t) = [4.2\cos(0.47t),4.2\sin(0.47t)]^\top$. Our goal is to enforce specific performance funnel constraints on tracking errors, defined as follows: $-\rho_i(t) < x_{1,i} - x_{1,i}^d(t) < \rho_i(t)$ for $i \in \I_1^2$ and all $t \geq 0$. Here, $\rho_1(t)$ and $\rho_2(t)$ are strictly positive, time-varying performance bounds. Without loss of generality, we assume $\rho_1(t) = \rho_2(t) = (1.75 - 0.3)\exp(-0.35t) + 0.3$.

To express these requirements analogously to the problem formulation in Section \ref{sec:problemformulation}, we consider $y = h(x_1) = x_1$ for the dynamics \eqref{eq:robot_transformed_dyn} under the following funnel constraints: $\rl_i(t) \coloneqq -\rho_i(t) + x_{1,i}^d(t) < x_{1,i} < \rho_i(t) + x_{1,i}^d(t) \eqqcolon \ru_i(t)$ for $i \in \I_1^2$ and all $t \geq 0$. Note that $h(x_1)$ readily satisfies Assumptions \ref{assum:output_map_jacob} and \ref{assum:output_map_elements}. Moreover, the above funnel constraints resemble a time-varying box constraint in the $x_1$ space (i.e., Assumption \ref{assum:coercive_alphabar} holds). As $\rho_1(t)$ and $\rho_2(t)$ are strictly positive, both funnel constraints are well-defined and feasible. Furthermore, these two funnel constraints are decoupled, as each one imposes time-varying upper and lower bounds on independent state variables, namely $x_{1,1}$ and $x_{1,2}$. This feature is also evident by verifying that Condition II of Lemma \ref{lem:global_max_suffi} holds. Satisfaction of Condition II of Lemma \ref{lem:global_max_suffi} also indicates that Assumption \ref{assu:alpha_globalmax} is valid. Now since the aforementioned funnel constraints are well-defined and decoupled, they are mutually satisfiable for all time. Therefore, the constrained set $\Ob(t)$ defined in \eqref{eq:omega_alpha_bar} is guaranteed to be feasible for all $t \geq 0$. 

In this specific example, one can reasonably assume that Assumption \ref{assu:feasible_output_constr} holds, given that $\Ob(t)$ remains feasible for all time and does not become excessively tight over certain time intervals (i.e., having overly stringent time-varying constraints). Indeed, by selecting a sufficiently large $\nu$ in \eqref{smooth_alph} one can get a closer under-approximation of $\Ob(t)$, which provides more confidence on feasibility of $\Omega(t)$. Satisfaction of  Assumption \ref{assu:feasible_output_constr} allows us to use the suggested design of $\ralph(t)$ in \eqref{eq:alpha_lower_bound} for consolidating constraint \eqref{eq:consuli_const}.

\begin{figure}[!tbp]
	\centering
	\begin{subfigure}[t]{0.49\linewidth}
		\centering
		\includegraphics[width=\linewidth]{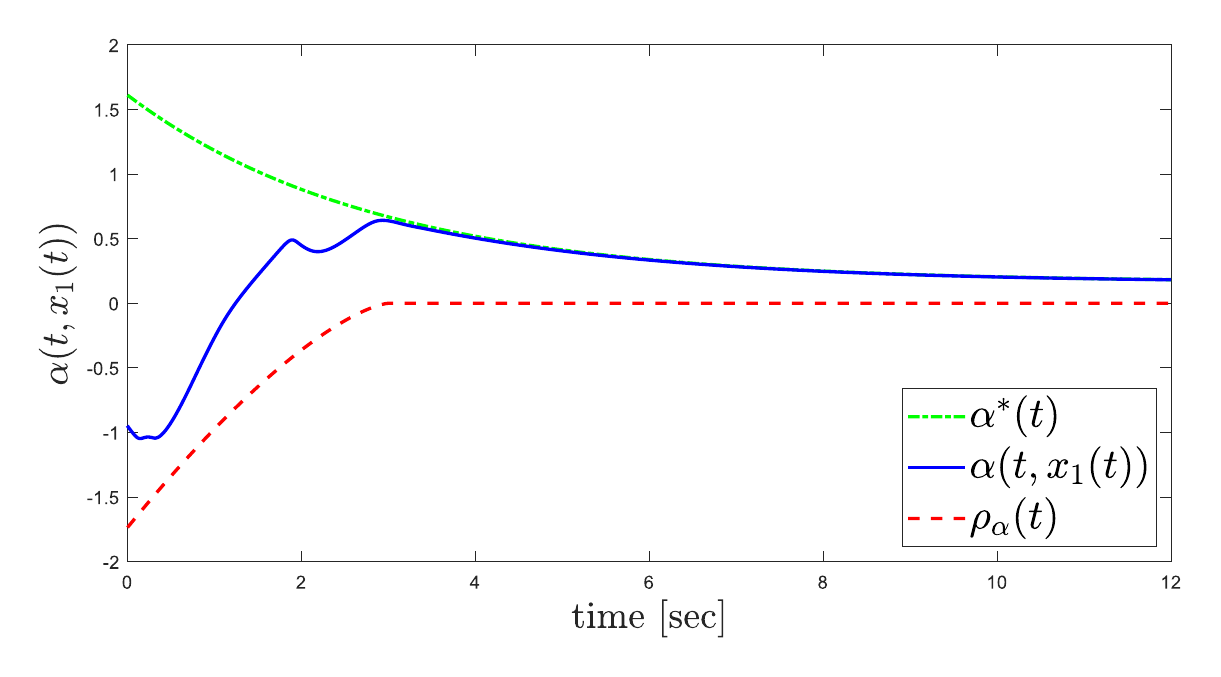}
	\end{subfigure}
	\vspace{0.1cm}
	\begin{subfigure}[t]{0.49\linewidth}
		\centering
		\includegraphics[width=\linewidth]{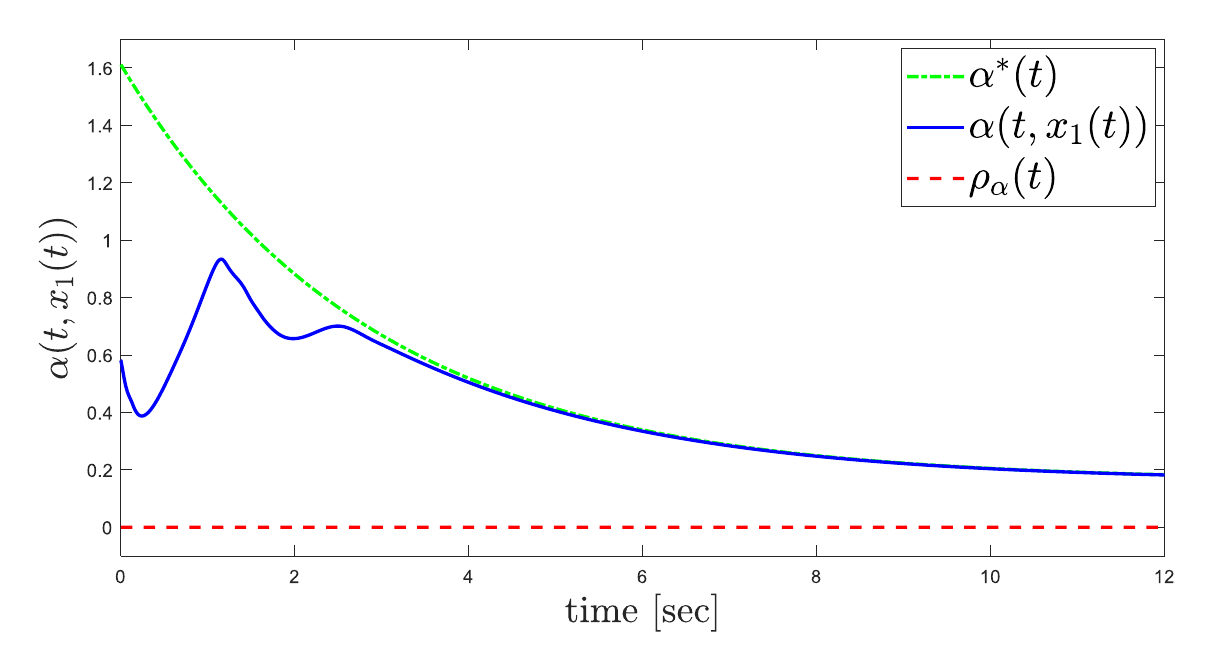}
	\end{subfigure}
	\vspace{0.1cm}
	\begin{subfigure}[t]{0.49\linewidth}
		\centering
		\includegraphics[width=\linewidth]{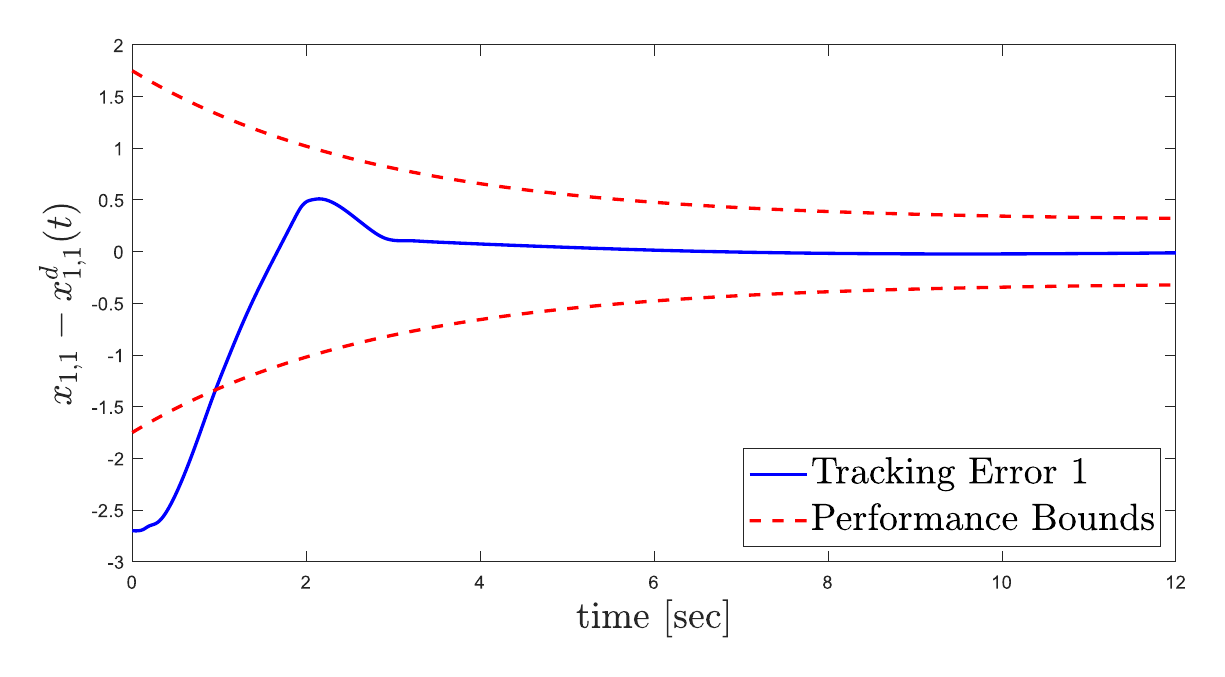}
	\end{subfigure}
	\vspace{0.1cm}
	\begin{subfigure}[t]{0.49\linewidth}
		\centering
		\includegraphics[width=\linewidth]{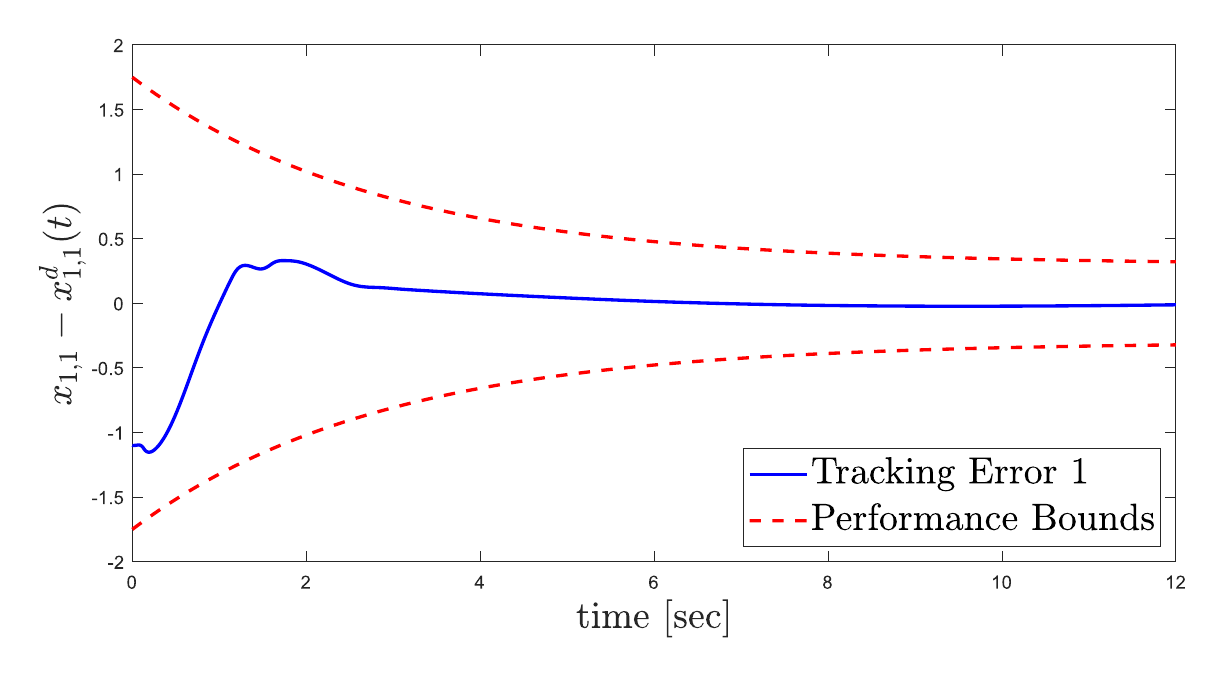}
	\end{subfigure}
	\vspace{0.1cm}
	\begin{subfigure}[t]{0.49\linewidth}
		\centering
		\includegraphics[width=\linewidth]{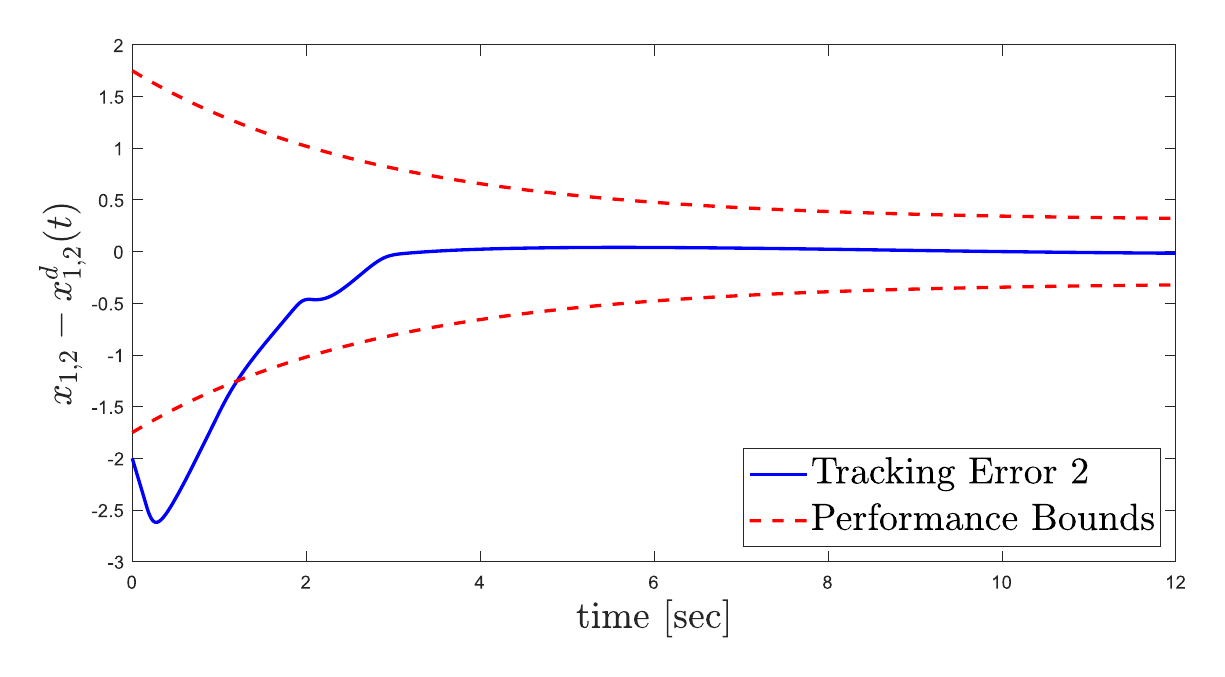}
	\end{subfigure}
	\begin{subfigure}[t]{0.49\linewidth}
		\centering
		\includegraphics[width=\linewidth]{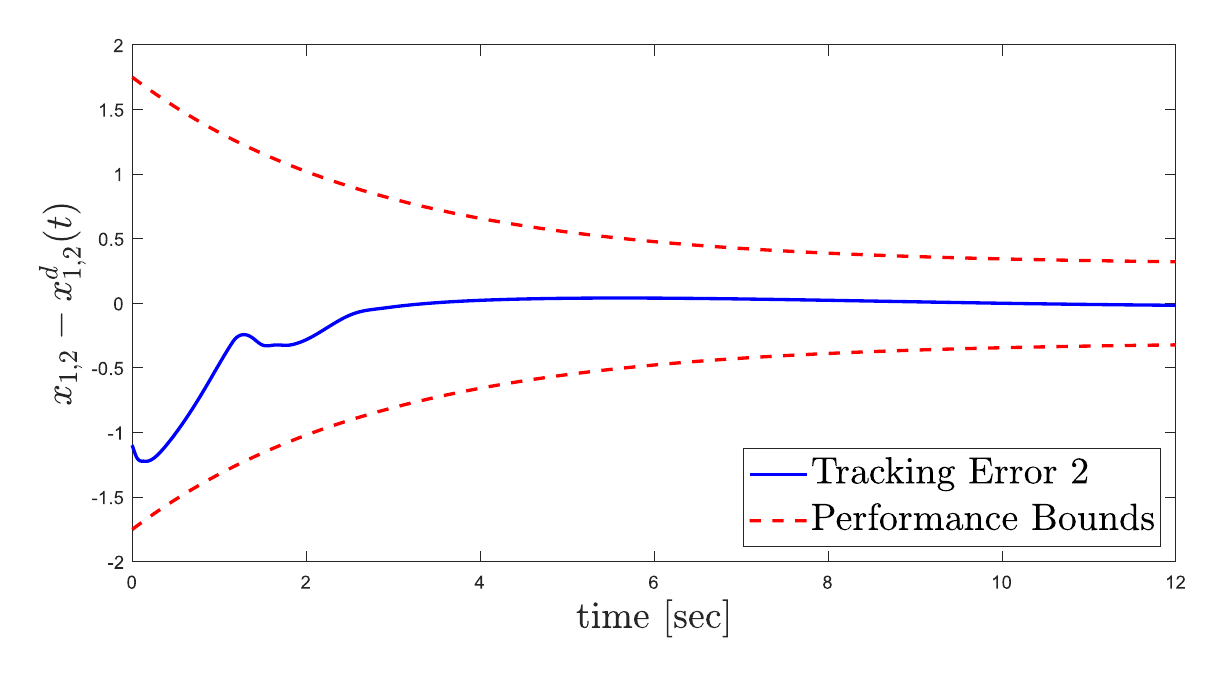}
	\end{subfigure}
	\begin{subfigure}[t]{0.49\linewidth}
		\centering
		\includegraphics[width=\linewidth]{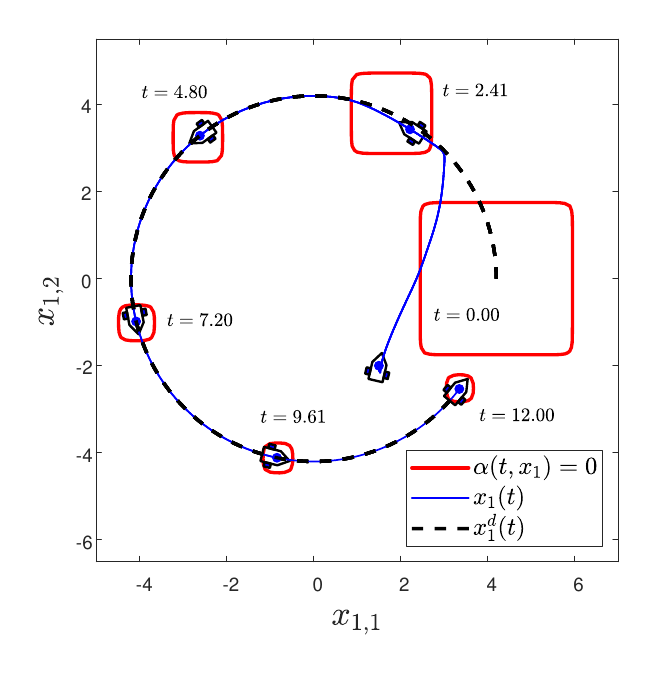}
		\caption{}
		\label{fig:PPC_scenario_out}
	\end{subfigure}
	\begin{subfigure}[t]{0.49\linewidth}
		\centering
		\includegraphics[width=\linewidth]{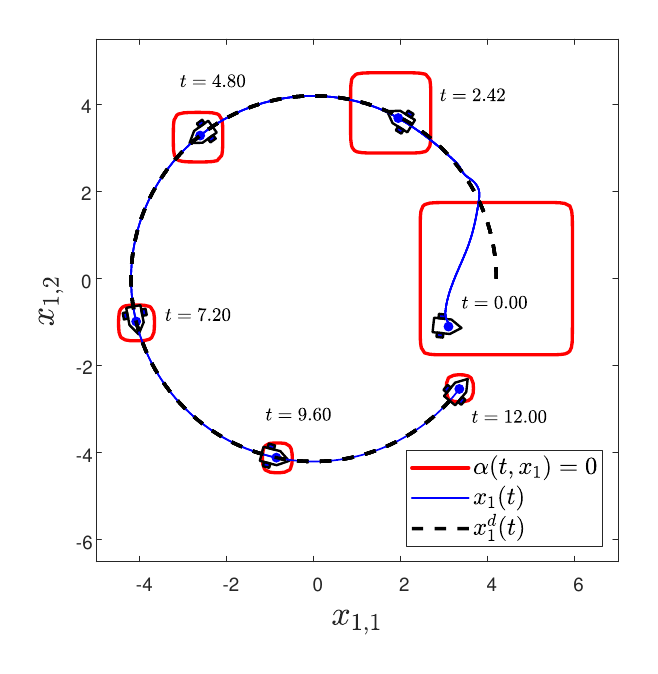}
		\caption{}
		\label{fig:PPC_scenario_in}
	\end{subfigure}
	\caption{Trajectory tracking of the mobile robot under prescribed performance conditions. (a) The scenario where the tracking errors performance specifications are not initially satisfied. (b) The situation in which the performance specifications are initially met. \vspace{-0.6cm}}
	\label{fig:PPC_Scenario}
\end{figure}

The numerical values of the parameters used to implement control law \eqref{eq:control_law} are provided in Table \ref{tab:contr_pram_simu1}. It is important to note that, as per the guidelines outlined in Subsections \ref{subsec:ralph_design_simpl} and \ref{subsec:control_design}, the values of $\rho_0$ and $\vartheta_{2,j}^0$, $j \in \I_1^2$ in Table \ref{tab:contr_pram_simu1}, are determined based on the initial condition $x(0)$ of the transformed mobile robot dynamics in \eqref{eq:robot_transformed_dyn}. Henceforth, we use $x_1(t)$ instead of $x_1(t;x(0))$ for brevity. Fig.~\ref{fig:PPC_Scenario} shows the evolution of $\alpha(t,x_1(t))$ and the tracking errors of the mobile robot's hand position under \eqref{eq:control_law} for two scenarios. 

\begin{table}[!tbp]
	\centering
	\begin{tabular}{c|c}
		Eq. no  & Parameter(s) \\ 
		\toprule
		\eqref{smooth_alph_def}  & $\nu = 10$ \\ 
		\hline
		\eqref{eq:alpha_lower_bound} & $T = 3$, $\beta = 0.3$, $\rho_{\infty} = 0$, $\rho_0 < \alpha(0,x_1(0))$ \\ 
		\hline
		\eqref{eq:mapped_alpha} &  $\upsilon = 8$ \\
		\hline
		\eqref{eq:1st_intermed_ctrl_explicit} & $k_1 = 1$ \\ 
		\hline
		\eqref{exponential_performance_fun} & $\vartheta_{2,j}^{\infty} = 0.1$, $l_{2,j} = 1$, $\vartheta_{2,j}^0 > |e_{2,j}(0,\bar{x}_{2}(0))|, j \in \I_1^2$ \\
		\hline
		\eqref{eq:i_th_intermed_ctrl_explicit} & $k_2 = 1$ \\
		\hline
	\end{tabular}
	\caption{Numerical values of the parameters involved in control law \eqref{eq:control_law}. \vspace{-0.3cm}}
	\label{tab:contr_pram_simu1}
\end{table}

In the first scenario (Fig.~\ref{fig:PPC_scenario_out}), since $\alpha(0,x_1(0)) < 0$ the robot's initial position does not initially satisfy the prescribed performance bounds on the tracking errors ($x_1(0) \notin \Ob(0)$). However, by enforcing \eqref{eq:consuli_const} under $\ralph(t)$ in \eqref{eq:alpha_lower_bound} and applying \eqref{eq:control_law}, we observe that $\alpha(t,x_1(t))$ becomes and remains positive within the user-defined finite time limit of $T = 3$ seconds. This signifies the achievement of the tracking error performance specifications within 3 seconds.

In the second scenario (Fig.~\ref{fig:PPC_scenario_in}), where $\alpha(0,x_1(0)) > 0$, the performance bounds on the tracking errors are initially satisfied ($x_1(0) \in \Ob(0)$). By maintaining $\alpha(t,x_1(t))$ positive, we ensure the continuous fulfillment of the tracking errors' specifications throughout the simulation. It is worth noting that, as anticipated, $\alphastr(t)$ remains positive for all time, which indicates the feasibility of $\Omega(t)$ for all time. Note that, $\alphastr(t)$ is unknown to the control system and is included in the figures solely for the purpose of verifying the simulation results. The value of $\alphastr(t)$ at each time step is obtained through solving optimization \eqref{eq:alpha_opt_TVO} offline for a dense set of time instances.

Lastly, Fig.~\ref{fig:PPC_Scenario} (bottom) provides a visual representation of the simulation results through snapshots of the mobile robot's hand position trajectory $x_1(t)$ along with the time-varying constrained set $\Omega(t)$ for both scenarios (recall that $\partial \cl(\Omega(t)) = \{x_1 \in \R^2 \mid \alpha(t, x_1) = 0\}$).

The simulation results presented above highlight a key advantage of our proposed control methodology. Unlike conventional PPC and TVBLF-based control design approaches, which necessitate the initial satisfaction of the (output) constraints for their effective implementation, our approach operates without such restrictions.

\subsection{Coupled Time-Varying Constraints}
For our second simulation example we consider coupled time-varying (output) constraints, for which previous approaches (FC, PPC, TVBLF-based control) are not applicable. Moreover, previous approaches cannot ensure a least violating solution (as per \eqref{eq:least_vio_def}) when constraint infeasibilities arise.

Consider the transformed mobile robot dynamics in \eqref{eq:robot_transformed_dyn} with the output map $y = h(t,x_1) = \left[h_1(t,x_1), h_2(t,x_1), h_3(t,x_1)\right]^\top$, for which we assume the following (coupled) time-varying constraints: $\rl_1(t) < h_1(t,x_1) < \ru_1(t)$ (funnel constraint), $\rl_2(t) < h_2(t,x_1)$ (LBO constraint), and $h_3(t,x_1) < \ru_3(t)$ (UBO constraint), where $\rl_1(t) = - 0.7 - \sin(0.4t)$, $\ru_1(t) = 1.1 + 3\sin(0.45t)$, $\rl_2(t) = -1 - 0.5\cos(0.3t)$, and $\ru_3(t) = 0.5+\sin(0.4t)$. Moreover, let $h_1(t,x_1) = x_{1,1} - o_1(t)$, $h_2(t,x_1) = c_1(t) (x_{1,1} - o_1(t))^2 + c_2(t) (x_{1,2} - o_2(t)) + c_3(t) (x_{1,1} - o_1(t))$, and $h_3(t,x_1) = c_4(t) (x_{1,1} - o_1(t))^2 + (x_{1,2} - o_2(t))$, in which $o_1(t) = 5\cos(0.28t), o_2(t) = 5\sin(0.28t)$, $c_1(t) = -2 + 2\cos(t)$ , $c_2(t) = 1 + 0.5 \sin(0.7t)$, $c_3(t) = \sin(0.4t)$, and $c_4(t) = 1 - \cos(0.5t)$ are all bounded continuously differentiable functions of time. The time-varying output map $h(t, x_1)$ satisfies Assumptions \ref{assum:output_map_jacob} and \ref{assum:output_map_elements}. Furthermore, due to the way the constraints are designed, the set $\Ob(t)$ (and consequently $\Omega(t)$) remains bounded for all time (Assumption \ref{assum:coercive_alphabar}).  Moreover, it can be verified that the constraints fulfill Condition I of Lemma \ref{lem:global_max_suffi}, thus confirming the validity of Assumption \ref{assu:alpha_globalmax}. In simple terms, these constraints define a bounded time-varying region that the mobile robot's (hand) position should enter and remain within for all time (i.e., a time-varying region tracking problem). Nevertheless, we did not assume that the constrained region is always feasible. Therefore, we utilize the proposed estimation scheme \eqref{eq:first_order_gradaccent} along with $\ralph(t)$ given by \eqref{eq:rho_lower_optim_schme} for the consolidating constraint \eqref{eq:consuli_const}.  

For the simulations of this subsection, all numerical values used for the control law \eqref{eq:control_law} match those in Table \ref{tab:contr_pram_simu1}, with the only difference being that $\ralph(t)$ follows \eqref{eq:rho_lower_optim_schme}. Specifically, we set $\mu = 0.2$ in \eqref{eq:rho_lower_optim_schme}, and the parameters for $\varrho(t)$ in \eqref{eq:alpha_lower_bound_nomi} are set to $\varrho_0 < \alpha(0,x_1(0))$, $\varrho_{\infty} = 0.5$, $T = 3$, and $\beta = 0.3$. Building on the discussion in Subsection \ref{subsec: estimator}, we conduct two simulations to highlight how the performance of the estimation scheme \eqref{eq:first_order_gradaccent} impacts constraint satisfaction under control law \eqref{eq:control_law}. We consider two cases: (\textbf{A}) setting $k_{\alpha} = 2, \epsilon_g = 1, \mch = 0.1$, and (\textbf{B}) setting $k_{\alpha} = 0.2, \epsilon_g = 10, \mch = 1$ in \eqref{eq:first_order_gradaccent_dyn} and \eqref{smooth_switch_fun_estimator}. Both simulations assume that the initial condition for the estimator \eqref{eq:first_order_gradaccent_dyn} is the same as the initial hand position of the mobile robot, i.e., $\xtil_1(0) = x_1(0)$, leading to $\alphah (0) = \alpha(t,x_1(0))$.

\begin{figure}[!tbp]
	\centering
	\begin{subfigure}[t]{0.49\linewidth}
		\centering
		\includegraphics[width=\linewidth]{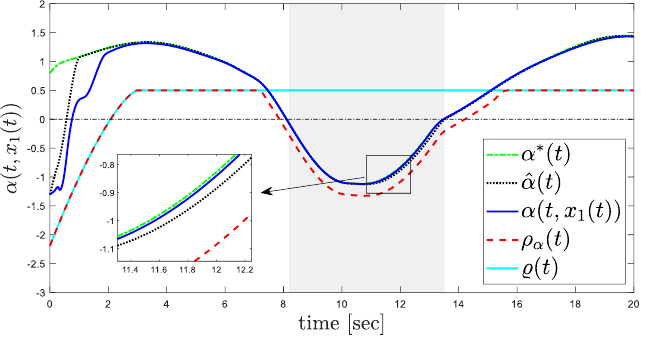}
	\end{subfigure}
	\vspace{0.15cm}
	\begin{subfigure}[t]{0.49\linewidth}
		\centering
		\includegraphics[width=\linewidth]{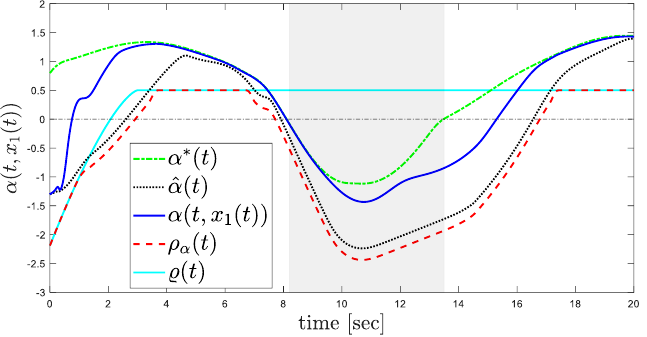}
	\end{subfigure}
	\vspace{0.15cm}
	\begin{subfigure}[t]{0.486\linewidth}
		\centering
		\includegraphics[width=\linewidth]{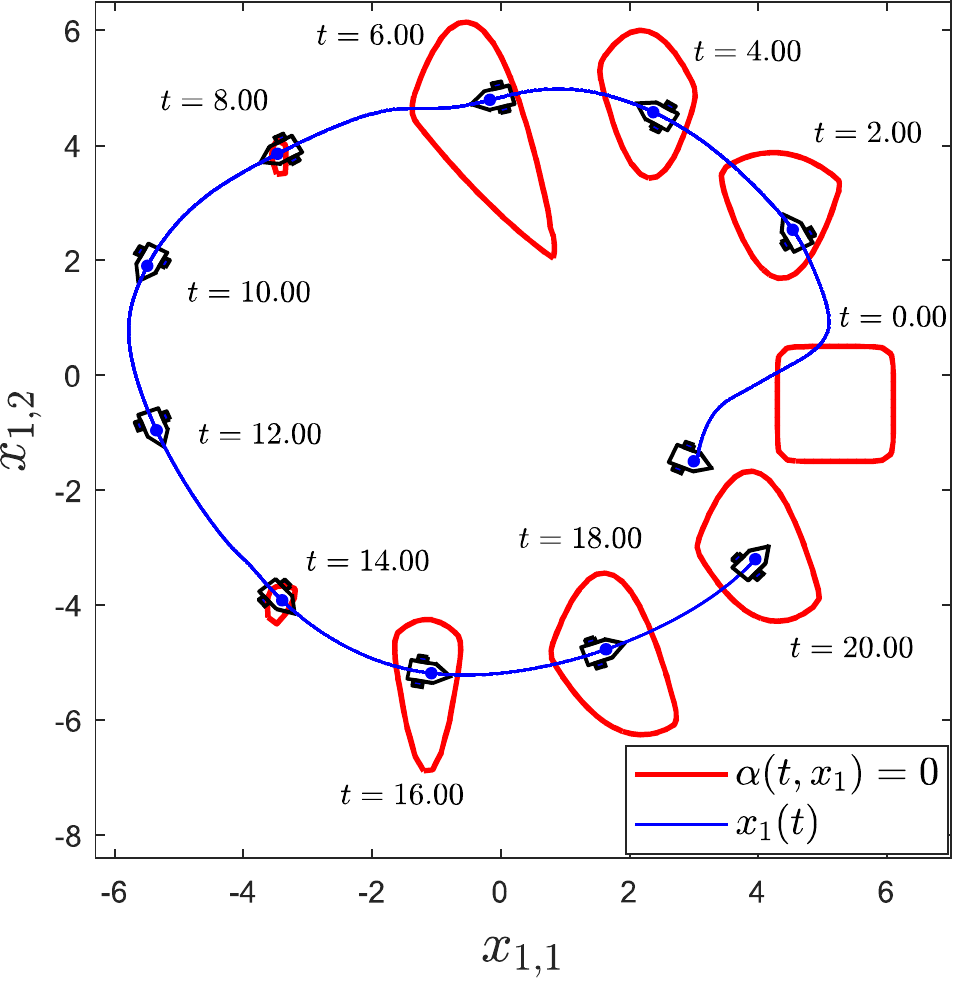}
		\caption{\vspace{-0.2cm}}
		\label{fig:coupled1}
	\end{subfigure}
	\begin{subfigure}[t]{0.49\linewidth}
		\centering
		\includegraphics[width=\linewidth]{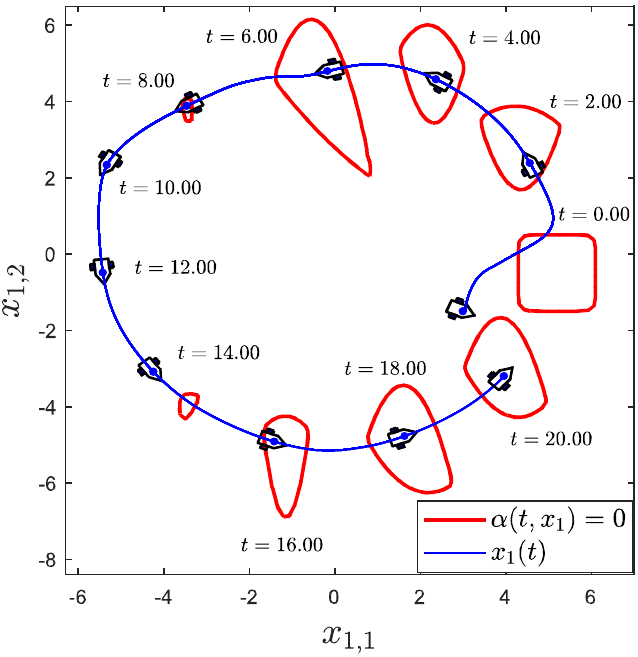}
		\caption{\vspace{-0.2cm}}
		\label{fig:coupled2}
	\end{subfigure}
	\caption{Time-varying region tracking of the mobile robot. (a) When the estimator's tuning parameters are set to $k_{\alpha} = 2, \epsilon_g = 1, \mch = 0.1$, a minimal conservative behavior in satisfaction of the time-varying constraints (or region tracking) is observed, owing to the estimator's good performance in estimating (unknown) $\alphastr(t)$. Moreover, a least violating solution is ensured with a small gap whenever the constraints become infeasible (empty region). (b) Setting the estimator's parameters to $k_{\alpha} = 0.2, \epsilon_g = 10, \mch = 1$  leads to a least violating solution with a larger gap, that adversely impacts the control law's performance, resulting in a weaker satisfaction of time-varying constraints. \vspace{-0.5cm}}
	\label{fig:PPC_Scenariosss}
\end{figure}

Fig.~\ref{fig:coupled1} (top) shows the evolution of $\alpha(t,x_1(t))$ under \eqref{eq:control_law} with the estimator's parameters tuned according to case (\textbf{A}). After a brief transient period, the estimator's output ($\alphah(t)$) closely follows $\alphastr(t)$, such that the estimation error remains small for all time. Additionally, thanks to the satisfaction of consolidating constraint \eqref{eq:consuli_const}, time-varying output constraints are guaranteed to be met with a margin of $\varrho_{\infty} = 0.5$ after a user-defined $T = 3$ seconds. However, for a time interval between $t = 8$ and $t = 14$ (shaded interval), we get $\alphastr(t) < 0$, indicating that the constraints become temporarily infeasible. During this time, the proposed $\ralph(t)$ \eqref{eq:rho_lower_optim_schme} diverts from the nominal lower bound $\varrho(t)$ to ensure a least violating solution. When the constraints become feasible again ($\alphastr(t) > 0$), $\ralph(t)$ quickly returns to the nominal constraint satisfaction requirement i.e., $\ralph(t) = \varrho_{\infty} = 0.5$. Finally, in Fig.~\ref{fig:coupled1} (bottom), snapshots of the mobile robot's hand position are shown along with the constrained region. Note that, the constrained region is shown only when it is feasible (nonempty).

The simulation scenario is repeated with the estimator's parameters adjusted according to case (\textbf{B}), and the results are presented in Fig.~\ref{fig:coupled2}. As discussed in Subsection \ref{subsec: estimator} and Remark \ref{rem:estimation_err_ultimate_bound}, this adjustment leads to a reduced performance in estimating $\alphastr(t)$. In Fig.~\ref{fig:coupled2} (top), we can see that the evolution of $\ralph(t)$ is influenced by $\alphah(t)$, deviating from the nominal lower bound $\varrho(t)$ (roughly) between $t = 1$ to  $t = 4$. However, as this deviation is not significant it turns out that the controller is still capable of meeting the user-defined specifications for constraints satisfaction by maintaining $\alpha(t,x_1(t))$ above the nominal lower bound $\varrho(t)$ for over 7 seconds (although only $\ralph(t) < \alpha(t,x_1(t))$ is guaranteed by the proposed controller). As the time-varying constraints tend to become infeasible (shaded interval), $\alphastr(t)$ rapidly decreases, which induces a significant divergence between $\ralph(t)$ and $\varrho(t)$ due to a large estimation error. As a result, the controller can only ensure a least violating solution with a considerably large gap. Recall that, as per Lemma \ref{lem:least_violating}, the gap for the least violating solution is given by $\mu^\ast = \etil + \mu$, where $\etil = \alphastr(t) - \alphah(t) \geq 0$ represents the estimation error. From Fig.~\ref{fig:coupled2} (top), it is evident that, even when the constraints become feasible again, owing to a rapid increase of $\alphastr(t)$ a large estimation error continues to persist for some time, which hinders $\ralph(t)$ from approaching $\varrho(t)$. This phenomenon makes the controller to present a weaker constraint satisfaction behavior. This is more evident in Fig.~\ref{fig:coupled2} (bottom), where the mobile robot is still out of the (feasible) constrained region at $t = 14$. This simulation underscores the direct impact of the estimator's performance on the control law. Therefore, if one expects that $\alphastr(t)$ can change rapidly, careful tuning of the estimator's parameters in \eqref{eq:first_order_gradaccent_dyn} and \eqref{smooth_switch_fun_estimator} becomes essential.

\begin{figure}[!tbp]
	\centering	
	\begin{subfigure}[t]{0.48\linewidth}
		\centering
		\includegraphics[width=\linewidth]{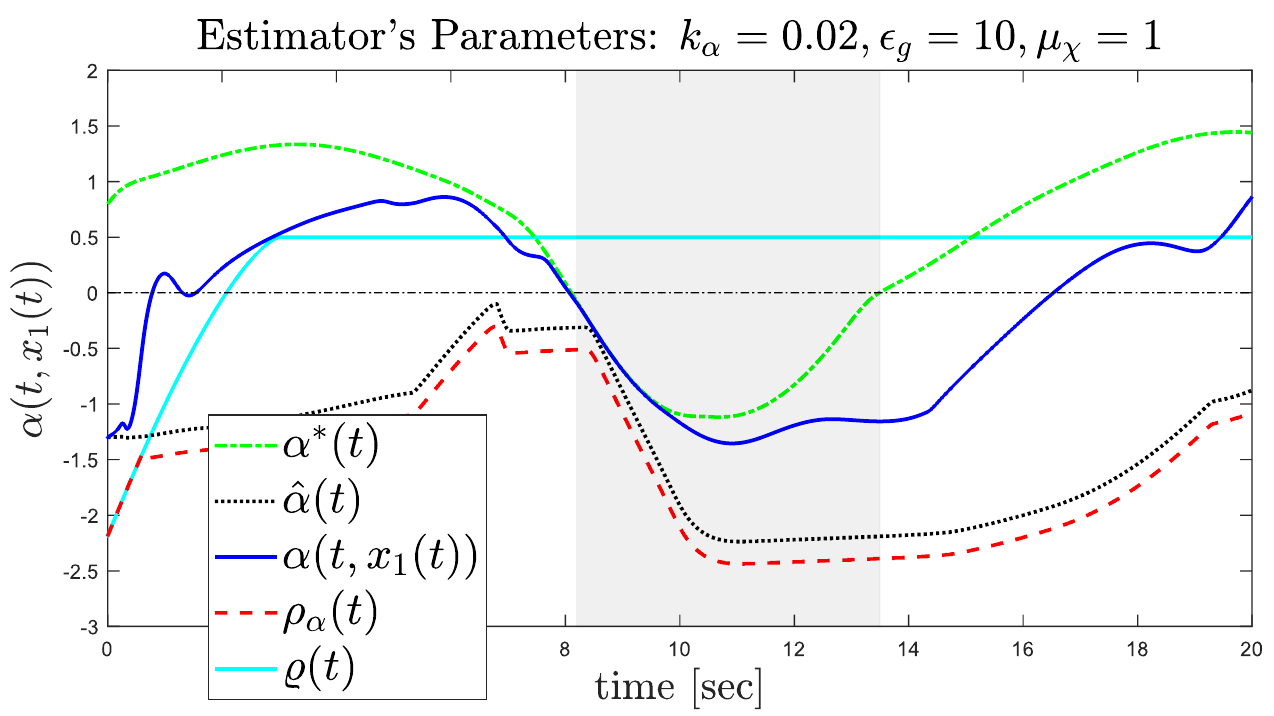}
		\caption{Smaller $k_{\alpha}$ \vspace{0.1cm}}
	\end{subfigure}
	~
	\begin{subfigure}[t]{0.48\linewidth}
		\centering
		\includegraphics[width=\linewidth]{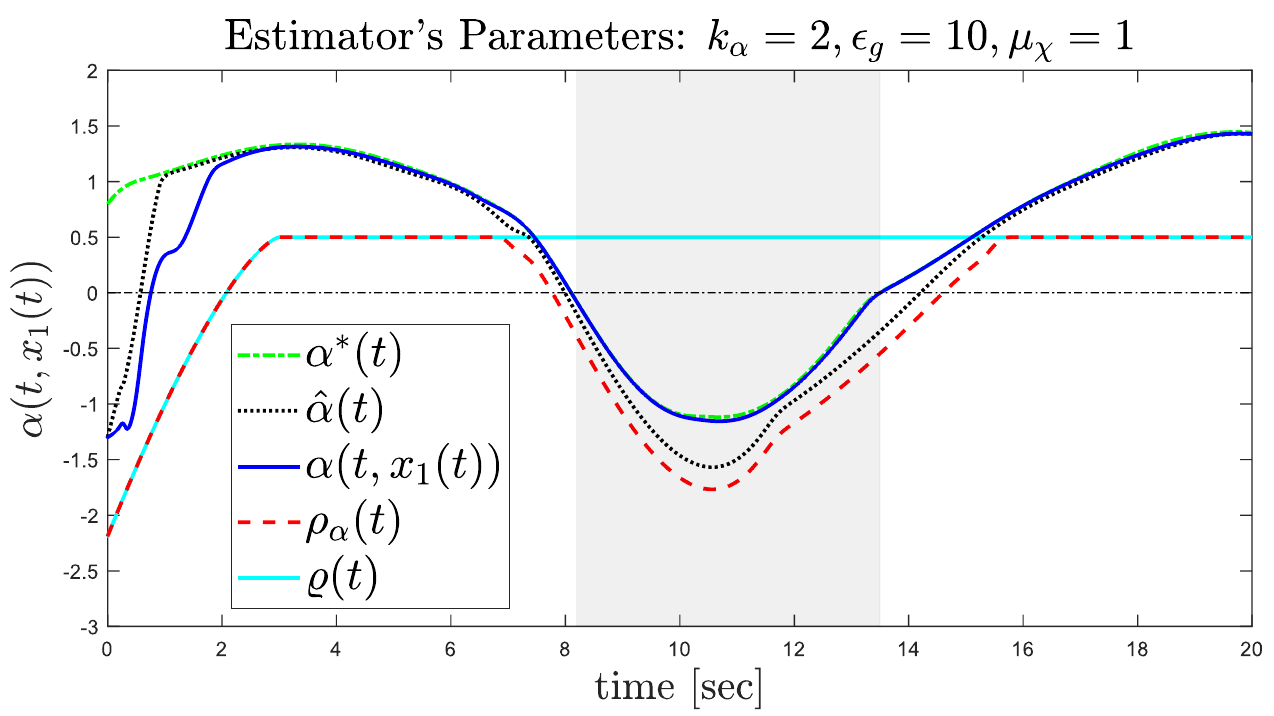}
		\caption{Larger $k_{\alpha}$  \vspace{0.1cm}}
		\label{fig:comparison_c}
	\end{subfigure}
	
	\begin{subfigure}[t]{0.48\linewidth}
		\centering
		\includegraphics[width=\linewidth]{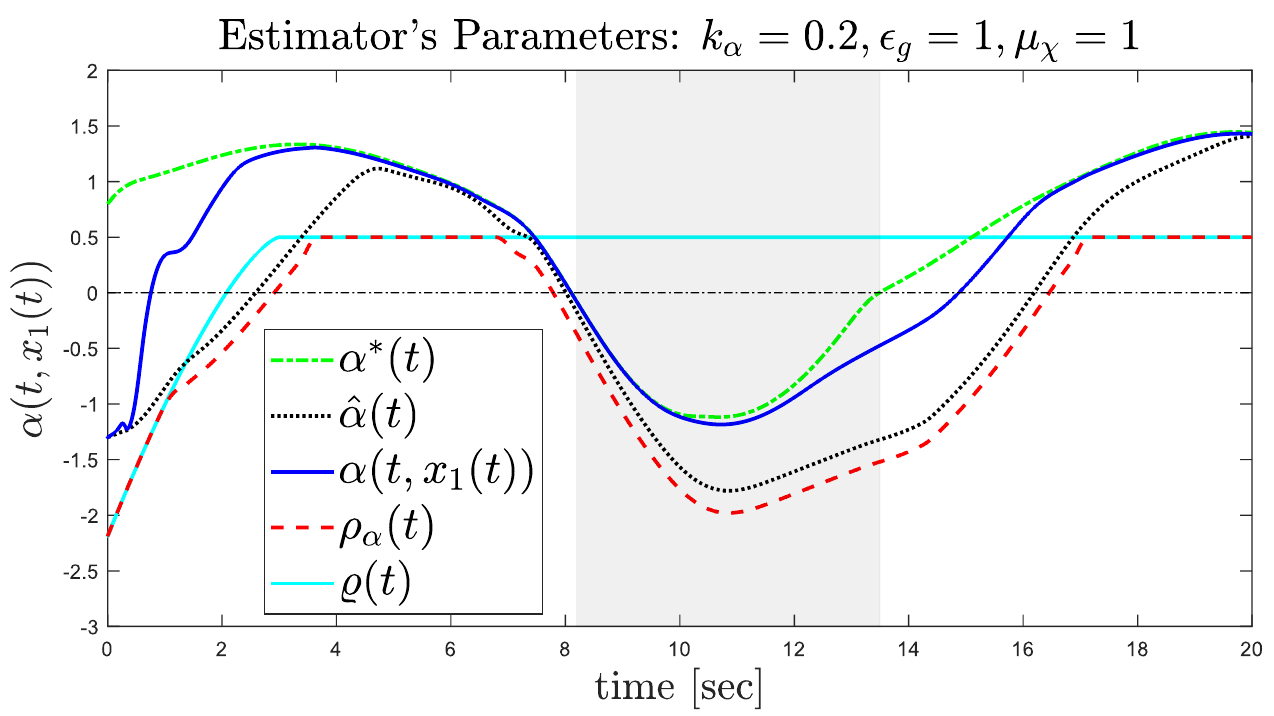}
		\caption{Smaller $\epsilon_g$  \vspace{0.1cm}}
		\label{fig:comparison_d}
	\end{subfigure}
	~
	\begin{subfigure}[t]{0.48\linewidth}
		\centering
		\includegraphics[width=\linewidth]{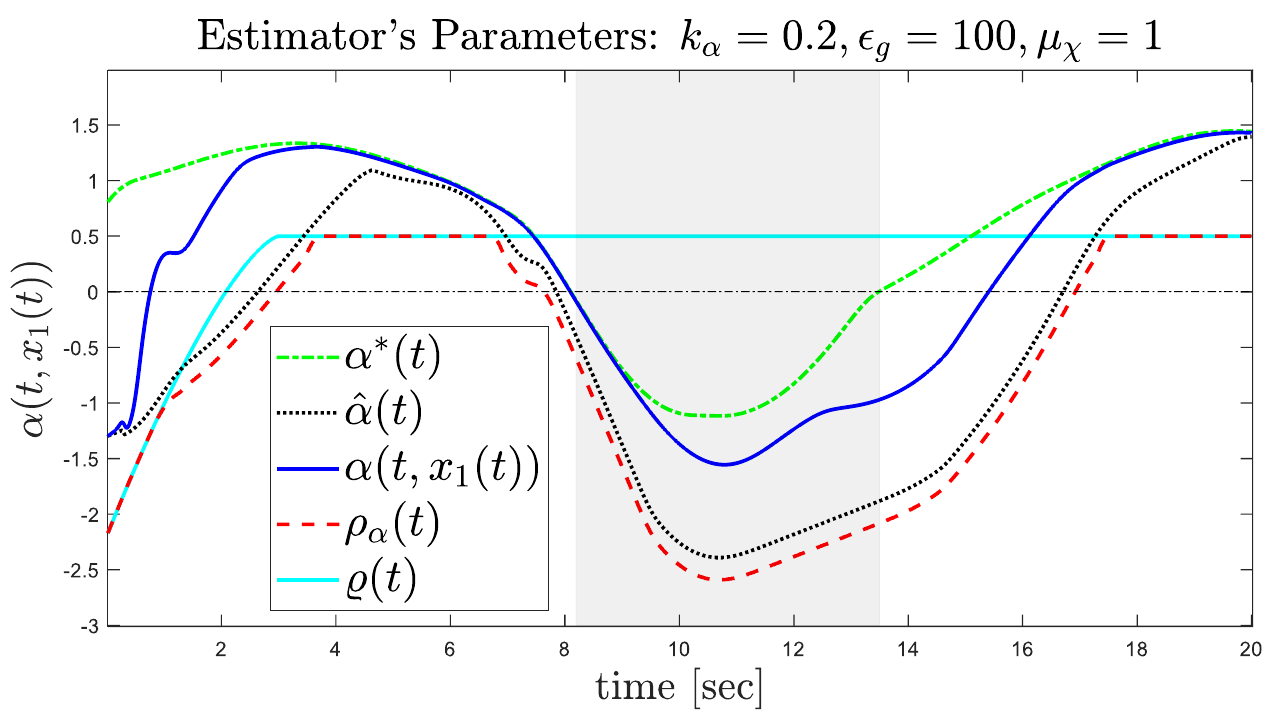}
		\caption{Larger $\epsilon_g$  \vspace{0.1cm}}
	\end{subfigure}
	
	\begin{subfigure}[t]{0.48\linewidth}
		\centering
		\includegraphics[width=\linewidth]{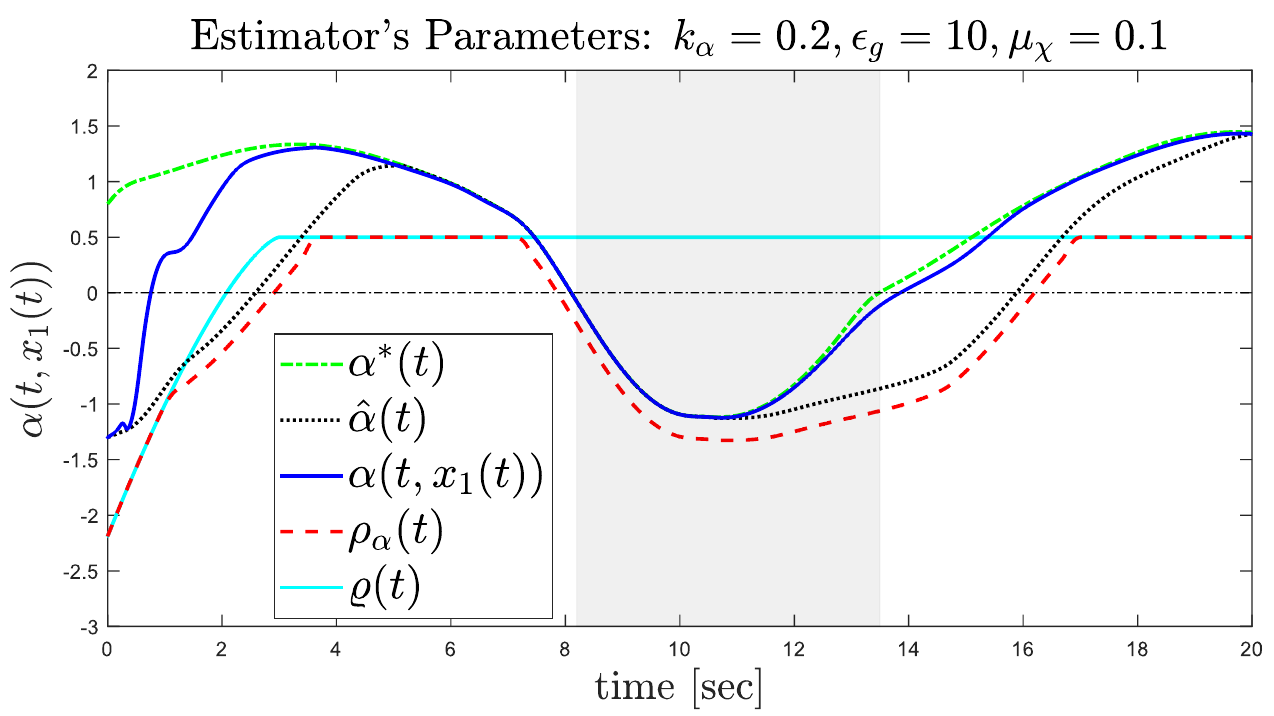}
		\caption{Smaller $\mu_\chi$}
		\label{fig:comparison_f}
	\end{subfigure}
	~
	\begin{subfigure}[t]{0.48\linewidth}
		\centering
		\includegraphics[width=\linewidth]{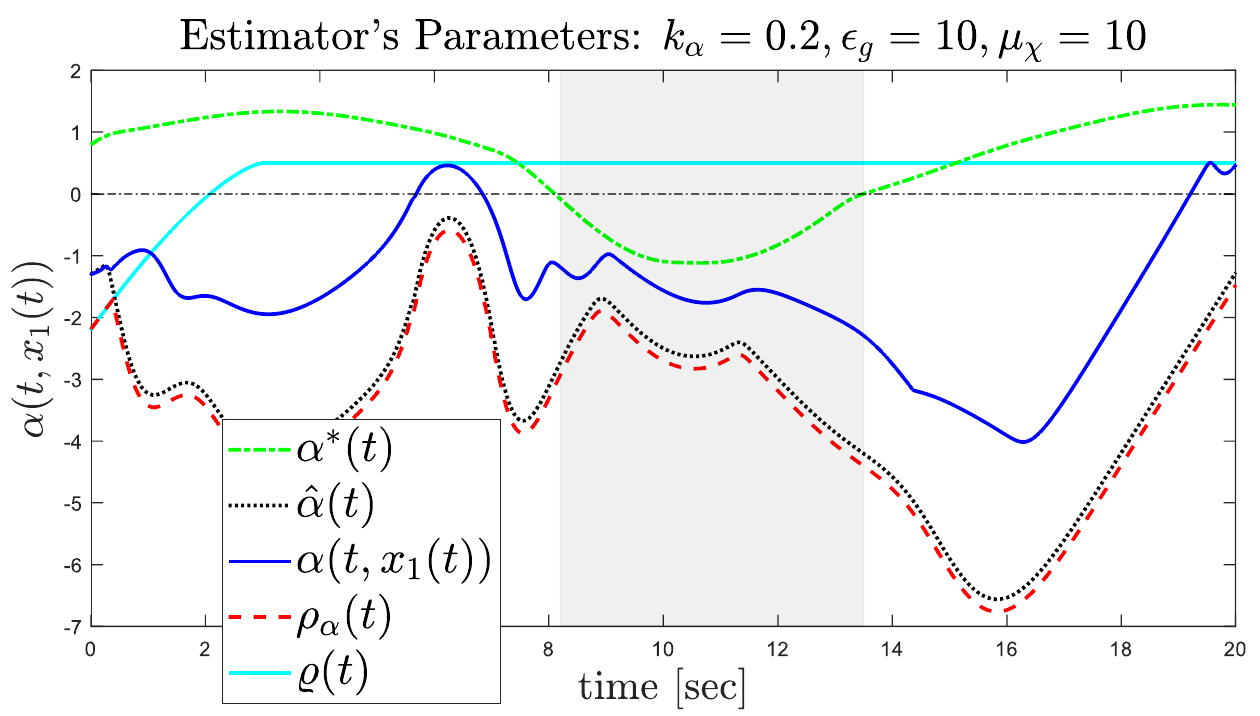}
		\caption{Larger $\mu_\chi$}
	\end{subfigure}
	\caption{Simulation results under various tuning of estimator parameters. \vspace{-0.4cm}}
	\label{fig:comparison}
\end{figure}

To further clarify the impact of individually adjusting each tuning parameter (\( k_{\alpha} \), \( \epsilon_g \), and \( \mu_\chi \)) on the estimator's performance, consider using the parameter settings from case (\textbf{B}) as the baseline (whose results are shown in Fig.\ref{fig:coupled2}). Fig.\ref{fig:comparison} displays the simulation outcomes for various parameter adjustments, where \( k_{\alpha} \), \( \epsilon_g \), and \( \mu_\chi \) are each increased and decreased independently, while all other parameters remain fixed at the baseline settings. As discussed in Section \ref{subsec: estimator}, increasing \( k_{\alpha} \) and decreasing \( \epsilon_g \) and \( \mu_\chi \) enhances the estimator’s performance. This improvement is evident by comparing Fig.\ref{fig:coupled2} (top) with Figs.\ref{fig:comparison_c}, \ref{fig:comparison_d}, and \ref{fig:comparison_f}. In particular, increasing \( k_{\alpha} \) and reducing \( \mu_\chi \) show a more pronounced enhancement in the estimation's performance . As noted earlier, improved estimation performance reduces the likelihood of unnecessary or overly conservative relaxations of the \( \ralph(t) \) in \eqref{eq:rho_lower_optim_schme}. This results in a more effective control for handling coupled time-varying output constraints, especially in situations where $\alphastr(t)$ can change rapidly and constraints may temporarily be infeasible.

\section{Conclusions}
\label{sec:conclu}

This work introduced a novel low-complexity feedback control design for high-order uncertain MIMO nonlinear systems with multiple (potentially coupled) time-varying output constraints. Our method addresses these constraints by interpreting their satisfaction as the fulfillment of a single consolidating constraint related to the signed distance with respect to the boundary of the time-varying constrained set. We have shown that by dynamically adjusting the lower bound of the consolidating constraint, our method ensures a least violating solution when the time-varying constraints become infeasible for an unknown time interval. Moreover, it overcomes the limitations of existing feedback control design approaches when dealing with coupled time-varying output constraints. Therefore, the developed control method can be applied to a broader range of applications. Future work includes applying this method to various applications and relaxing Assumption \ref{assu:alpha_globalmax}. 

\bibliographystyle{ieeetr}
\bibliography{Refs}

\appendices
\section{Proof of Lemma \ref{lem:Omegab_bounded}}
\label{appen:proof_lemma_Omegab_bounded}
To start with, notice that all $\ru_i(t), \rl_i(t)$ in \eqref{eq:predicate_const_rep} are bounded, and based on Assumption \ref{assum:output_map_elements}, $|h_i(t,x_1)| \leq \bh_i(x_1), i \in \I_1^m$ holds. Therefore, $\psi_i(t,x_1), i \in \I_1^{m+p}$ are bounded for all $t\geq 0$ and any fixed $x_1$. As a result, $\alphb: \Rpos \times \R^n \rightarrow \R$ in \eqref{eq:metric} is bounded for all $t\geq 0$ and any fixed $x_1$. Let $\alphb_{t}(x_1) \coloneqq \alphb(t,x_1)$. According to Assumption \ref{assum:coercive_alphabar}, $-\alphb_{t}(x_1)$ is coercive in $x_1$ for each $t$. Therefore, by \cite[Proposition 2.9]{Grippo2023intro}, all super-level sets $\alphb_{t}(x_1) \geq c$, where $c \in \R$, are bounded for each $t$. Furthermore, based on Assumption \ref{assum:coercive_alphabar} and \cite[Theorem 1.4.4, p. 27]{peressini1988mathematics}, we can infer that there exists a time-dependent constant $\bar{c}(t) \in \R$ such that the super-level sets $\alphb_{t}(x_1) > \bar{c}(t)$ are empty. This implies that $\Ob(t)$ in \eqref{eq:omega_alpha_bar} is bounded, and in particular, it is empty if $\bar{c}(t) < 0$ at time $t$. Moreover, from \eqref{smooth_alph_ineq} we can verify that $-\alphb(t,x_1)$ is coercive if and only if $-\alpha(t,x_1)$ is coercive. As a result, we can apply similar arguments as above to establish the boundedness of $\Omega(t)$.

\section{Proof of Lemma \ref{lem:alphb_coercive_h}}
\label{appen:proof_lemma_alphb_coercive_h}
Using \eqref{eq:metric}, we can determine whether $-\alphb(t,x_1)$ is coercive by verifying that, for each time instant $t$, at least one of the functions $\psi_i(t,x_1)$ in \eqref{eq:predicate_const_rep} approaches $-\infty$ as $\|x_1\| \rightarrow +\infty$ (along any path on $\R^n$). Note that $-\alpha(t,x_1)$ in \eqref{smooth_alph} is also coercive under the same condition. Since the functions $\psi_i(t,x_1), i \in \I_1^{m+p}$ in \eqref{eq:predicate_const_rep} are bounded for all $t\geq 0$ and any fixed $x_1$ (see proof of Lemma \ref{lem:Omegab_bounded}), we can interpret this requirement in terms of $h_i(t,x_1)$. Specifically, if there exists an $i \in \I_1^p$ such that $h_i(t,x_1) \rightarrow \pm \infty$ for each time instant $t$, then it ensures that $\psi_{2i-1}(t,x_1) \rightarrow -\infty$ or $\psi_{2i}(t,x_1) \rightarrow -\infty$ in \eqref{eq:predicate_funnel_rep} and vice versa. To simplify the verification process, we only need to check whether $\|h_f(t,x_1)\| \rightarrow +\infty$ for each time instant $t$ and along a path on $\R^n$ as $\|x_1\| \rightarrow +\infty$. From \eqref{eq:predicate_onesided_rep}, we can also see that if there exists a $j \in \I_{p+1}^{p+q}$ such that $h_j(t,x_1) \rightarrow - \infty$ for each time instant $t$, then there exists an $i \in \I_{2p+1}^{2p+q}$ such that $\psi_{i}(t,x_1) \rightarrow - \infty$ and vice versa. Similarly, if there exists a $j \in \I_{p+q+1}^{m}$ such that $h_j(t,x_1) \rightarrow + \infty$ for each time instant $t$, then there exists an $i \in \I_{2p+q+1}^{p+m}$ such that $\psi_{i}(t,x_1) \rightarrow - \infty$ and vice versa. In summary, if along any path on $\R^n$ as $\|x_1\| \rightarrow +\infty$ at least one of the conditions I-III in the lemma holds, then $-\alphb(t,x_1)$ (resp. $-\alpha(t,x_1)$) is coercive and vice versa. 

\section{Proof of Lemma \ref{lem:global_max_suffi}}
\label{appen:proof_lemma_globalmax}

\textbf{Case I:} Consider $\alpha(t,x_1)$ in \eqref{smooth_alph}. First, note that since $\psi_{i}(t,x_1), i \in \I_1^{m+p}$ are concave functions in $x_1 \in \R^n$ at time $t$ then as $\nu > 0$, $-\nu \, \psi_{i}(t,x_1), i \in \I_1^{m+p}$ are convex at time $t$. Moreover, from \cite[Section 3.5]{boyd2004convex} it is known that $e^{-\nu \, \psi_{i}(t,x_1)}, i \in \I_1^{m+p}$ are log-convex functions. Hence, $\sum_{i=1}^{m+p} e^{- \nu  \, \psi_i(t,x_1)}$ is log-convex. Consequently, $\alpha(t,x_1)$ in \eqref{smooth_alph} is a concave function at time $t$. Furthermore, since $\alpha(t,x_1)$ has bounded level sets, from Assumption \ref{assum:coercive_alphabar}, it attains a well-defined global maximum (i.e., the global maximum exists). Therefore, one can conclude that every critical point of $\alpha(t,x_1)$ is a global maximizer at time $t$.

\textbf{Case II:} Here, we first establish that under the given conditions $\alpha(t,x_1)$ attains only one critical point and then we show that the critical point is the (unique) global maximizer of $\alpha(t,x_1)$. Recall that the critical points of $\alpha(t,x_1)$ are obtained by solving $\gradxalph(t,x_1) = \mathbf{0}_n$. Given the assumed ordering of constraint types in \eqref{eq:predicate_const_rep} one can write $\alpha(t,x_1)$ in \eqref{smooth_alph} as follows:
\begin{flalign} \label{smooth_alph_alt} 
	&\alpha(t,x_1) = -\dfrac{1}{\nu} \ln \Big( \! \sum_{i=1}^{p}  e^{- \nu  \, ( h_i(t,x_1) - \rl_i(t) )} \!+ \! e^{- \nu  \, (\ru_i(t) - h_i(t,x_1))} \nonumber \\
	 & + \!\! \sum_{i=p+1}^{p+q} \! e^{- \nu  \, ( h_i(t,x_1) - \rl_i(t) )} + \!\!\! \sum_{i=p+q+1}^{m} \!\!\! e^{- \nu  \, (\ru_i(t) - h_i(t,x_1))} \Big). \!\!\!\!\!\!
\end{flalign}
Using \eqref{smooth_alph_alt} and \eqref{smooth_alph}, and after some calculations, we can obtain $\gradxalph(t,x_1)$ in a compact form as:
\begin{equation} \label{eq:grad_alph}
	\gradxalph(t,x_1) = J^{\top}(t,x_1) \, \gamma(t,x_1) \, e^{\nu \alpha(t,x_1)},
\end{equation}
where $J(t,x_1) = \frac{\partial h(t,x_1)}{\partial x_1} \in \R^{m \times n}$ is the Jacobian of $y = h(t,x_1)$, and $\gamma(t,x_1) \coloneqq  \col(\gamma_i(t,x_1)) \in \R^{m}$, in which $\gamma_i(t,x_1), i \in \I_1^m$ are given by: 
\begin{subnumcases}{\label{eq:gamma_vec}} 
	e^{- \nu  \, ( h_i(t,x_1) - \rl_i(t))} - e^{- \nu  \, \left(\ru_i(t) - h_i(t,x_1)\right)}, & \hspace{-0.33cm} $i \in \I_1^p$ \label{eq:gamma_i_funnel} \\
	e^{- \nu  \, ( h_i(t,x_1) - \rl_i(t))},  & \hspace{-0.35cm} $i \in \I_{p+1}^{p+q}$ \label{eq:gamma_i_lower} \\
	- e^{- \nu  \, \left(\ru_i(t) - h_i(t,x_1)\right)}, & \hspace{-0.35cm}  $i \in \I_{p+q+1}^m$ \label{eq:gamma_i_upper} 
\end{subnumcases}

Notice that in \eqref{eq:grad_alph} $e^{\nu \alpha(t,x_1)} > 0$, therefore, $\gradxalph(t,x_1) = \mathbf{0}_n$ if and only if $J^{\top}(t,x_1) \, \gamma(t,x_1) = \mathbf{0}_n$. If $m = n$ and each output constraint is a funnel constraint (i.e., $p = m = n$) then all $\gamma_i(t,x_1)$ will be given by \eqref{eq:gamma_i_funnel}. In this case, $J(t,x_1) \in \R^{n \times n}$ in \eqref{eq:grad_alph} is a square matrix. If at time instance $t$ we have $\rank(J) = n$ for all $x_1 \in \R^n$, then $J^{\top}(t,x_1) \, \gamma(t,x_1) = \mathbf{0}_n$ holds if and only if $\gamma(t,x_1) = \mathbf{0}_n$ at time $t$. Therefore, under the above conditions, at time instant $t$, we get $\gradxalph(t,x_1) = \mathbf{0}_n$ if and only if $\gamma_i(t,x_1)  = 0, \forall i \in \I_1^p$. Owing to \eqref{eq:gamma_i_funnel} this leads to having $h_i(t,x_1) = 0.5 (\ru_i(t) + \rl_i(t)), \forall i \in \I_1^p$, hence, we get the following system of nonlinear equations:
\begin{equation} \label{eq:nonlin_equation}
	F(t,x_1) \coloneqq h(t,x_1) - 0.5\left( \ru(t) + \rl(t) \right) = \mathbf{0}_n,
\end{equation}
with $\ru(t) \coloneqq \col(\ru_i(t)) \in \R^{n}$, $\rl(t) \coloneqq \col(\rl_i(t)) \in \R^{n}$. Now for each time instant $t$ define $F_t(x_1) \coloneqq F(t,x_1)$ and recall that in \eqref{eq:nonlin_equation} $0.5\left( \ru(t) + \rl(t) \right)$ is bounded for all time and the elements of $h(t,x_1)$ do not grow unbounded by the variation of $t$ (Assumption \ref{assum:output_map_elements}). We are interested in checking the existence and uniqueness of the solution to \eqref{eq:nonlin_equation} at each time instant $t$, which boils down to checking the existence and uniqueness of the solution to $F_t(x_1) = \mathbf{0}_n$ for each $t$. Since $h(t,x_1)$ is norm-coercive (i.e., $\|h(t,x_1)\| \rightarrow +\infty \quad \text{as} \quad \|x_1\| \rightarrow +\infty$) then $F_t(x_1)$ is norm-coercive as well. Moreover, from \eqref{eq:nonlin_equation} $F_t(x_1)$ has the same Jacobian matrix as $h(t,x_1)$, which is invertible by assumption. Consequently, all conditions of the global inverse function theorem \cite[Collorary]{wu1972global} are met, and thus $F_t(x_1)$ is a diffeomorphism at each time instant $t$. Therefore, $F_t(x_1) = \mathbf{0}_n$ or equivalently ${F}(t,x_1) = \mathbf{0}_n$ has a (single) unique solution $x_1^\ast(t)$ for each $t$, which is the unique critical point of $\alpha(t,x_1)$ at time $t$. Note that, since ${F}(t,x_1)$ is continuous, $x_1^\ast(t)$ depends continuously on time.

Next, we will show that the unique critical point of $\alpha(t,x_1)$ at time $t$, i.e., $x_1^\ast(t)$, is indeed the global maximum point of $\alpha(t,x_1)$ at time $t$. In this regard, we consider the second derivative test on the critical point's trajectory of $\alpha(t,x_1)$. From \eqref{eq:grad_alph} and followed by matrix differentiation rules \cite{van2010consistent}, we obtain the Hessian matrix of $\alpha(t,x_1)$, i.e., $\Hes(t,x_1) \coloneqq \frac{\partial}{\partial x_1} \left( \nabla_{x_1} \alpha(t,x_1) \right)$ as:
\begin{align}\label{eq:second_deriv_alph}
	&\resizebox{.8\hsize}{!}{$\Hes(t,x_1) =   \dfrac{\partial}{\partial x_1} \left(  J^\top(t,x_1) \right) \left( \left[ \gamma(t,x_1) \, e^{\nu \alpha(t,x_1)} \right] \otimes I_n  \right)$}   \\
	& \resizebox{.95\hsize}{!}{$+ J^\top(t,x_1) \;  \dfrac{\partial \, \gamma(t,x_1) }{\partial x_1} \,  e^{\nu \alpha(t,x_1)} + J^\top(t,x_1) \;   \gamma(t,x_1) \; \dfrac{\partial \, e^{\nu \alpha(t,x_1)} }{\partial x_1}.$} \nonumber 
\end{align}
Recall that on the critical point's trajectory we have $\gamma(t,x_1^\ast(t))  = 0$. Hence, evaluating \eqref{eq:second_deriv_alph} on $x_1^\ast(t)$ gives: 
\begin{flalign} \label{eq:hess_at_critical_point}
	&\resizebox{.92\hsize}{!}{$\Hes(t,x_1^\ast(t)) = J^\top(t,x_1^\ast(t)) \;  \left. \dfrac{\partial}{\partial x_1} \left(\gamma(t,x_1)\right) \right|_{x_1=x_1^\ast(t)}   e^{\nu \alpha(t,x_1^\ast(t))}.$}\!\!\!\!\!\!\!& 
\end{flalign}
From \eqref{eq:gamma_i_funnel}, one can get: 
\begin{equation}\label{gamma_x_at_critical_point}
	\left. \dfrac{\partial}{\partial x_1} \left(\gamma(t,x_1)\right) \right|_{x_1=x_1^\ast(t)} = \Gamma(t,x_1^\ast(t)) \;  J (t,x_1^\ast(t)),
\end{equation}
where $\Gamma(t,x_1^\ast(t)) \in \R^{n \times n}$ is a negative definite diagonal matrix whose diagonal entries are given by:
\begin{equation}
	-\nu \left[ e^{-\nu (h_i(t,x_1^\ast(t)) - \rl_i(t))} +  e^{-\nu ( \ru_i(t) - h_i(t,x_1^\ast(t)) )} \right].
\end{equation}  Therefore:
\begin{flalign} \label{eq:hessian_final}
	&\resizebox{.92\hsize}{!}{$ \Hes(t,x_1^\ast(t)) = J^\top(t,x_1^\ast(t))  \;  \Gamma(t,x_1^\ast(t)) \; J(t,x_1^\ast(t))   \;   e^{\nu \alpha(t,x_1^\ast(t))}.$} \!\!\!\!\! \!\!&
\end{flalign} 
Notice that $e^{\nu \alpha(t,x_1^\ast(t))}>0$ for all time. Since $J(t,x_1^\ast(t))$ is a full rank square matrix and $\Gamma(t,x_1^\ast(t))$ is a negative definite matrix, one can infer that the Hessian matrix $\Hes(t,x_1^\ast(t))$ is negative definite \cite{horn2012matrix}. Therefore, the critical point $x_1^\ast(t)$ is a local maximizer. Since $x_1^\ast(t)$ is the unique critical point of $\alpha(t,x_1)$ at time $t$, we conclude that $x_1^\ast(t)$ is indeed the (unique) global maximizer of $\alpha(t,x_1)$ at time $t$.\footnote{Note that under Condition II of Lemma \ref{lem:global_max_suffi}, $-\alpha(t,x_1)$ is radially unbounded in $x_1$ (see also Remark \ref{rem:boundedness_invexity} and Assumption \ref{assum:coercive_alphabar}). Hence, if $\alpha(t,x_1)$ admits a unique critical point, that point must be its global maximizer. This inference can be used in place of the presented Hessian-based analysis. The observation is particularly useful when $r=1$ in \eqref{eq:sys_dynamics_highorder} and $h(t,x_1)$ is only $\C^1$ in $x_1$, in which case the Hessian may not exist.}

\section{Proof of Theorem \ref{th:main}}
\label{appen:proof_theorem}

First, note that from \eqref{eq:intermediate_err} we have $x_2 = e_2 + s_1(t,x_1)$, $x_3 = e_3 + s_2(t,\xb_2)$ and $x_i = e_i + s_{i-1}(t,\xb_{i-1}), i \in \I_4^r$. Therefore, from \eqref{eq:normalized_err_vec} and with a slight abuse of notation, we can recursively obtain:
\begin{subequations} \label{eq:x_rewritten}
	\begin{align}
		x_2 &= \varTheta_2(t) \, \eh_2 + s_1(t,x_1), \label{eq:x_2_rewritten} \\
		x_3 &= \varTheta_3(t) \, \eh_3 + s_2(t,x_1, \eh_2), \label{eq:x_3_rewritten} \\
		x_i &= \varTheta_i(t) \, \eh_i + s_{i-1}(t,x_1,\eh_2,\ldots,\eh_{i-1}), \quad i \in \I_4^r.   
	\end{align}
\end{subequations}
From the system dynamics \eqref{eq:sys_dynamics_highorder} and \eqref{eq:x_2_rewritten} one can write:
\begin{align} \label{eq:x_1_dot}
	\dot{x}_1 &\coloneqq \phi_1(t,x_1,\eh_2) \\ 
	&= f_1(t,x_1) + G_1(t,x_1) \left(\varTheta_2(t) \, \eh_2 + s_1(t,x_1)\right).  \nonumber 
\end{align} 
Taking the time derivative of \eqref{eq:e_alph} and utilizing \eqref{eq:sys_dynamics_highorder} and \eqref{eq:x_2_rewritten} yields:
\begin{align} \label{eq:e_alph_dot}
	\dealph &\coloneqq   \phi_{\alpha}(t,x_1,\eh_2) = \frac{\partial \alpha(t,x_1)}{\partial x_1} \dot{x}_1 + \frac{\partial \alpha(t,x_1)}{\partial t} - \dralph(t) \nonumber\\
	& \!\!\! \! = \frac{\partial \alpha(t,x_1)}{\partial x_1} \Big[f_1(t,x_1) + G_1(t,x_1) \left(\varTheta_2(t) \, \eh_2 + s_1(t,x_1)\right) \Big] \nonumber \\
	&  \quad  + \frac{\partial \alpha(t,x_1)}{\partial t} - \dralph(t). 
\end{align} 
Moreover, differentiating \eqref{eq:normalized_err_vec} with respect to time and substituting equations \eqref{eq:sys_dynamics_highorder}, \eqref{eq:x_rewritten}, and \eqref{eq:control_law} results in: 
\begin{subequations}
	\begin{flalign}
		\dot{\eh}_2  &\coloneqq \phi_2(t,x_1,\eh_2, \eh_3) & \label{eq:dot_ehat_2} \\
		&= \varTheta_2^{-1}(t) \Big[ f_2(t,x_1,\eh_2) + G_2(t,x_1,\eh_2) \big(\varTheta_3(t) \, \eh_3 & \nonumber \\
		& \quad + s_2(t,x_1, \eh_2)\big) - \dot{s}_1(t,x_1) - \dot{\varTheta}_2(t) \eh_2 \Big], & \nonumber
	\end{flalign} 
	\begin{flalign}
		\dot{\eh}_i  &\coloneqq \phi_{i}(t,x_1,\eh_2,\ldots,\eh_i) & \\ 
		&= \varTheta_i^{-1}(t) \Big[ f_i(t,x_1,\eh_2,\ldots,\eh_i) + G_i(t,x_1,\eh_2,\ldots,\eh_i) & \nonumber \\ 
		& \quad  \times \big(\varTheta_{i+1}(t) \, \eh_{i+1} + s_i(t,x_1, \eh_2,\ldots,\eh_i) \big) \nonumber \\ 
		& \;- \dot{s}_{i-1}(t,x_1,\eh_2,\ldots,\eh_{i-1}) - \dot{\varTheta}_i(t) \eh_i \Big], \quad i \in \I_3^{r-1}, \nonumber
	\end{flalign}
	\begin{flalign}
		\dot{\eh}_r  &\coloneqq \phi_{r}(t,x_1,\eh_2,\ldots,\eh_r) & \\ 
		&= \varTheta_r^{-1}(t) \Big[ f_r(t,x_1,\eh_2,\ldots,\eh_r) + G_r(t,x_1,\eh_2,\ldots,\eh_r) & \nonumber \\ 
		& \; \times u(t,x_1,\eh_2,\ldots,\eh_r) - \dot{s}_{r-1}(t,x_1,\eh_2,\ldots,\eh_{r-1})  \nonumber \\
		& \quad - \dot{\varTheta}_r(t) \eh_r \Big], \nonumber 
	\end{flalign}
\end{subequations}

Next, define the following time-varying set:
\begin{equation}\label{eq:Omega_x}
	\Ox(t) \coloneqq \{ x_1 \in \R^n \mid   \alpha(t,x_1) > \ralph(t) \}.
\end{equation}
Owing to Assumption \ref{assum:coercive_alphabar}, $-\alpha(t,x_1)$ is coercive (see the proof of Lemma \ref{lem:Omegab_bounded}), and then from \cite[Proposition 2.9]{Grippo2023intro}, $\Ox(t)$ is bounded at each time instance $t$ ($\Ox(t)$ is a time-dependent super level-set of $\alpha(t,x_1)$). Therefore, one can infer that $\Ox(t)$ is bounded and open for all $t\geq 0$. In addition, notice that $\Ox(t)$ is nonempty for all $t \geq 0$, since by Property (i) of $\ralph(t)$ in Subsection \ref{subsec:single_const}, $\ralph(t) < \alphastr(t)$ holds for all time. Now define:
\begin{equation}
	\Oxs \coloneqq \bigcup_{t=0}^{+\infty} \Ox(t) \subset \R^n, 
\end{equation}  
which is the (time-invariant) super set containing $\Ox(t), \forall t \geq 0$. Owing to the properties of $\Ox(t)$ established above, $\Oxs$ is nonempty, bounded, and open. 

Now, let us define $z \coloneqq [x_1^\top, \ealph, \eh_2^\top, \ldots, \eh_r^\top]^\top \in \R^{nr+1}$ and consider the dynamical system:
\begin{equation} \label{eq:z_dynamics}
	 \dot{z} = \phi(t,z) \coloneqq
	 \begin{bmatrix}
	 	\phi_1(t,x_1,\eh_2) \\
	 	\phi_{\alpha}(t,x_1,\eh_2) \\
	 	\phi_2(t,x_1,\eh_2, \eh_3) \\
	 	\vdots \\
	 	\phi_{r}(t,x_1,\eh_2,\ldots,\eh_r)
	 \end{bmatrix},
\end{equation} 
as well as the (nonempty) open set:
\begin{equation}\label{eq:Omega_z}
	\Oz \coloneqq \Oxs \times (0,+\infty) \times \underset{(r-1)\mathrm{\;times}}{\underbrace{(-1,1)^n \times \ldots \times (-1,1)^n}}.
\end{equation}

In the sequel, we proceed in three phases. First, we show that there exists a unique and maximal solution $z: [0, \taum) \rightarrow \R^{nr+1}$ for \eqref{eq:z_dynamics} over the set $\Oz$ (i.e., $z(t;z(0)) \in \Oz, \forall t \in [0, \taum)$). Next, we prove that the proposed control scheme guarantees, for all $t \in [0, \taum)$: (a) the boundedness of all closed loop signals in \eqref{eq:z_dynamics} as well as that (b) $z(t;z(0))$ remains strictly within a compact subset of $\Oz$ for all $t \in [0, \taum)$, which leads by contradiction to $\taum = +\infty$ (i.e., forward completeness) in the last phase. Recall that, the latter means the signals $\ealph$ and $\eh_i, i \in \I_2^r$, remain within some strict subsets of $(0,+\infty)$ and $(-1,1)^n$, respectively, which in turn leads to the satisfaction of \eqref{eq:consuli_const} and \eqref{eq:i-th_inter_funnels}.

\textbf{\textit{Phase I.}} The set $\Oz$ is nonempty, open and independent of time. In addition, note that for a given initial condition $x(0)$ in \eqref{eq:sys_dynamics_highorder} we know $\ralph(0) < \alpha(0,x_1(0))$ holds by construction of $\ralph(t)$. Consequently, we have $x_1(0) \in \Oxs$ and from \eqref{eq:e_alph} one can verify that $\ealph(0,x_1(0)) \in (0, +\infty)$. Moreover, as mentioned at \textit{Step i-a} in Subsection \ref{subsec:control_design}, $\vartheta_{i,j}^0$ in \eqref{exponential_performance_fun} are selected such that $\vartheta_{i,j}^0 > |e_{i,j}(0,\bar{x}_{i}(0))|$, which ensures $\eh_{i,j}(0,\bar{x}_{i}(0)) \in (-1, 1)$ for all $j \in \I_1^n$ and $i \in \I_2^r$. Therefore, for all $i \in \I_2^r$ we have $\eh_i(0,\xb_i(0)) \in (-1,1)^n$. Overall, one can infer that $z(0) \in \Oz$. Additionally, recall that for all $i \in \I_1^r$, the system nonlinearities $f_i(t,\xb_i)$ and $G_i(t,\xb_i)$ are locally Lipschitz in $\xb_i$ and piece-wise continuous in $t$, and the intermediate control laws $s_i(t,\xb_i)$ and $u(t,x)$ are smooth. Consequently, one can verify that $\phi(t,z)$ on the right hand side of \eqref{eq:z_dynamics} is locally Lipschitz in $z$ over the set $\Oz$ and is piece-wise continuous in $t$. Therefore, the hypotheses of Theorem 54 in \cite[p.~476]{sontag1998mathematical} hold and the existence and uniqueness of a maximal solution $z(t;z(0)) \in \Oz$ for a time interval $t \in [0, \taum)$ is guaranteed.

\textbf{\textit{Phase II.}} We have proven in \textit{Phase I} that $z(t;z(0)) \in \Oz, \forall t \in [0, \taum)$, which implies:
\begin{equation*}
	\begin{cases}
		x_1(t;z(0)) \in \Oxs, \\
		\ealph(t;z(0)) \in (0,+\infty), \\
		\eh_i(t;z(0)) \in (-1,1)^n, \, \forall i \in \I_2^r,
	\end{cases} \quad \text{for all} \; t \in [0, \taum).
\end{equation*}
Therefore, $\epialph$ in \eqref{eq:mapped_alpha} and all $\epi_{i,j}$ in \eqref{eq:mapping_fun} (i.e., $\epi_{i} \in \R^n, i \in \I_2^r$) are well-defined for all $t \in [0, \taum)$.

Henceforth, for the sake of brevity, we omit dependencies in some of the notations when there is no ambiguity. Taking the time derivative of \eqref{eq:mapped_alpha} and using \eqref{eq:e_alph_dot} gives:
\begin{flalign}\label{eq:epialph_dot}
	\depialph = \frac{\partial \epialph}{\partial \ealph} \dealph &=  \frac{1}{\ealph} \Big[\frac{\partial \alpha}{\partial x_1} \Big( f_1(t,x_1) + G_1(t,x_1)& \nonumber \\
	& \hspace{-0.2cm} \times \left( \varTheta_2 \, \eh_2 + s_1(t,x_1) \right) \Big) + \frac{\partial \alpha}{\partial t} - \dralph \Big]. \hspace{-0.5cm}&
\end{flalign}
Moreover, differentiating $\epi_i \in \col(\epi_{i,j})$ with respect to time and using \eqref{eq:mapping_fun}, \eqref{eq:intermediate_err},  \eqref{eq:sys_dynamics_highorder}, and \eqref{eq:x_rewritten} results in:
\begin{flalign} \label{eq:epi_i_dot}
	\depi_i &= \Xi_i \Big[f_i(t,x_1,\eh_2,\ldots,\eh_i) + G_i(t,x_1,\eh_2,\ldots,\eh_i)& \nonumber \\
	& \quad \times \Big(\varTheta_{i+1} \, \eh_{i+1} + s_{i}(t,x_1, \eh_2,\ldots,\eh_i) \Big)&  \\
	& \quad - \dot{s}_{i-1}(t,x_1, \eh_2,\ldots,\eh_{i-1}) - \dot{\varTheta}_i \eh_i \Big], \quad i \in \I_2^{r-1}, \!\!\!\!\!\!& \nonumber
\end{flalign}
where for $i=r$, the term $\varTheta_{i+1} \, \eh_{i+1} + s_{i}$ should be replaced by $u$ in \eqref{eq:control_law}. Recall that, $\varTheta_i \coloneqq \diag(\vartheta_{i,j})$ and $\Xi_i \coloneqq \diag(\xi_{i,j})$, in which $\xi_{i,j}$ are given in \eqref{eq:xi_i,j}.

\textit{Step 1.} To ensure the satisfaction of \eqref{eq:consuli_const}, we are interested in establishing the boundedness of $|\epialph|$. We begin by considering the implicit upper bound property $\alpha(t,x_1) \leq \alphastr(t)$ stated in \eqref{eq:consuli_const}. Combining this property with \eqref{eq:e_alph} and \textit{Phase I}, we obtain:
\begin{equation} \label{eq:implicit_bound_ealph}
	\ealph(t) \in (0,b), \quad \forall t \in [0, \taum),
\end{equation}
where $b \coloneqq \sup_{\forall t\geq 0} (\alphastr(t) - \ralph(t)) > 0$. It is important to note that although $b$ can be arbitrarily large, it remains bounded due to the boundedness of $\alphastr(t)$ and $\ralph(t)$. Next, by examining \eqref{eq:mapped_alpha} and \eqref{eq:implicit_bound_ealph}, we observe that $|\epialph|$ can only grow unbounded when $\ealph(t) \rightarrow 0$ or equivalently when $\alpha(t,x_1(t;z(0))) \rightarrow \ralph(t)$. Note that Property (ii) of $\ralph(t)$ in Subsection \ref{subsec:single_const} ensures $\alphastr(t) - \ralph(t) \geq \varsigma > 0$ for all $t \geq 0$. Now, let us consider the following two cases: 

\emph{Case (1.a):} When $\ealph \in [\frac{\varsigma}{2},b)$ holds, from \eqref{eq:e_alph} and \eqref{eq:mapped_alpha}, it is evident that $|\epialph|$ is bounded by a positive constant $\epialphbo > 0$, which is given by:
\begin{equation} \label{eq:trivialbound}
	\epialphbo \coloneqq \max \left\{ \left| \ln \left( \frac{\varsigma}{2 \upsilon} \right) \right| , \left| \ln \left( \frac{b}{\upsilon} \right) \right| \right\}.
\end{equation}
Recall that according to Assumption \ref{assu:alpha_globalmax}, $\|\gradxalph(t,x_1)\| = \mathbf{0}_n$ if and only if $\alpha(t,x_1) = \alphastr(t)$. Therefore, $\|\gradxalph(t,x_1)\| = \mathbf{0}_n$ can only occur for values of $\ealph$ within the interval $[\varsigma,b)$. Consequently, even when $\|\gradxalph\| = \mathbf{0}_n$, the bound in \eqref{eq:trivialbound} still holds, which ensures the boundedness of $|\epialph|$. 

\emph{Case (1.b):} When $\ealph \in (0,\frac{\varsigma}{2})$ holds, due to the continuity of $\gradxalph(t,x_1)$, there exists a positive constant $\epsilon_{\alpha}$ such that $\|\gradxalph\| \geq \epsilon_{\alpha}$. Now consider the barrier function $V_1(\epialph) = \frac{1}{2} \epialph^2$ (introduced in Subsection \ref{subsec:control_design}) as a positive definite and radially unbounded Lyapunov function candidate with respect to $\epialph$. Taking the time derivative of $V_1$, substituting \eqref{eq:epialph_dot} and \eqref{eq:1st_intermed_ctrl_explicit}, and exploiting the fact that $G_1^s(t,x_1)$ is uniformly positive definite (see Assumption \ref{assum:symm_g}), we obtain:
\begin{align} \label{eq:lyap1}
	\dot{V}_1 &= \frac{\epialph}{\ealph} \left[\eta_1 + \frac{\partial \alpha}{\partial x_1} G_1(t,x_1) s_1(t,x_1) \right] \nonumber \\
	&= -k_1 \frac{\epialph^2}{\ealph^2} \gradxalph^\top G_1^s(t,x_1) \gradxalph + \frac{\epialph}{\ealph} \eta_1 \nonumber \\
	&\leq -k_1 \underline{\lambda}_1 \frac{|\epialph|^2}{\ealph^2} \|\gradxalph\|^2 + \frac{|\epialph|}{\ealph} |\eta_1| \nonumber \\
	&= - \frac{|\epialph|}{\ealph} \left[k_1 \underline{\lambda}_1 \|\gradxalph\|^2 \frac{|\epialph|}{\ealph}  - |\eta_1|  \right],
\end{align}
where 
\begin{equation*}
	\eta_1 \coloneqq \frac{\partial \alpha}{\partial x_1} \left(f_1(t,x_1) + G_1(t,x_1) \, \varTheta_2 \, \eh_2  \right) +  \frac{\partial \alpha}{\partial t} - \dralph.
\end{equation*}
In the following, we show that $|\eta_1|$ is bounded for all $t \in [0, \taum)$. Firstly, it is important to note that $|\dralph(t)|$ and $\|\varTheta_2(t)\|$ are bounded by construction for all time. Moreover, due to Assumptions \ref{assum:uncertain_f} and \ref{assum:symm_g}, we know that $\|f_1(t,x_1)\| \leq \bar{f}_1(x_1)$ and $\|G_1(t,x_1)\| \leq \bar{g}_1(x_1)$. Owing to the continuity of $\bar{f}_1(x_1)$ and $\bar{g}_1(x_1)$, and the fact that $x_1(t) \in \Oxs$ for all $t \in [0, \taum)$,  by employing the Extreme Value Theorem, we conclude that $\|f_1(t,x_1)\|$ and $\|G_1(t,x_1)\|$ are bounded for all $t \in [0, \taum)$. Similarly, under Assumptions \ref{assum:output_map_jacob} and \ref{assum:output_map_elements}, and by considering \eqref{eq:grad_alph} while acknowledging the continuity of $\alpha(t,x_1)$, and the boundedness of $\rl_i(t)$ and $\ru_i(t)$, we conclude that $\|\frac{\partial \alpha(t,x_1)}{\partial x_1}\|$ is bounded for all $t \in [0, \taum)$ using the Extreme Value Theorem. Likewise, by taking the time derivative of \eqref{smooth_alph_alt}, we can straightforwardly establish the boundedness of $|\frac{\partial \alpha(t,x_1)}{\partial t}|$ for $t \in [0, \taum)$ under Assumption \ref{assum:output_map_elements}. Lastly, we recall that $\eh_2 \in (-1,1)^n$ for all $t \in [0, \taum)$, as established in \textit{Phase I}. Consequently, considering all the arguments presented above, we conclude that, for all $t \in [0, \taum)$, there exists an unknown positive constant $\bar{\eta}_1$ such that $|\eta_1| < \bar{\eta}_1$. Now we can verify from \eqref{eq:lyap1} and $\|\gradxalph\| \geq \epsilon_{\alpha}$ that $\dot{V}_1$ is negative if:
\begin{equation}
	|\epialph| > \frac{\bar{\eta}_1 \, \varsigma}{2 \, k_1 \, \underline{\lambda}_1 \, \epsilon_{\alpha}^2 },
\end{equation}
and consequently:
\begin{equation} \label{eq:nontrivialbound}
	|\epialph(t)| \leq \epialphbt \coloneqq \max \left\{ |\epialph(0)| ,\frac{\bar{\eta}_1 \, \varsigma}{2 \, k_1 \, \underline{\lambda}_1 \, \epsilon_{\alpha}^2 } \right\}, \; \forall t \in [0, \taum).
\end{equation}

Now based on the results of Case (1.a) and Case (1.b),  combining \eqref{eq:trivialbound} and \eqref{eq:nontrivialbound} leads to:
\begin{equation} \label{eq:finalbound}
	|\epialph(t)| \leq \epialphb \coloneqq \max \{\epialphbo,\epialphbt\}, \; \forall t \in [0, \taum), \forall\ealph \in (0,b),
\end{equation}
where $\epialphb$ is independent of $\taum$. Furthermore, by taking the inverse logarithmic function in \eqref{eq:mapped_alpha} and utilizing \eqref{eq:finalbound}, we obtain:
\begin{equation} \label{eq:ealph_strictsubset}
	\upsilon e^{-\epialphb} \eqqcolon \ealphl \leq \ealph(t) \leq \ealphb \coloneqq \upsilon e^{\epialphb}, \quad \forall t \in [0, \taum).
\end{equation}
As a result, considering \eqref{eq:finalbound} and \eqref{eq:ealph_strictsubset} as well as the boundedness of $\|\gradxalph(t,x_1)\|$ for all $t \in [0, \taum)$, the first intermediate control signal $s_1$ in \eqref{eq:1st_intermed_ctrl_explicit} is well-defined (since $\ealph(t)$ remains strictly positive) and bounded for all $t \in [0, \taum)$. Additionally, using \eqref{eq:x_rewritten} we also conclude the boundedness of $x_2$ for all $t \in [0, \taum)$. Finally, differentiating $s_1(t,x_1)$ with respect to time and substituting  \eqref{eq:x_1_dot}, \eqref{eq:e_alph_dot}, and \eqref{eq:epialph_dot} yields:
\begin{flalign} \label{s_1_dot}
	\dot{s}_1 &= -k_1 \frac{\epialph}{\ealph} \Hes(t,x_1) \left[ f_1(t,x_1) + G_1(t,x_1) \left(\varTheta_2(t) \, \eh_2 + s_1 \right)\right]& \nonumber \\ 
	& \quad -k_1 \frac{\epialph}{\ealph} \frac{\partial}{\partial t} \big(\nabla_{x_1} \alpha \big) - k_1 \nabla_{x_1} \alpha(t,x_1) \frac{(1-\epialph)\dealph}{\ealph^2}, \!\!&
\end{flalign}
where $\Hes(t,x_1)$ denotes the Hessian of $\alpha(t,x_1)$. It is straightforward to deduce the boundedness of $|\dealph|$ for all $t \in [0, \taum)$ using \eqref{eq:e_alph_dot}. Furthermore, from \eqref{eq:grad_alph} and \eqref{eq:second_deriv_alph} and due to Assumptions \ref{assum:output_map_jacob} and \ref{assum:output_map_elements} as well as continuity of $\alpha(t,x_1)$, one can establish that $\|\Hes(t,x_1)\|$ and $\|\frac{\partial}{\partial t} \big(\nabla_{x_1} \alpha \big)\|$ are bounded for all $t \in [0, \taum)$. Consequently, since the boundedness of all other terms on the right-hand side of \eqref{s_1_dot} has already been proved for all $t \in [0, \taum)$, it can be concluded that $\dot{s}_1$ remains bounded for all $t \in [0, \taum)$.

\textit{Step 2.} Similarly to \textit{Step 1}, we can consider the barrier function $V_2(\epi_2) = \frac{1}{2} \epi_2^\top \epi_2$ as a positive definite and radially unbounded Lyapunov function candidate with respect to $\epi_2$. By taking the time derivative of $V_2$ and substituting \eqref{eq:epi_i_dot} and \eqref{eq:i_th_intermed_ctrl_explicit}, while also incorporating the fact that $G_2^s(t,x_1,\eh_2)$ is uniformly positive definite, we obtain the following expression:
\begin{align} \label{eq:lyap2}
	\dot{V}_2 &= \epi_2^\top \Xi_2 \left( \eta_2 + G_2(t,x_1,\eh_2) \, s_2(t,x_1,\eh_2) \right) \nonumber \\
	&= -k_2 \, \epi_2^\top \, \Xi_2 \,  G_2^s(t,x_1,\eh_2) \, \Xi_2 \, \epi_2  + \epi_2^\top \, \Xi_2 \, \eta_2 \nonumber \\
	&\leq -k_2 \, \underline{\lambda}_2 \, \|\epi_2\|^2 \, \|\Xi_2\|^2 + \|\epi_2\| \, \|\Xi_2\| \, \|\eta_2\| \nonumber \\
	&= -\|\epi_2\| \, \|\Xi_2\| \left( k_2 \, \underline{\lambda}_2 \, \|\epi_2\| \, \|\Xi_2\| -   \|\eta_2\|  \right),
\end{align}
where:
\begin{equation*}
	\eta_2 \coloneqq f_2(t,x_1,\eh_2) + G_2(t,x_1,\eh_2) \varTheta_3 \, \eh_3  -\dot{s}_1 - \dot{\varTheta}_2 \eh_2. 
\end{equation*}
Akin to the analysis provided in Step 1, under Assumptions \ref{assum:uncertain_f} and \ref{assum:symm_g}, and the application of the Extreme Value Theorem, it is straightforward to establish the existence of a positive (unknown) constant $\bar{\eta}_2$ such that $\| \eta_2 \| \leq \bar{\eta}_2$ for all $t \in [0, \taum)$. Furthermore, it was previously shown in Phase I that $\eh_2 \in (-1,1)^n, \forall t \in [0,\taum)$, which implies $\eh_{2,j} \in (-1,1), \forall t \in [0,\taum), \forall j \in \I_1^n$. Consequently, from \eqref{eq:xi_i,j} and \eqref{exponential_performance_fun} we deduce $\xi_{2,j} \geq \frac{2}{\vartheta_{2,j}^\infty} > 0$ for all $j \in \I_1^n$ and all $t \in [0,\taum)$. As a result, since $\Xi_2 = \diag(\xi_{2,j})$, there exists a positive constant $\epsilon_{\xi_2} \coloneqq \max_j|\frac{2}{\vartheta_{2,j}^\infty}|$ such that $\|\Xi_2\| \geq \epsilon_{\xi_2}, \forall t \in [0,\taum)$.

Now, considering \eqref{eq:lyap2} and the aforementioned facts, it is evident that $\dot{V}_2$ is negative under the condition:
\begin{equation}
	\|\epi_2\| > \frac{\bar{\eta}_2 }{k_2 \, \underline{\lambda}_2 \, \epsilon_{\xi_2} },
\end{equation}
which implies an upper bound on $\|\epi_2\|$ as follows:
\begin{equation} \label{eq:nontrivialbound_epi2}
	\|\epi_2(t)\| \leq \epibar_2 \coloneqq \max \left\{ \|\epi_2 (0)\| ,\frac{\bar{\eta}_2 }{ k_2 \, \underline{\lambda}_2 \, \epsilon_{\xi_2} } \right\}, \; \forall t \in [0, \taum),
\end{equation}
where $\epibar_2 > 0$ is independent of $\taum$. Moreover, taking the inverse of \eqref{eq:mapping_fun} and using the upper bound in \eqref{eq:nontrivialbound_epi2} reveals that:
\begin{equation} \label{eq:epi2_strictsubset}
	\!\!\! -1 < \tfrac{e^{-\epibar_2} -1}{e^{-\epibar_2} + 1} \eqqcolon -\sigma_{2,j}  \leq \eh_{2,j}(t) \leq \sigma_{2,j} \coloneqq \tfrac{e^{\epibar_2} -1}{e^{\epibar_2} + 1} < 1,
\end{equation}
for all $t \in [0, \taum)$ and all $j \in  \I_1^n$. By \eqref{eq:epi2_strictsubset} and \eqref{eq:xi_i,j}, it becomes evident that $\xi_{2,j}, j \in \I_1^n$ remain bounded for all $t \in [0, \taum)$. Consequently, considering \eqref{eq:nontrivialbound_epi2}, we can establish that the second intermediate control signal $s_2(t,x_1,\eh_2)$ in \eqref{eq:i_th_intermed_ctrl_explicit} remains bounded for all $t \in [0, \taum)$. Moreover, invoking \eqref{eq:x_3_rewritten} we also conclude the boundedness of $x_3$ for all $t \in [0, \taum)$. 

Finally, differentiating $s_2(t,x_1,\eh_2)$ with respect to time and substituting \eqref{eq:epi_i_dot} gives:
\begin{align} \label{s_2_dot}
	\dot{s}_2 &= -k_2 \, \dot{\Xi}_2 \, \epi_2 - k_2 \, \Xi_2 \, \depi_2  \nonumber \\
	&= -k_2 \, \dot{\Xi}_2 \, \epi_2 - k_2 \Big[f_2(t,x_1,\eh_2) + G_2(t,x_1,\eh_2)  \nonumber \\
	& \quad \; \times \Big(\varTheta_{3} \, \eh_{3} + s_{2}(t,x_1, \eh_2) \Big) - \dot{s}_{1} - \dot{\varTheta}_2 \eh_2 \Big].
\end{align}
Note that, by taking the time derivative of \eqref{eq:xi_i,j} one can obtain the diagonal elements of $\dot{\Xi}_i = \diag(\dot{\xi}_{i,j}), i \in \I_2^r$, as follows:
\begin{equation}\label{eq:xi_dot_ij}
	\dot{\xi}_{i,j} = -0.5 \, \xi_{i,j}^2 \, \dot{\vartheta}_{i,j} \, \left( 1 - 2 \, \eh_{i,j} \, \dot{\eh}_{i,j} \right), \quad j \in \I_1^n. 
\end{equation}
In particular, from \eqref{eq:xi_dot_ij} and \eqref{eq:dot_ehat_2} and using the aforementioned results, it is straightforward to infer the boundedness of $\xi_{2,j}, j =\I_1^n$. Accordingly, since the boundedness of all terms on the right-hand side of \eqref{s_2_dot} are already established for all $t \in [0, \taum)$, we conclude that $\dot{s}_2$ remains bounded for all $t \in [0, \taum)$.

\textit{Step i ($3 \leq i \leq r$).} Applying the same analysis described in Step 2 iteratively to the subsequent steps, while considering $V_i (\epi_i) = \frac{1}{2} \epi_i^\top \epi_i$, we can draw the following conclusion:
\begin{equation} \label{eq:nontrivialbound_epi_i}
	\|\epi_i(t)\| \leq \epibar_i \coloneqq \max \left\{ \|\epi_i (0)\| ,\frac{\bar{\eta}_i }{ k_i \, \underline{\lambda}_i \, \epsilon_{\xi_i} } \right\}, \; \forall t \in [0, \taum),
\end{equation}
in which $\epibar_i > 0$ is independent of $\taum$ and $\epsilon_{\xi_i} \coloneqq \max_j|\frac{2}{\vartheta_{i,j}^\infty}| > 0$, and there exist (unknown) constants $\bar{\eta}_i > 0, i \in \I_3^r$, which satisfy $\|\eta_i\| < \bar{\eta}_i, \forall t \in [0, \taum)$, where:
\begin{subequations}
	\begin{align}
		\eta_i \coloneqq& f_i(t,x_1,\eh_2,\ldots,\eh_{i}) + G_i(t,x_1,\eh_2,\ldots,\eh_{i}) \varTheta_{i+1} \, \eh_{i+1}  \nonumber \\
		&-\dot{s}_{i-1} - \dot{\varTheta}_i \eh_i, \quad i=\I_3^{r-1}, \\
		\eta_r \coloneqq& f_r(t,x_1,\eh_2,\ldots,\eh_{r}) -\dot{s}_{r-1} - \dot{\varTheta}_r \eh_r.
	\end{align}
\end{subequations}
Correspondingly, \eqref{eq:mapping_fun} and \eqref{eq:nontrivialbound_epi_i} also lead to:
\begin{equation} \label{eq:epi_i_strictsubset}
	-1 < \tfrac{e^{-\epibar_i} -1}{e^{-\epibar_i} + 1} \eqqcolon -\sigma_{i,j}  \leq \eh_{i,j}(t) \leq \sigma_{i,j} \coloneqq \tfrac{e^{\epibar_i} -1}{e^{\epibar_i} + 1} < 1,
\end{equation}
for $i \in \I_3^r, j \in \I_1^n$, and all $t \in [0, \taum)$. As a result, we can show that all intermediate control signals $s_i$ and system states $x_{i+1}, i=\I_3^{r-1}$, as well as the control law $u$ remain bounded for all $t \in [0, \taum)$.

\textbf{\textit{Phase III.}} Now we shall establish that $\taum = \infty$. In this direction, firstly, consider inequalities \eqref{eq:ealph_strictsubset}, \eqref{eq:epi2_strictsubset}, and \eqref{eq:epi_i_strictsubset}, and accordingly define: 
\begin{subequations}
	\begin{flalign}
		&\Omega_{\ealph}^\prime \coloneqq [\ealphl,\ealphb], \\
		&\Omega_{\eh_{i}}^\prime \coloneqq [-\sigma_{i,1}, \sigma_{i,1}] \times \ldots \times [-\sigma_{i,n}, \sigma_{i,n}], \quad  i \in \I_2^r,\!\!\!\!\!\!&  \\ 
		&\Omega_{\eh}^\prime \coloneqq \Omega_{\eh_{2}}^\prime \times \ldots \times \Omega_{\eh_{r}}^\prime \subset (-1,1)^n \times \ldots \times (-1,1)^n.\!\!\!\!\!\!& 
	\end{flalign}
\end{subequations}
In addition, owing to \eqref{eq:ealph_strictsubset}, from \eqref{eq:Omega_x} it is straightforward to infer that $x_1(t) \in \Ox^\prime(t) \subset \Ox(t)$ for all $t \in [0, \taum)$, where:
\begin{equation}
	\Ox^\prime(t) \coloneqq \{x_1 \in \R^n \mid  \ealphl \leq \alpha(t,x_1) - \ralph(t) \leq \ealphb \},
\end{equation}
from which we can define $\Oxsp \coloneqq \bigcup_{t=0}^{+\infty} \Ox^\prime(t) \subset \Oxs$ and claim that $x_1(t) \in \Oxsp, \forall t \in [0, \taum)$. Secondly, define $\Oz^\prime = \Oxsp \times \Omega_{\ealph}^\prime \times \Omega_{\eh}^\prime$, which is a nonempty and compact subset of $\Oz$ given in \eqref{eq:Omega_z}. Note that, from \eqref{eq:ealph_strictsubset}, \eqref{eq:epi2_strictsubset}, and \eqref{eq:epi_i_strictsubset} we have $z(t;z(0)) \in \Oz^\prime, \forall t \in [0, \taum)$. Now assuming a finite $\taum < \infty$, since $\Oz^{\prime} \subset \Oz$, Proposition C.3.6 in \cite[p.~481]{sontag1998mathematical} dictates the existence of a time instant $t^{\prime} \in [0,\taum)$ such that $z(t^\prime;z(0)) \notin \Oz^{\prime}$, which is a contradiction. Therefore, $\taum = \infty$. As a result, all closed-loop control signals remain bounded $\forall t \geq 0$. Finally, recall that since $\ealph(t) \in [\ealphl,\ealphb] \subset (0,+\infty)$ for all $t \geq 0$, invoking \eqref{eq:e_alph} ensures the satisfaction of the consolidating constraint in \eqref{eq:consuli_const} for all time, which completes the proof.

\section{Proof of Theorem \ref{th:refined_rho}}
\label{appen:proof_th_leastviolating}

We begin by establishing that $\ralph(t)$ given by \eqref{eq:rho_lower_optim_schme}, along with its derivative, remain bounded for all time. Next, we further show that $\ralph(t)$ attains Properties (i) and (ii) outlined in Subsection \ref{subsec:single_const}, which allows us to conclude that the specific design of $\ralph(t)$ in \eqref{eq:rho_lower_optim_schme} fulfills the prerequisites stipulated in Theorem \ref{th:main}, see Remark \ref{rem:feasibility_assumption_theorem}. Consequently, the proposed control law in \eqref{eq:control_law} effectively ensures the satisfaction of the consolidating constraint \eqref{eq:consuli_const}, as well as guaranteeing the boundedness of all closed-loop signals for all time.

Firstly, consider the dynamics of the estimation by taking the time derivative of \eqref{eq:first_order_gradaccent_out} and substituting \eqref{eq:first_order_gradaccent_dyn}:
\begin{equation} \label{eq:alpha_hat_dyn}
	\dot{\alphah} = \frac{\partial \alpha}{\partial t} \!+\! k_{\alpha} \| \gradxtilalph \|^2 \!- \!\frac{\| \gradxtilalph \|^2   }{\| \gradxtilalph \|^2 + \epsilon_g \chi(\| \gradxtilalph \|)}  \frac{\partial \alpha}{\partial t}.\!
\end{equation} 
Recall that $\alphah(t)$ is upper-bounded by its maximum value, i.e., $\alphah(t) =  \alpha(t,\xtil_1(t)) < \alphastr(t) = \alpha(t,\xtil_1^\ast(t))$, where $\xtil_1^\ast(t)$ represents the time-varying optimum of $\alpha(t,\xtil_1)$. Therefore, to ensure the boundedness of $\alphah(t)$, we only need to show that it is lower-bounded. Under Assumption \ref{assu:estimation_technical}, outside of the compact set $\Og$, i.e., when $\| \gradxtilalph \| > \mch$, the right-hand side of \eqref{eq:alpha_hat_dyn} reduces to: $
\dot{\alphah} = k_{\alpha} \| \gradxtilalph \|^2$, which is strictly positive. Therefore, within the set $\R^n / \Og$, $\alphah(t)$ is increasing and thus $\alphah(t)$ does not approach $-\infty$, meaning that $\alphah(t)$ is lower-bounded. On the other hand, inside the compact set $\Og$, the right-hand side of \eqref{eq:alpha_hat_dyn} is generally sign-indefinite, so $\alphah(t)$ may either decrease or increase. However, since $\alphah(t,\xtil_1)$ is continuous and $\Og$ compact, $\alphah(t,\xtil_1(t))$ cannot approach $-\infty$ for all $\xtil_1(t) \in \Og$. As a result, we conclude that $\alphah(t) = \alpha(t,\xtil_1(t))$ remains bounded for all time. Moreover, owing to the compactness of the level curves of $\alpha(t,\xtil_1)$ and the boundedness of $\alphah(t)$, we conclude that $\xtil_1(t)$ remains bounded for all time. 

Taking the time derivative of \eqref{eq:rho_lower_optim_schme} gives: 
$\dralph(t) = \dot{\iota}(t) \left( \varrho(t) - \alphah(t) + \mu \right) + \iota(t) \dot{\varrho}(t)  + (1-\iota(t)) \dot{\hat{\alpha}}(t)$. Note that $\iota(t)$, $\varrho(t)$, $\dot{\varrho}(t)$, and $\alphah(t)$ are all bounded. Additionally, from \eqref{smooth_switch_fun}, it can be seen that $\dot{\iota}(t)$ is bounded if $\dot{\hat{\alpha}}(t)$ is bounded. Therefore, the boundedness of $\dralph(t)$ is ensured by establishing the boundedness of $\dot{\hat{\alpha}}(t)$ in \eqref{eq:alpha_hat_dyn}. Under Assumptions \ref{assum:output_map_jacob} and \ref{assum:output_map_elements}, and the boundedness of $\ru_i(t)$, $\rl_i(t)$, $\drl_i(t)$, and $\dru_i(t)$ in \eqref{eq:predicate_const_rep}, it can be deduced that for any fixed $\xtil_1$, the continuous functions $\frac{\partial \alpha(t,\xtil_1)}{\partial t}$ (given in \eqref{eq:partial_alpha_partial_time}) and $\| \gradxtilalph(t,\xtil_1) \|$ remain bounded for all time. Therefore, owing to the boundedness of $\xtil_1(t)$, we can deduce that $\dot{\hat{\alpha}}(t)$ also remains bounded for all time, concluding the boundedness of $\dralph(t)$.

Secondly, recall that $\alphah(t) = \alpha(t,\xtil_1) \leq \alphastr(t)$ always holds. Now for the case that $\iota(t) = 0$ from \eqref{eq:rho_lower_optim_schme} we have $\ralph(t) = \alphah(t) - \mu$, and thus $\alphastr(t) - \ralph(t) \geq \mu$. In addition, when $\iota(t) = 1$ from \eqref{eq:rho_lower_optim_schme} we get $\ralph(t) = \varrho(t)$ and from \eqref{smooth_switch_fun} it also holds that $\varphi = \alphah - \varrho(t) > \mu$. Hence, one can verify that $\alphastr(t) - \ralph(t) = \alphastr(t) - \varrho(t)  > \alphastr(t) + \mu - \alphah(t) \geq \mu$. When $\iota(t) \in (0,1)$, from \eqref{smooth_switch_fun} we know that $ 0 \leq \varphi(t) \leq \mu$, from which we get $0 \leq \alphah(t) - \varrho(t) \leq \mu$. Now from \eqref{eq:rho_lower_optim_schme} and under the worst case scenario that is $\alphastr(t) = \alphah(t)$ we obtain:
\begin{align*}
	\alphastr(t) - \ralph(t) &= \alphastr(t) - \iota(t) \varrho(t) - (1 -\iota(t)) ( \alphah(t) - \mu) \\
	&\geq  \iota(t) (\alphastr(t) - \varrho(t)) + (1 -\iota(t)) \mu \\
	& > (1 -\iota(t)) \mu.
\end{align*}      
Consequently, one can infer that for any value $\iota \in [0,1]$ there must exist a constant $\varsigma$ ($0 < \varsigma \leq \mu$) such that $\alphastr(t) - \ralph(t) \geq \varsigma > 0$ for all $t \geq 0$. Hence, Property (i) in Subsection \ref{subsec:single_const} holds for $\ralph(t)$ given by \eqref{eq:rho_lower_optim_schme}. 

Finally, if $\varrho_0 < \alpha(0,x_1(0))$ in \eqref{eq:alpha_lower_bound_nomi}, one can ensure that $\ralph(0) < \alpha(0,x_1(0))$ holds (i.e., Property (ii) in Subsection \ref{subsec:single_const} holds) for any initialization $\xtil_1(0)$ in \eqref{eq:first_order_gradaccent}. To this end, assume $\varrho(0) = \varrho_0 < \alpha(0,x_1(0))$ and consider $\xtil_1(0)$ is such that: (a) $\varphi(0) = \alphah(0) - \varrho(0) > \mu$, (b) $0 \leq \varphi(0) \leq \mu$, and (c) $ \varphi(0) < 0$. For case (a), from  \eqref{eq:rho_lower_optim_schme} and \eqref{smooth_switch_fun} it is obvious that  $\ralph(0) = \varrho(0) < \alpha(0,x_1(0))$. Considering case (b) since $\varrho(0) - \mu \leq \ \alphah(0) - \mu \leq \varrho(0)$ and $0 \leq \iota(0) \leq 1$ one can infer that the convex combination $\ralph(0) = \iota(0) \varrho(0) + (1 -\iota(0)) ( \alphah(0) - \mu)$ can only take a value less than or equal to $\varrho(0)$, hence, we get $\ralph(0) < \alpha(0,x_1(0))$. For case (c) it is straightforward to verify that $\ralph(0) = \alphah(0) - \mu < \varrho(0) - \mu < \alpha(0,x_1(0))$. Therefore, Property (ii) in Subsection \ref{subsec:single_const} holds for $\ralph(t)$ given by \eqref{eq:rho_lower_optim_schme}.

Overall, owing to the above analysis $\ralph(t)$ in \eqref{eq:rho_lower_optim_schme} satisfies the conditions of Theorem \ref{th:main}, thereby, applying the control law \eqref{eq:control_law} in \eqref{eq:sys_dynamics_highorder} leads to the satisfaction of $\ralph(t) < \alpha(t,x_1(t;x(0)))$, as well as boundedness of all closed-loop signals for all time.

\end{document}